\documentclass[fleqn,usenatbib]{rasti}
\usepackage{newtxtext,newtxmath}

\usepackage[T1]{fontenc}

\DeclareRobustCommand{\VAN}[3]{#2}
\let\VANthebibliography\thebibliography
\def\thebibliography{\DeclareRobustCommand{\VAN}[3]{##3}\VANthebibliography}

\usepackage{graphicx}	
\usepackage{amsmath}	

\usepackage{tikz}				
\usepackage{tikz-imagelabels} 	

\newcommand{\Msun}{\ensuremath{~{\rm M}_\odot}}                   
\newcommand{\Teff}{\ensuremath{T_{\rm eff}}}                      
\newcommand{\logg}{\ensuremath{\log g}}                           
\newcommand{\corot}{CoRoT}
\newcommand{\kepler}{{\it Kepler}}
\newcommand{\ktwo}{K2}

\newcommand{\gaia}{{\it Gaia}}
\newcommand{\cheops}{CHEOPS}
\newcommand{\tess}{TESS}
\newcommand{\plato}{PLATO}
\newcommand{\ebop}{{\sc ebop}}
\newcommand{\jktebop}{{\sc jktebop}}
\newcommand{\python}{{\tt python}}
\newcommand{\skl}{{\tt Scikit-Learn}}
\newcommand{\tensorflow}{{\tt TensorFlow}}
\newcommand{\keras}{{\tt Keras}}

\newcommand{\astropy}{{\tt Astropy}}
\newcommand{\lightkurve}{{\tt LightKurve}}
\newcommand{\numpy}{{\tt NumPy}}
\newcommand{\uncertainties}{{\tt Uncertainties}}
\newcommand{\ebopmaven}{{\sc ebop maven}}

\newcommand{\rAplusrB}{\ensuremath{r_{\rm A}+r_{\rm B}}}
\newcommand{\ecosw}{\ensuremath{e\cos{\omega}}}
\newcommand{\esinw}{\ensuremath{e\sin{\omega}}}
\newcommand{\bP}{\ensuremath{b_{\rm P}}}
\newcommand{\qphot}{\ensuremath{q_{\rm phot}}}

\newcommand{\reff}[1]{{#1}}

\title[Machine learning estimation of dEB parameters]{EBOP MAVEN: A machine learning model to estimate the input parameters for analytic fitting of detached eclipsing binary light curves}

\author[S. Overall and J. Southworth]{
	Stephen Overall,$^{1}$\thanks{E-mail: s.p.overall@keele.ac.uk}
	John Southworth$^{1}$
	\\
	$^{1}$Astrophysics Group, Keele University, Staffordshire, ST5 5BG, UK
}

\date{Accepted XXX. Received YYY; in original form ZZZ}

\pubyear{\the\year{}}

\begin{document}
	\label{firstpage}
	\pagerange{\pageref{firstpage}--\pageref{lastpage}}
	\maketitle
	
	\begin{abstract}
		Detached eclipsing binary stars (dEBs) are a key source of data on fundamental stellar parameters. \reff{Within the light curve databases of survey missions such as \kepler\ and \tess\ are a wealth of new systems awaiting characterisation. We aim to improve the scalability of efforts to process these data by developing a Convolutional Neural Network (CNN) machine learning model to assist in the automation of their analysis. From a phase-folded and binned dEB light curve the model predicts system parameters relating to stellar fractional radii, orbital inclination and eccentricity, and the stellar brightness ratio, for use as input values in subsequent formal analysis with the established \jktebop\ analytic code. We find the model able to predict these parameters for a previously unseen test dataset of 20\,000 synthetic dEB systems with a mean error of 14.1\% when compared with the label values, improving to 8.6\% against a subset representative of real systems. When tested with the \tess\ light curves of a set of real well-characterised systems, the model's predictions yield a mean error of $8.7\pm0.7\%$ when compared with label values derived from existing published analyses. Subsequent fitting of the \tess\ light curves with the \jktebop\ analytic code while using the model predictions as input values finds 27 of the 28 systems achieving a good fit. On the strength of these results, we plan to build a new characterisation pipeline based on the machine learning model and \jktebop\ code with the intention of producing a target catalogue of dEB systems for potential observation with the forthcoming \plato\ mission.}
	\end{abstract}
	
	
	
	
	\section{Introduction}
	Eclipsing binary stars are a key source of stellar parameters. With a combination of a photometric light curve,  spectroscopic radial velocity (RV) observations and relatively simple mathematics, it is possible to measure the component stars’ masses, radii and orbital parameters to great precision \citep{Russell12apjI, Hilditch01book}. Subsequently the stars' luminosity and absolute magnitudes are easily calculated from the radii and effective temperature measurements taken from spectroscopic observations. Detached eclipsing binaries (dEBs) are those where the orbital separation is sufficient for the components to have evolved independently from a common origin. These systems present an ideal source of fundamental stellar parameters for the construction and ongoing refinements of stellar models \citep{Andersen91aarv, TorresAndersen+10aarv, Southworth21univ}.
	
	Beyond their contribution to stellar modelling, dEBs have been used more directly to investigate the structure and evolution of stars. They have been used in investigations into radius inflation in low-mass main sequence stars \citep{SpadaDemarque+13apj, SwayneMaxted+21mnras, JenningsSouthworth+23mnras}, the mass discrepancy in high-mass stars \citep{TkachenkoPavlovski+20aa}, convective core overshooting \citep{ClaretTorres16aa, ClaretTorres18apj}, the mixing length \citep{Kirkby-KentMaxted+16aa, GraczykSmolec+16aa} and star spots \citep{WangFu+21mnras:starspots, WangFu+22mnras:starspots}. Their usefulness as test subjects for stellar theory is expanded further when one or more components exhibit intrinsic variability, such as stochastic low-frequency (SLF) variability \citep{SouthworthBowman22mnras} or $\beta$ Cepheid \citep{SouthworthBowman+21mnras}, $\delta$ Scuti \citep{ChenDing+22apjs,JenningsSouthworth+24mnras} and $\gamma$ Doradus \citep{SouthworthVanReeth22mnras} pulsations, as these give insights into the star's internal structure. They also make a useful contribution to the cosmological distance scale, where the well-defined luminosity of dEBs can be used to determine distances to high precision \citep{PietrzynskiGraczyk+19nat:distance}.
	
	The number of binary systems available for characterisation has expanded greatly with the advent of recent space-based photometric exoplanet surveys whose sensitivity make them highly successful at detecting other forms of stellar variability. Missions such as \corot\ (Convection, Rotation and planetary Transits; \citealt{AuvergneBodin+09aa, Deleuil+18aa:CoRoT}), \kepler\ (\citealt{Borucki+16rpf:Kepler}), \ktwo\ \citep{Howell+14pasp:K2} and \tess\ (Transiting Exoplanet Survey Satellite; \citealt{Ricker+15jatis:Tess}) have \reff{published vast photometric time-series datasets which may contain many previously undetected eclipsing binary systems (e.g. \citealt{PrsaKochoska+22apjs, MowlaviHoll+23aa, KostovPowell+25apjs}).} The \tess\ dataset is particularly fertile as its initial mission to perform a whole sky survey over two years will have allowed it to potentially capture a vast number of bright, short-period eclipsing systems. This will be expanded further when ESA's \plato\ (PLAnetary Transits and Oscillations of stars; \citealt{RauerAerts+24}) mission is launched which, while not a whole sky survey like \tess, will be able to capture longer-period variability than \tess\ over a larger field of view than \kepler.
	
	Other missions of interest include ESA \gaia\ \citep{GaiaCollab16aa:Gaia} which is publishing a vast dataset of astrometric measurements of previously unmatched precision, complemented by photometry and spectroscopy of selected targets. \cheops\ (CHaracterising ExOPlanets Satellite; \citealt{BenzBroeg+21expastron}) is another ESA mission whose main science goal is to improve the characterisation of known exoplanets, however this has not prevented its use for dEB systems (e.g. \citealt{SwayneMaxted+21mnras}).
	
	Recent years have witnessed a growing use of Machine Learning (ML) techniques, often as a means to address the vast volumes of data that are produced by modern systematic survey programmes. In describing the concept of computer learning, \cite{Mitchell97mlbook} states that "a computer program is said to learn from experience E with respect to some task T and some performance measure P, if its performance on T, as measured by P, improves with experience E". As a way of interacting with data it differs from a traditional approach as a ML model's algorithm adapts to a given task based on its experience gained from the data, or a representative subset, rather than requiring the researcher to predefine the algorithm to be used.
	
	Broadly, machine learning algorithms fall into two main categories, known as unsupervised and supervised learning. Unsupervised algorithms let "the data speak for themselves" \citep{BallBrunner10ijmp} in that they derive information directly from the source data without the need for prior training. This approach finds uses in clustering, where objects are assigned to clusters or groups based on similarities in shared features (e.g. \citealt{Garcia-Dias+18aa:kmeans, Castro-Ginard+18aa:DBSCAN, Prisinzano+22aa:DBSCAN}), and conversely in anomaly or outlier detection, where objects are highlighted by their dissimilarity to other objects (e.g. \citealt{GilesWalkowicz19mnras, GilesWalkowicz20mnras}). Another common task is dimensionality reduction (e.g. \citealt{Matchev+22psj:PCA}) which has uses in compression, data visualisation (such as projecting high-dimensional data onto a 2-d plane for printing) and in preparing data for supervised algorithms where fewer features may allow the use of a simpler model.
	
	Supervised algorithms are trained on known representative datasets accompanied by curated "ground truth" \citep{Baron19} values, known as labels, for those features which they are being trained to predict. The training is an iterative process whereby the model optimises its internal configuration to better achieve the results indicated by the labels. Supervised learners are generally used to address two main problem types: classification and regression. A classification problem is one which assigns previously unseen objects to a discrete class based on measured features, for example stars may be classified based on their light curve features (e.g. \citealt{BarbaraBedding+22mnras}) or their \gaia\ astrometry data (e.g. \citealt{TorresCantero+19mnras}). A regression problem is one where output variables are expected to represent continuous data, for example the determination of stellar rotation periods from \kepler\ observations (e.g. \citealt{BretonSantos+21aa}) or the proposed prediction of dEB stellar parameters from \tess\ light curves discussed here.
	
	Further applications of machine learning techniques to binary systems include \cite{Pashchenko+18mnras} who evaluated a range of classification algorithms applied to features extracted from the light curves of various classes of variable stars including eclipsing binary systems. A regression neural network was used in the compilation of the \tess\ Eclipsing Binary Star catalogue (TESS-EBS; \citealt{PrsaKochoska+22apjs}) to assign each member a numeric morphology parameter to indicate how widely detached the system is \citep{MatijevicPrsa+12aj}. Most recently, \cite{WangDing+24apjs} and \cite{WronaPrsa25apjs} have separately employed forward modelling neural networks as a means of reducing the computational cost of generating the vast number of EB light curves required for formal posterior analysis.
	
	Finally, we discuss the Eclipsing Binaries via Artificial Intelligence (EBAI) project of \cite{PrsaGuinan+08apj} who developed a regression neural network to predict EB parameters from light curve data. This was integrated into a larger processing pipeline, with its predictions being used as input parameters for the subsequent formal fitting stage with the PHysics of Eclipsing BinariEs fitting code (PHOEBE; \citealt{PrsaZwitter05apj}). Phase-folded light curve data were pre-processed with a newly developed smoothing algorithm, known as {\tt polyfit}, from which the neural network's input data were sampled at 200 equidistant phase points. With these input data the model would predict system parameters covering the ratio of the components' effective temperature, the sum of their fractional radii, and components of the system's orbital eccentricity and inclination, with test results accurate to within 10\% across 90\% of a synthetic test dataset.
	
	In this work we seek to follow a similar strategy to the EBAI project, specifically to aid the analysis of \tess\ light curves when fitted with the \jktebop\footnote{\url{https://www.astro.keele.ac.uk/jkt/codes/jktebop.html}} code \citep{Southworth+04a:mnras:jktebop}. \jktebop\ is an extensive development of the \ebop\ code of P.\ B.\ Etzel \citep{NelsonDavis72apj:ebop, PopperEtzel81apj:ebop}, where the stars are modelled as biaxial spheroids between eclipses and simplified to overlapping spheres during eclipses \citep{Southworth+04a:mnras:jktebop}. Models are iteratively optimised, with the Levenberg-Marquardt algorithm (MRQMIN; \citealt{Press+92book}) used to perform a least squared fit of the phase-folded light curve. The strength of the code's approach is that it is computationally inexpensive and will quickly minimise the residuals to find the best fitting model, however the likelihood of this being the global minimum and therefore the optimum model are greatly improved if the starting position is nearby within the parameter space. This requirement for an informed starting position stands in the way of using the code at scale in the characterisation of large datasets, and is one we seek to address by training a machine learning model to set the starting position parameters from minimally processed light curve data. The development of this model, which we have named \ebopmaven\ (Model Analysis input Value Estimation Neural network), is a key step in an ongoing project to produce a catalogue of potential dEB targets for future observation with the \plato\ mission.
	
	\section{Method}
	
	\subsection{Parameter selection}
	\reff{The set of predicted parameters was chosen to be those required for input to \jktebop. These are closely related to features within light curves so are suitable for training a machine learning model to make reliable predictions.} It is not possible to determine absolute values for the size and separation of the components of a dEB system from its light curve alone and fitting codes respond to this by fitting for relative values. With \jktebop\ the quantities fitted are the fractional radii of the two stars, defined as their absolute radii in units of the system's semi-major axis ($r_{\rm A}=R_{\rm A}/a$ and $r_{\rm B}=R_{\rm B}/a$). While it can fit for these values directly it supports the alternative scheme of fitting for the sum (\rAplusrB) and ratio ($k \equiv r_{\rm B}/r_{\rm A}$) of the fractional radii. The latter scheme was chosen as the parameters are less correlated than the individual fractional radii \citep{Southworth08mnras} and \rAplusrB\ can be estimated directly from the width of the eclipses.
	
	As with the fractional radii, \jktebop\ allows for the system's orbital eccentricity ($e$) and argument of periastron ($\omega$) to be fitted directly or combined as the Poincaré elements (\ecosw\ and \esinw). The latter approach was preferred as \ecosw\ and \esinw\ may be estimated from the phase of the secondary eclipse relative to the primary and the ratio of the eclipse durations respectively. The luminosity ratio of a system's stars is fitted with the central surface brightness ratio ($J$) which may be estimated from the ratio of the eclipse depths. \reff{However eclipse depths are also affected by the inclination and the ratio of the radii, with the degeneracy difficult to break unless the eclipses are total (as a result of the stellar disks fully overlapping) or a spectroscopic light ratio is available \citep{TorresLacy+00aj}.}
	
	The orbital inclination ($i$) is a key fitting parameter giving the angle between a system's orbital plane and the plane of the sky. This parameter affects the overall scale and shape of the eclipses \citep{PrsaGuinan+08apj}, \reff{with systems where $i=90\degr$ having typically U-shaped total eclipses trending towards shallower V-shaped partial eclipses as the inclination decreases.} This is a difficult parameter to estimate as the inclination has an inconsistent effect on the light curve; systems with the same inclination where only the scale differs may produce light curves with different eclipse shapes. As an alternative to directly predicting the inclination, we chose to predict the primary impact parameter (\bP) which is the minimum distance between the centres of the stellar disks, in units of the radius of star A, as star A is eclipsed by star B. This relates to the orbital inclination through $\bP=\cos{i}/r_{\rm A}$ where the orbit is circular, and more consistently reflects the eclipse shapes seen in light curves. This approach adds a complication as \jktebop\ does not accept the primary impact parameter as an input making it is necessary to calculate the inclination with 
	\begin{equation}\label{eqn:inc-from-predictions}
		i = \arccos{\left( \bP \cdot r_{\rm A}\,\frac{1+\esinw}{1-e^2} \right)}
	\end{equation}
	where $r_{\rm A} = (\rAplusrB)/(1+k)$ and $e^2 = (\ecosw)^2 + (\esinw)^2$, with any error being accumulated from that of the individual quantities.

	\subsection{Training and testing data}
	\subsubsection{Synthetic training dataset}\label{sec:synth-train-ds}
	In order to train a regression model able to make useful predictions for a broad range of dEB configurations, a large labelled training dataset is required to cover the expected parameter space. Existing catalogues of dEB systems struggle to meet this need, with inherent observational bias and limited data volumes constraining the parameter space. For example, the Detached Eclipsing Binary Catalogue (DEBCat\footnote{\url{https://www.astro.keele.ac.uk/jkt/debcat/}}; \citealt{Southworth15debcat}) currently lists fewer than 400 well-characterised systems. With the requirement for high volumes of well-distributed data, a strategy of using synthetic training data was adopted.
	
	The training dataset was created by randomly sampling parameters from the distributions given in Table~\ref{tab:synth-train-ds-sampling-strategy}. \reff{From these were derived the instance labels and synthetic phase-folded and binned light curves were generated with \jktebop\ for use as input features. The range for the stars' fractional radii is restricted to a maximum of 0.23 to concentrate the training on well-detached systems suitable for fitting with \jktebop\ \citep{SouthworthSmalley+05mnras} as its simplified model makes it unsuitable for the tidally deformed stars of close binaries \citep{NorthStuder+97aa}. The central surface brightness ratio ($J$) was calculated from two equivalent uniform distributions, one for each of the stars.}
	
	It was found that sampling \ecosw\ and \esinw\ directly from uniform distributions tended to overly favour highly eccentric systems, which are more likely to exhibit eclipses. The alternative approach of sampling $e$ and $\omega$ in order to calculate the Poincaré elements was found to offer intuitive control over their distribution. Again using uniform distributions for both $e$ and $\omega$ tended to overly favour highly eccentric systems which was countered by adopting a half-normal distribution for the former.
	
	\begin{table}\centering
		\caption{The parameter sampling strategy for the full synthetic training dataset.}
		\label{tab:synth-train-ds-sampling-strategy}
		\setlength\tabcolsep{3pt} 
		\begin{tabular}{lccc}
			\hline
			Parameter & Symbol & Distribution & Values\\
			\hline
			\reff{fractional radius} & $r_{\rm A}$, $r_{\rm B}$ & uniform & [0.001, 0.23)\\
			\reff{relative surface brightness} & $J_{\rm A}$, $J_{\rm B}$ & uniform & [0.001, 1.0)\\
			orbital inclination ($\degr$) & $i$ & \reff{uniform in $\cos{i}$} & $[50, 90]$\\
			\reff{orbital eccentricity} & $e$ & half-normal & $\mu=0, \sigma=0.5$\\
			argument of periastron ($\degr$) & $\omega$ & uniform & $[0, 360)$\\ 	
			\hline
		\end{tabular}
	\end{table}
	
	The effect of stellar limb darkening on the generated light curves was handled with the quadratic law. This is given by
	\begin{equation}
		\frac{I(\mu)}{I(1)} = 1-a(1-\mu)-b(1-\mu)^2
	\end{equation}
	where $\mu=\cos{\gamma}$ with $\gamma$ being the angle between the stellar surface normal and the observer's line of sight, and with the linear and quadratic coefficients, $a$ and $b$, being selected from the \tess\ specific table published by \cite{Claret18aa}. In the absence of physical stellar parameters with which to select coefficients, trials were conducted by testing models trained on datasets generated using various sets of coefficients selected from plausible locations within the table. The use of coefficients as both single sets and by alternate picking from lists was tested, with the latter approach found to offer no consistent benefit. The best performing coefficients were found to be a single set, with $a=0.38$ and $b=0.22$ for both stars, drawn from a densely populated region of the coefficient parameter space around $\logg\approx4.0$ and $\Teff\approx5600$~K. 
	
	The ratio of the stellar masses also has an effect on the generated light curve. This manifests itself through the ellipsoidal effect, where the distorted shape of stars in close proximity increases their projected surface when the orbit is near quadrature, leading to an increase in apparent brightness. With the ellipsoidal effect being correlated with the stellar masses and separation, we chose to generate a value for mass ratio from the stars' fractional radii using $\qphot=k^{1.4}$ derived from the mass-radius relations of \cite{DemircanKahraman91apss}. The \jktebop\ code also accepts input values for modelling the reflection effect and gravity darkening. For the former we left \jktebop\ to calculate the coefficients and for the latter we fixed the coefficients at zero.
	
	\begin{figure}
		\centering
		\includegraphics[width=\linewidth]{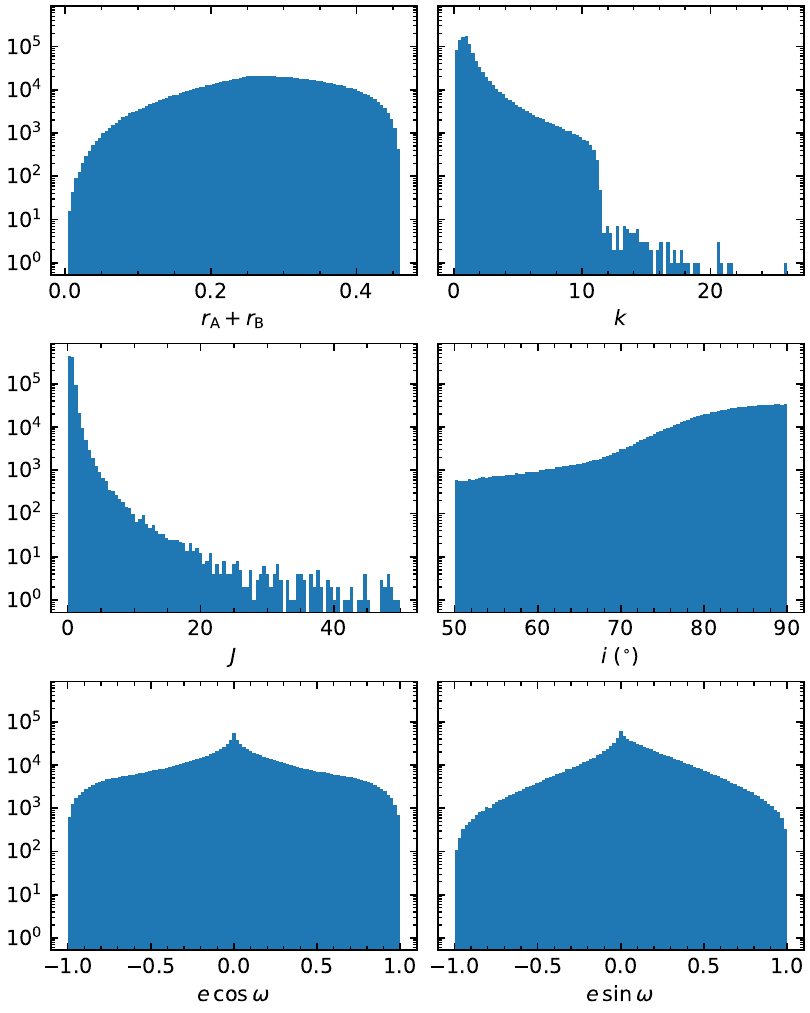}
		\caption{Histograms showing the distribution of label values within the 1 million instances the full synthetic training dataset.}
		\label{fig:synth-train-ds-histograms}
	\end{figure}
	
	\reff{The set of parameters for each candidate instance was evaluated to ensure that the membership of the training data set consisted of instances that were compatible with \jktebop\ parameter restrictions and the resulting primary and secondary eclipses had depths of at least 0.01~mag.} This final criterion was imposed to exclude instances with extremely shallow eclipses where a reliable characterisation is unlikely without the benefit of additional constraints such as a spectroscopic light ratio. Phase-folded light curves were generated for those instances meeting these criteria; however, the components were swapped and related parameters revised where the secondary eclipse was found to be the deeper. This reflects our usual practice of positioning the deeper eclipse at phase 0 when analysing dEB light curves.
	
	The final distribution of the training labels for the \reff{1 million training} instances are shown in the histograms of Fig.~\ref{fig:synth-train-ds-histograms}. These diverge from the sampling distributions through the imposition of the previously discussed criteria. The downward trend shown in $i$, despite the uniform sampling distribution, reflects the naturally decreasing likelihood that candidate systems produce eclipses as the inclination decreases. The selection of the deeper eclipse as the primary eclipse, with the stellar components swapped and parameters revised as necessary, affects the distributions of $k$, $J$ and \esinw. \reff{For $k$ and $J$ this contributes to skew seen in the distributions.} The asymmetric distribution of \esinw\ stems from the deepening of secondary eclipses in eccentric systems when $\omega$ is in the third quadrant, which leads to an increased likelihood of the components being swapped when in this configuration. Finally, the minimum required eclipse depth contributes to the cut-off in $k$ near 11 reflecting the \reff{high} likelihood of extremely shallow eclipses where the stellar radii are so widely differing. \reff{The data set is not without biases, some naturally occurring such as that seen in the histogram of $i$. Those arising from the choice of the deeper eclipse as the primary and the minimum eclipse depth criterion reflect the intended use of the model.}

	\subsubsection{Synthetic testing dataset}\label{sec:synth-test-ds}
	While the training dataset ensures extensive coverage across the range of values supported by the \jktebop\ models, no effort was made to ensure physical plausibility. This was considered an appropriate strategy for training, with the regression model required to learn the full extent of the function which correlates input features to their corresponding output labels with minimal restriction. A different approach was adopted for the testing dataset so as not to repeat and potentially reinforce any assumptions made during training. In order to test with sufficient data representative of the model's use case, test data were derived from physically plausible synthetic star systems based on theoretical \reff{models} from the MESA Isochrones and Stellar Tracks (MIST) project \citep{Dotter16apjs, ChoiDotter+16apj}.
	
	\begin{table}\centering
		\caption{The orbital parameter sampling strategy for the synthetic test dataset based on MIST stellar models.}
		\label{tab:synth-test-ds-sampling-strategy}
		\setlength\tabcolsep{5pt} 
		\begin{tabular}{lccc}
			\hline
			Parameter & Symbol & Distribution & Values\\
			\hline
			primary mass (\Msun) & $M_{\rm A}$ & \multicolumn{2}{|c|}{\reff{see section \ref{sec:synth-test-ds}}} \\
			secondary mass (\Msun) & $M_{\rm B}$ & \multicolumn{2}{|c|}{see section \ref{sec:synth-test-ds}} \\
			orbital period (d) & $P$ & uniform & $[2.5, 25.0]$\\
			orbital inclination ($\degr$) & $i$ & \reff{uniform in $\cos{i}$} & $[50, 90]$\\
			orbital eccentricity & $e$ &  \multicolumn{2}{|c|}{see section \ref{sec:synth-test-ds}} \\
			argument of periastron ($\degr$) & $\omega$ & uniform & $[0, 360)$ \\
			\hline
		\end{tabular}
	\end{table}
	
	Each star system was generated based on an initial random selection of metallicity and system age (constrained within the main-sequence to red-giant phases) from those published in MIST 1.2 pre-packaged isochrones\footnote{\url{http://waps.cfa.harvard.edu/MIST/data/tarballs_v1.2/MIST_v1.2_vvcrit0.4_basic_isos.txz}}. The initial mass of the primary star was chosen from those published in the MIST dataset within the range of 0.4 to 20.0\Msun\ based on a probability function defined by the normalised product of the initial mass function of \cite{Chabrier03pasp}, as summarised by \cite{Maschberger13mnras}, given by
	
	\begin{equation}\label{eqn:chabrier-imf}
		p(M) = \left\{
		\begin{array}{ll}
			0.0443 M^{-2.3} & \text{for }M\geq1\Msun \\
			\frac{0.158}{M}\exp{\left[-\frac{1}{2} \left(\frac{\log_{10}{M}-\log_{10}{0.079}}{0.69}\right)^2 \right]} & \text{for }M<1\Msun
		\end{array}\right.
	\end{equation}
	and the multiplicity function used by \cite{WellsPrsa21apjs}
	
	\begin{equation}\label{eqn:wells-prsa-multiplicity-function}
		p(M) = \tanh{\left(0.31M + 0.18\right)}
	\end{equation}
	
	The initial mass of the secondary star was chosen with \reff{a probability function based on the mass-ratio distributions of \cite{MoeDiStefano17apjs}}. These values were used with the metallicity and age to lookup the current stellar masses, radii, luminosities, effective temperature and surface gravities of the system's stars. 
	
	Each system's orbital period was randomly chosen from a uniform distribution between 25 days and a minimum value calculated with Kepler's third law, the stellar masses and a minimum separation given by
	\begin{equation}
		a_{\rm min} = \frac{\max{(R_{\rm A}, R_{\rm B})}}{r_{\rm max}}
	\end{equation}
	with $r_{\rm max} = 0.23$ to ensure the systems were well-detached while covering a similar range to the training data. With the period defined, Kepler's third law was used once more to yield the orbital semi-major axis. The remaining orbital parameters were chosen at random from uniform distributions, with the inclination and the argument of periastron subject to the fixed limits shown in Table~\ref{tab:synth-test-ds-sampling-strategy} and the orbital eccentricity subject to a maximum value given by
	\begin{equation}\label{eqn:e-max-test}
		e_{\rm max} = \min{\left( 1-1.5(\rAplusrB), \,\,e_{\rm max, mds} \right)}
	\end{equation}
	as used by \cite{WellsPrsa21apjs}, where
	\begin{equation}\label{eqn:e-max-mds}
		e_{\rm max, mds} = 
		\left\{\begin{array}{ll}
			0 & \text{for }P\leq2\,\textrm{d} \\
			1-\left(\frac{P}{2\,\textrm{d}}\right)^{-2/3} & \text{for }P>2\,\textrm{d}
		\end{array}\right.
	\end{equation}
	is taken from \cite{MoeDiStefano17apjs}.
	
\begin{figure}
		\centering
		\includegraphics[width=\linewidth]{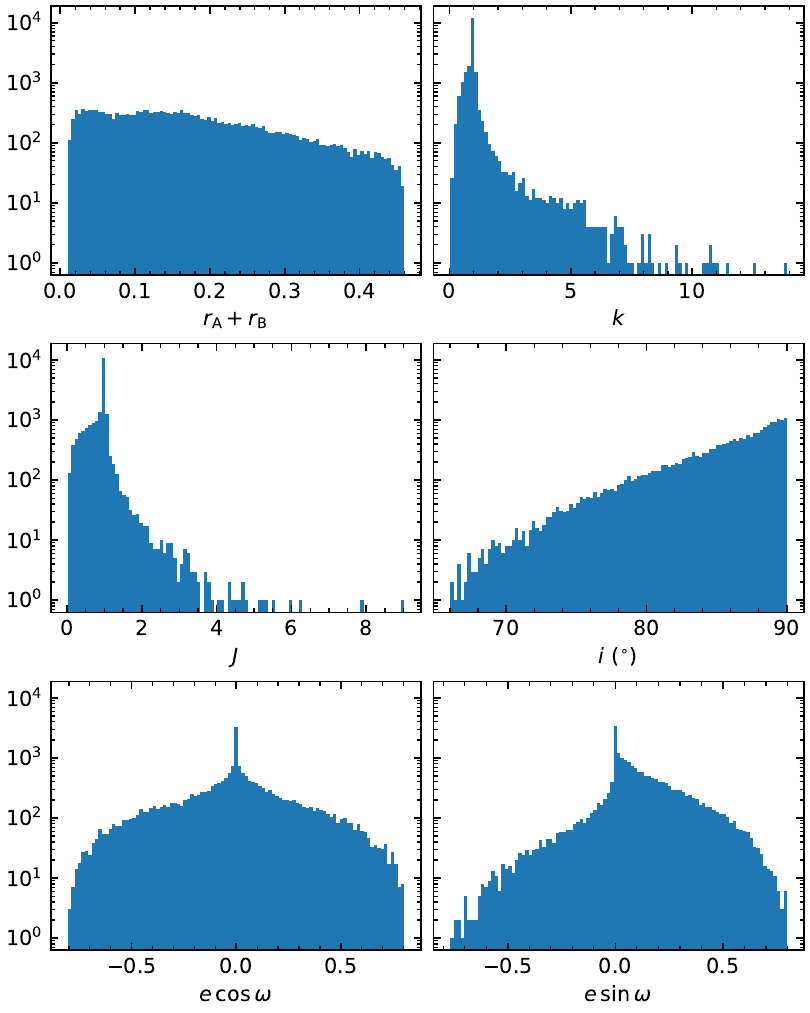}
		\caption{Histograms showing the distributions of label values within the 20\,000 instances of the synthetic testing dataset.}
		\label{fig:synth-test-ds-histograms}
	\end{figure}

	With the parameters from the MIST models, the surface brightness ratio of the two stars was calculated over the \tess\ bandpass using the published instrument response function\footnote{\url{https://heasarc.gsfc.nasa.gov/docs/tess/data/tess-response-function-v2.0.csv}} and power-2 limb darkening coefficients were read from tables published by \cite{ClaretSouthworth23aa}. A dataset of 20\,000 fully labelled test instances was assembled from those systems where both eclipse types were predicted to have a depth of at least 0.01~mag (see Fig.~\ref{fig:synth-test-ds-histograms}). Those instances where the secondary eclipse was seen to be deeper were updated to reassign the eclipses by swapping the components and their related parameters. For each instance a phase-folded light curve feature was generated with \jktebop\ to which Gaussian noise was added consistent with a randomly chosen apparent magnitude in the range $[6, 18]$ and the \tess\ photometric performance \citep{Ricker+15jatis:Tess}. Each light curve feature was further degraded with randomly chosen phase and vertical magnitude shifts to mimic imperfect data pre-processing.
	
	\subsubsection{The test dataset of real systems}\label{sec:real-test-ds}
	A final set of test data was based on real eclipsing binary systems selected from DEBCat which publishes a list of such systems where the masses and radii have been measured to within 2\% precision. The selection was constrained to those systems with at least one sector of \tess\ time-series photometry available and a published characterisation from which could be derived a full set of test labels. \reff{Where possible, analyses based on \jktebop\ were favoured as the published parameters may be used directly as labels.}
	
	A simple characterisation pipeline was created for the known systems with \tess\ time-series data downloaded from the Mikulski Archive for Space Telescopes (MAST\footnote{\url{https://mast.stsci.edu/portal/Mashup/Clients/Mast/Portal.html}}) and subsequently processed using the \lightkurve\ \citep{Lightkurve+18}, \astropy\ \citep{Astropy22apj}, \numpy\ \citep{Harris+2020numpy} and \uncertainties\ \citep{Lebigot2016uncertainties} \python\ packages. Where possible, the choice of sectors and the use of simple aperture photometry (SAP; \citealt{JenkinsTwicken+16}) or pre-search data conditioning SAP (PDCSAP) flux data was informed by the published works from which the labels were taken. For those works that pre-date \tess\ the best available \tess\ time-series data were used with SAP fluxes preferred to avoid the additional processing of PDCSAP pipeline, which may be detrimental to stellar eclipse signals \citep{Southworth20obsR1}. Fluxes were converted to magnitudes then de-trended and rectified to zero by fitting and subtracting low order polynomials and, where necessary, the out-of-eclipse light curves were flattened to reduce the nuisance effect of stellar variability from pulsations or star spots.
	
	The downloaded and processed data were fitted with \jktebop\ version 43 using input parameters derived from the published characterisation, in order to create a set of control fits for the test systems. Where necessary, published steps to discard poor quality data and the use of fitting parameters such as limb-darkening coefficients and spectroscopic brightness ratios were reproduced to achieve stable fitted results comparable to the published values. \reff{Several candidate systems were discarded} at this stage where it was not possible to achieve a stable control fit with the simplified general purpose pipeline used.
	
	\begin{figure}
		\centering
		\includegraphics[width=\linewidth]{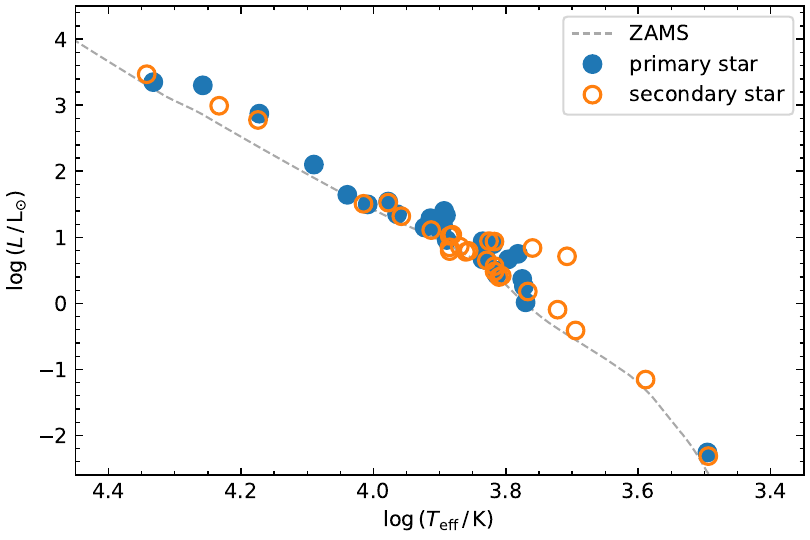}
		\caption{The members of the test dataset of real systems plotted on a Hertzsprung-Russell diagram with a zero-age main-sequence (ZAMS) line derived from MIST 1.2 evolutionary models \citep{Dotter16apjs, ChoiDotter+16apj}.}
		\label{fig:real-test-ds-hrd}
	\end{figure}
	
	The systems making up the test dataset are listed in Table~\ref{tab:real-test-ds-details} with the component stars shown plotted on a Hertzsprung-Russell diagram in Fig.~\ref{fig:real-test-ds-hrd}. The mass range extends from V436 Per with B-type components of $6.88\Msun$ and  $7.35\Msun$ \citep{SouthworthBowman22mnras} to the twin M-dwarf system CM Dra with components of $0.23\Msun$ and $0.21\Msun$ \citep{MartinSethi+24mnras}. Orbital periods range from the $25.936$ d of V436 Per to $0.891$ d for V456 Cyg \citep{VanReethSouthworth+22aa}.
	
	A number of the systems present specific challenges to the characterisation process. The characterisation of V436 Per has three potential solutions with values of $k$ near 0.69, 0.84 and 1.1 \citep{SouthworthBowman22mnras} therefore accurate predictions are required to avoid converging on the wrong result. V456 Cyg is a very compact system with fractional radii of 0.27 and 0.26, both significantly higher than would normally be considered for modelling with \jktebop\ and outside the range of values within the training dataset. However, in this case the published characterisation by \cite{VanReethSouthworth+22aa} was produced with \jktebop\ and a control fit can be made which agrees with their results. The secondary star of AI Phe is significantly larger than the primary ($k\simeq1.6$) despite the secondary eclipses being shallower due to the orientation of this eccentric system \citep{Maxted+20mnras}. Both V1229 Tau and V570 Per have shallow partial eclipses of 0.11 mag or less which may be challenging for predictions and fitting, especially for the ratio of the radii $k$ \citep{SouthworthMurphy+23mnrasl}. In addition to the specific cases mentioned, many members of the test set consist of stars with similar radii giving light curves with similar primary and secondary eclipses, which can lead to difficulty in determining the ratio of the radii \citep{TorresLacy+00aj, BarochGimenez+22aa}. \reff{Systems where one or both of the eclipses are total, those with similar eclipse depths, similar components or shallow eclipses are indicated within the T, E, C and S columns of Table~\ref{tab:real-test-ds-details}.} Plots of the phase-folded light curve feature for each of these systems is shown in Appendix A.
	
	\begin{table*}\centering
		\caption{The set of known well-characterised systems used for final testing, in descending order of the mass of star A. The spectral type given is that published in the referenced works.}
		\label{tab:real-test-ds-details}
		\defcitealias{SouthworthBowman22mnras}{1} \defcitealias{Southworth25obsR22}{2} \defcitealias{Southworth24obsR21}{3} \defcitealias{LacyFrueh85apj}{4} \defcitealias{Southworth25obsR25}{5} \defcitealias{SouthworthMurphy+23mnrasl}{6} \defcitealias{AbtLevato78pasp}{7} \defcitealias{BarochGimenez+22aa}{8} \defcitealias{Southworth25obsR24}{9} \defcitealias{SouthworthSmalley+05mnras}{10} \defcitealias{GraczykPietrzynski+22aa}{11} \defcitealias{Southworth21obsR7}{12} \defcitealias{VanReethSouthworth+22aa}{13} \defcitealias{BakisHensberge+14aj}{14} \defcitealias{Southworth24obsR18}{15} \defcitealias{OverallSouthworth24obsR17}{16} \defcitealias{Southworth23obsR12}{17} \defcitealias{Southworth25obsR23}{18} \defcitealias{Southworth23obsR14}{19} \defcitealias{Southworth21obsR4}{20} \defcitealias{Southworth23obsR13}{21} \defcitealias{Southworth21obsR5}{22} \defcitealias{Southworth21obsR6}{23} \defcitealias{Maxted+20mnras}{24} \defcitealias{Kirkby-KentMaxted+16aa}{25} \defcitealias{Southworth22obsR11}{26} \defcitealias{TorresSandbergLacy+14apj}{27} \defcitealias{MartinSethi+24mnras}{28} \defcitealias{MoralesRibas+09apj}{29} 
		\begin{tabular}{lrrrrrrrcccccl}
			\hline
			& & & & & & & & & \multicolumn{4}{|c|}{Flags} & \\ 
			& $r_{\rm A}+r_{\rm B}$ & $k$ & $J$ & $e\cos{\omega}$ & $e\sin{\omega}$ & $i~(\degr)$ & $L_{\rm 3}$ & Spectral Type & T & E & C & S & Reference \\ 
			\hline
			V436 Per & 0.080 & 1.1 & 1.0 & -0.13 & 0.36 & 88.0 & -0.0030 & B1.5 V & & & & & \citetalias{SouthworthBowman22mnras} \\ 
			V539 Ara & 0.40 & 0.79 & 0.96 & -0.046 & -0.031 & 85.2 & 0.033 & B3 V + B4 V & & & & & \citetalias{SouthworthBowman22mnras} \\ 
			MU Cas & 0.19 & 0.89 & 1.0 & 0.19 & 0.042 & 87.1 & 0.033 & B5 V + B5 V & & \checkmark & \checkmark & & \citetalias{Southworth25obsR22} \\ 
			IQ Per & 0.38 & 0.61 & 0.45 & -0.064 & -0.023 & 88.6 & 0.054 & B8 V + A6 V & \checkmark & & & & \citetalias{Southworth24obsR21, LacyFrueh85apj} \\ 
			AR Aur & 0.20 & 0.96 & 0.90 & 0.0 & 0.0 & 88.6 & 0.015 & B9 V + B9.5 V & & & & & \citetalias{Southworth25obsR25} \\ [3pt] 
			V1229 Tau & 0.27 & 0.78 & 0.56 & 0.0 & 0.0 & 78.6 & 0.029 & A0 Vp(Si) + Am & & & & \checkmark & \citetalias{SouthworthMurphy+23mnrasl, AbtLevato78pasp} \\ 
			V889 Aql & 0.11 & 0.99 & 0.97 & -0.22 & 0.30 & 89.1 & 0.19 & B9.5 V & & & \checkmark & & \citetalias{BarochGimenez+22aa} \\ 
			V596 Pup & 0.23 & 0.98 & 1.0 & -0.037 & 0.089 & 88.1 & 0.20 & A1 V + A1 V & & \checkmark & \checkmark & & \citetalias{Southworth25obsR24} \\ 
			WW Aur & 0.31 & 1.0 & 0.87 & 0.0 & 0.0 & 87.6 & 0.0 & A4 m + A5 m & & & \checkmark & & \citetalias{SouthworthSmalley+05mnras} \\ 
			V788 Cen & 0.23 & 0.60 & 0.90 & 0.0 & 0.0 & 82.8 & 0.021 & A2mA5-F2 & & & & & \citetalias{GraczykPietrzynski+22aa} \\ [3pt] 
			RR Lyn & 0.14 & 0.63 & 0.82 & -0.078 & -0.0016 & 87.5 & 0.036 & A3/A7V/F2 + F0 V & & & & & \citetalias{Southworth21obsR7} \\ 
			V362 Pav & 0.26 & 0.38 & 0.25 & 0.00030 & -0.0014 & 84.3 & -0.054 & A2mA5-A9 & \checkmark & & & \checkmark & \citetalias{GraczykPietrzynski+22aa} \\ 
			V456 Cyg & 0.52 & 0.94 & 0.77 & 0.0 & 0.0 & 83.2 & -0.034 & A2hA7mA4V + F3V & & & \checkmark & & \citetalias{VanReethSouthworth+22aa, BakisHensberge+14aj} \\ 
			GW Eri & 0.24 & 0.98 & 0.98 & 0.0 & 0.0 & 83.9 & 0.028 & A1mA2-A8 & & \checkmark & \checkmark & & \citetalias{GraczykPietrzynski+22aa} \\ 
			OO Peg & 0.31 & 0.90 & 0.97 & 0.0 & 0.0 & 83.6 & 0.0 & F2 V & & \checkmark & & & \citetalias{Southworth24obsR18} \\ [3pt] 
			CW Eri & 0.31 & 0.70 & 0.93 & 0.0050 & -0.012 & 86.4 & -0.00020 & F2 V & & & & & \citetalias{OverallSouthworth24obsR17} \\ 
			ZZ Boo & 0.24 & 1.1 & 0.98 & 0.0 & 0.0 & 88.6 & -0.00010 & F3 V & & \checkmark & \checkmark & & \citetalias{Southworth23obsR12} \\ 
			RZ Cha & 0.37 & 1.1 & 0.98 & 0.0 & 0.0 & 83.3 & 0.016 & F5 IV-V + F5 IV-V & & \checkmark & \checkmark & & \citetalias{Southworth25obsR23} \\ 
			V570 Per & 0.32 & 0.88 & 0.88 & 0.0 & 0.0 & 77.3 & 0.0 & F3 V + F5 V & & & & & \citetalias{Southworth23obsR14} \\ 
			AN Cam & 0.11 & 1.2 & 0.84 & 0.45 & 0.12 & 89.2 & 0.0 & F8 & \checkmark & & & & \citetalias{Southworth21obsR4} \\ [3pt] 
			IT Cas & 0.22 & 0.99 & 1.0 & 0.081 & -0.037 & 89.7 & 0.022 & F3 V + F3 V & & \checkmark & \checkmark & & \citetalias{Southworth23obsR13} \\ 
			V455 Aur & 0.22 & 0.95 & 0.95 & -0.0068 & 0.0071 & 85.0 & 0.028 & F5 V + F6 V & & \checkmark & \checkmark & & \citetalias{Southworth21obsR5} \\ 
			V505 Per & 0.17 & 0.98 & 0.98 & 0.0 & 0.0 & 87.9 & 0.0 & F5 V + F5 V & & \checkmark & \checkmark & & \citetalias{Southworth21obsR6} \\ 
			AI Phe & 0.099 & 1.6 & 0.50 & -0.065 & 0.18 & 88.4 & 0.0053 & G3 V & \checkmark & & & & \citetalias{Maxted+20mnras, Kirkby-KentMaxted+16aa} \\ 
			ZZ UMa & 0.27 & 0.75 & 0.55 & 0.0 & 0.0 & 89.4 & 0.0 & G0 V + G8 V & \checkmark & & & & \citetalias{Southworth22obsR11} \\ [3pt] 
			QR Hya & 0.16 & 0.94 & 0.94 & -8.0e-05 & -6.0e-05 & 86.1 & 0.0 & G1 V & & \checkmark & \checkmark & & \citetalias{GraczykPietrzynski+22aa} \\ 
			V530 Ori & 0.095 & 0.60 & 0.18 & -0.056 & 0.066 & 89.8 & 0.0 & G1 V + M1 V & \checkmark & & & \checkmark & \citetalias{TorresSandbergLacy+14apj} \\ 
			CM Dra & 0.13 & 0.94 & 0.98 & -0.0016 & 0.0050 & 89.6 & 0.0 & M4.5 & & \checkmark & \checkmark & & \citetalias{MartinSethi+24mnras, MoralesRibas+09apj} \\ 
			\hline
		\end{tabular}
		\\
		\vspace{1ex}
		{\raggedright \textbf{Notes.} \\
			\emph{Flags:} indicate whether the system has total eclipses (T), similar eclipse depths (E), similar sized components (C) or shallow eclipses of <0.1 mag (S) \\
			\emph{Reference:} (1) \cite{SouthworthBowman22mnras}, (2) \cite{Southworth25obsR22}, (3) \cite{Southworth24obsR21}, (4) \cite{LacyFrueh85apj}, (5) \cite{Southworth25obsR25}, (6) \cite{SouthworthMurphy+23mnrasl}, (7) \cite{AbtLevato78pasp}, (8) \cite{BarochGimenez+22aa}, (9) \cite{Southworth25obsR24}, (10) \cite{SouthworthSmalley+05mnras}, (11) \cite{GraczykPietrzynski+22aa}, (12) \cite{Southworth21obsR7}, (13) \cite{VanReethSouthworth+22aa}, (14) \cite{BakisHensberge+14aj}, (15) \cite{Southworth24obsR18}, (16) \cite{OverallSouthworth24obsR17}, (17) \cite{Southworth23obsR12}, (18) \cite{Southworth25obsR23}, (19) \cite{Southworth23obsR14}, (20) \cite{Southworth21obsR4}, (21) \cite{Southworth23obsR13}, (22) \cite{Southworth21obsR5}, (23) \cite{Southworth21obsR6}, (24) \cite{Maxted+20mnras}, (25) \cite{Kirkby-KentMaxted+16aa}, (26) \cite{Southworth22obsR11}, (27) \cite{TorresSandbergLacy+14apj}, (28) \cite{MartinSethi+24mnras}, (29) \cite{MoralesRibas+09apj}.\par
		}
	\end{table*}

	\subsection{The machine learning model}
	\subsubsection{Neural Networks}
\begin{figure}
		\centering
		\includegraphics[width=0.9\linewidth]{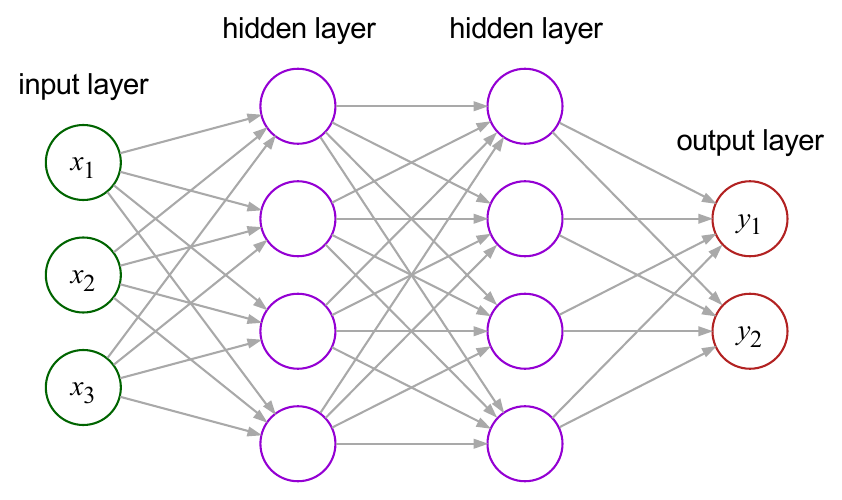}
		\caption{A fully connected neural network with three input neurons, two hidden layers and two output neurons.}
		\label{fig:generic-ann-diagram}
	\end{figure}

	\reff{A neural network (NN) is a flexible machine learning algorithm inspired by the working of biological neural networks and capable of learning regression and classification tasks. It is composed of neurons, each of which will compute the weighted sum of its input values and an optional bias term, outputting the result via a non-linear \emph{activation function} as
	\begin{equation}\label{eqn:ann-neuron-activation}
		y = \phi \left( \vec{w} \odot \vec{y}_{\scriptscriptstyle l-1} + b \right)
	\end{equation}
	where $\odot$ is the element-wise product operator, $\vec{w}$ is a vector of the neuron's input weight coefficients, $\vec{y}_{\scriptscriptstyle l-1}$ is a vector of its input values, $b$ is its bias value and $\phi$ is the activation function, such as the sigmoid function
	\begin{equation}\label{eqn:sigmoid-function}
		\sigma(x) = \frac{1}{1+e^{-x}}
	\end{equation}
	Neurons are typically organized as a network of layers (see Fig.~\ref{fig:generic-ann-diagram}) with an input layer consisting of one neuron per input feature and an output layer having one neuron per predicted value. Sandwiched between are zero or more \emph{hidden layers} consisting of one or more neurons, each connected to the output of every neuron in the preceding layer in a so called \emph{fully connected} configuration. Such a network, with multiple hidden layers is often referred to as a deep neural network (DNN).}
	
	\reff{A network is trained by optimising the neurons' parameters (the input weights and bias values) to minimise the value of a chosen loss function when it is used to evaluate the model's predictions against the training data label values. The parameters are optimized with a gradient descent (GD) algorithm which works by calculating the gradient of the loss function with respect to the model's parameters with
		\begin{equation}\label{eqn:cost-gradient}
		\nabla C(\vec{\Theta}) \equiv
		\left(
		\begin{matrix}
			\frac{\partial C}{\partial \theta_1} \\
			\vdots \\
			\frac{\partial C}{\partial \theta_n}
		\end{matrix}
		\right)
	\end{equation}
	where $\vec{\Theta}$ is the model's parameters and $C$ is the loss function. Over repeated training iterations the parameters are updated by descending the steepest gradient, with the step size being set by the learning rate $\eta$, giving the update as
	\begin{equation}\label{eqn:gd-parameter-update}
		\vec{\Theta} \leftarrow \vec{\Theta} - \eta \nabla C(\vec{\Theta})
	\end{equation}
	As GD converges on a minimum, the size of parameter updates tend to zero indicating when training is complete. The training leaves the learned behaviour encoded into the network of weights and biases.}
	
	\subsubsection{Convolutional neural networks}
	\reff{Each neuron of a fully connected NN has a global view of the entire output from the preceding layer, however this is at the expense of any local spatial or temporal relationship between data points. A different approach is required to learn and interpret these local features within the data and this is addressed by convolutional layers.}
	
	\reff{Within a convolutional layer each neuron, known as a \emph{filter}, operates within a restricted receptive field within the input data. Rather than input weights, the filter applies a convolution kernel to the contents of its receptive field enabling it to pick out features which may be spread across multiple data points. Like the input weights of a neuron, it is the convolution kernel of a filter that is optimized during training. The filter's output is a \emph{feature map} which is the product of its convolution kernel and receptive field taken at regular intervals over the entire input data. The output of a convolutional layer consists of one feature map per filter in a shape derived from the input data. If the case of a 1-D convolutional layer which has $K$ filters and a preceding layer which has $J$ filters, the $k$th filter of this layer will output a feature map given by
	\begin{equation}
		\vec{y}^{\scriptscriptstyle (k)} = \phi \left( \vec{b}^{\scriptscriptstyle (k)} + \sum^{\scriptscriptstyle J}_{\scriptscriptstyle j=1} \vec{w}^{\scriptscriptstyle (j,k)} * \vec{y}^{\scriptscriptstyle (j)}_{\scriptscriptstyle l-1} \right)
	\end{equation}
	where $\vec{b}^{\scriptscriptstyle (k)}$ is the bias values of the $k$th filter, $\vec{w}^{\scriptscriptstyle (j,k)}$ is the kernel weights to be applied to the $j$th input feature map by the $k$th filter, $\vec{y}^{\scriptscriptstyle (j)}_{\scriptscriptstyle l-1}$ is the input values taken from the $j$th feature map of the preceding layer, $\phi$ is the activation function and $*$ is the convolution operator.}
	
	\reff{The use of convolutional layers is of particular benefit with inputs containing a large number of potentially correlated features, such as images or timeseries data. With each filter able to detect features throughout the input data with a single convolution kernel, a layer typically requires a small number of filters, and therefore trainable parameters, to produce a dense feature map from its input. Conversely the layer may increase the output data volume when compared with the input, as the output extends into the feature map dimension. This can be somewhat mitigated with the layer's strides hyperparameter which controls the spatial interval at which the filter is applied to the input data, so a strides value of 2 will halve each spatial dimension of the layer's output.}
	
	\reff{Another approach to reducing the data volume produced by convolutional layers is to follow blocks of one or more layers with a pooling layer. These are non-trainable layers that derive their feature map output values from a simple aggregation function, such as the mean value or the max value found, when applied to the receptive field. With appropriate values for the pool size and strides hyperparameters, a pooling layer will output feature maps downsampled over the input's spatial dimensions.}
	
	\reff{Typically convolutional layers are employed in blocks, optionally interspersed with pooling layers, to derive a deep feature map from input data. The feature map is then used as the input to a deep neural network which learns to make predictions from the features detected. This structure of convolutional layers feeding a neural network is often referred to as a convolutional neural network (CNN) and is the approach adopted for \ebopmaven.}
	
	\subsubsection{Model development and hyperparameter tuning}
	The estimator was implemented in \python\ using the \skl\ \citep{Pedregosa+2011skl}, \tensorflow\ \citep{Abadi+2015tensorflow} and \keras\ \citep{Chollet+2015keras} packages. We found the structure of the convolutional layers to have a significant impact on the performance so a variety of model structures were evaluated, some adapted for 1-d data from well proven sources such as LeNet \citep{LeCun+98LeNet-5} and AlexNet \citep{Krizhevsky+12alexnet} or from existing works with astrophysical time-series data \citep{ShallueVanderburg18aj, BlancatoNess+22apj} and others from our own initial investigations into this work.
	
	An optimisation process was carried out to find the combination of hyperparameters and model structure \reff{giving the lowest prediction loss when evaluated against a test dataset}. This was carried out with the \texttt{hyperopt} package \citep{Bergstra+2013hyperopt} using the Tree-structured Parzen Estimator (TPE) algorithm \citep{Bergstra+11algos}. The TPE algorithm starts with an initial set of random searches after which the trial parameters for each subsequent iteration are drawn from a Gaussian density estimator (aka Parzen estimator) fitted to the distribution of those prior parameter values found to have produced the best-performing models. This sequential approach improves on simple brute force optimisation algorithms, such as grid or random searches, by avoiding regions of the parameter space found to produce poorly performing models, thus reducing the processing load to converge on a recommendation.
	
	\reff{In order to explore potential model structure, searches included candidate models with differing quantities and configurations of convolutional, pooling and neural network layers. Additionally hyperparameters were varied, including: the number of filters and kernel size of convolutional layers; the type, strides, pool size and padding of pooling layers; the number of neurons, initialization and activation function of the neural network layers; the inclusion and rate of any dropout layers; and the optimiser, schedule and the loss function used for training. Each candidate model was trained with the synthetic training dataset from which 10\% was held back for use a validation set with which the model was evaluated during training. A further 10\% of the training dataset was held back for use as test data for the final evaluation of each fully trained model}.
	
	\subsubsection{Training and optimization}
\begin{figure*}
		\includegraphics[width=0.9\linewidth, trim={3.4cm 3.5cm 0 0}, clip]{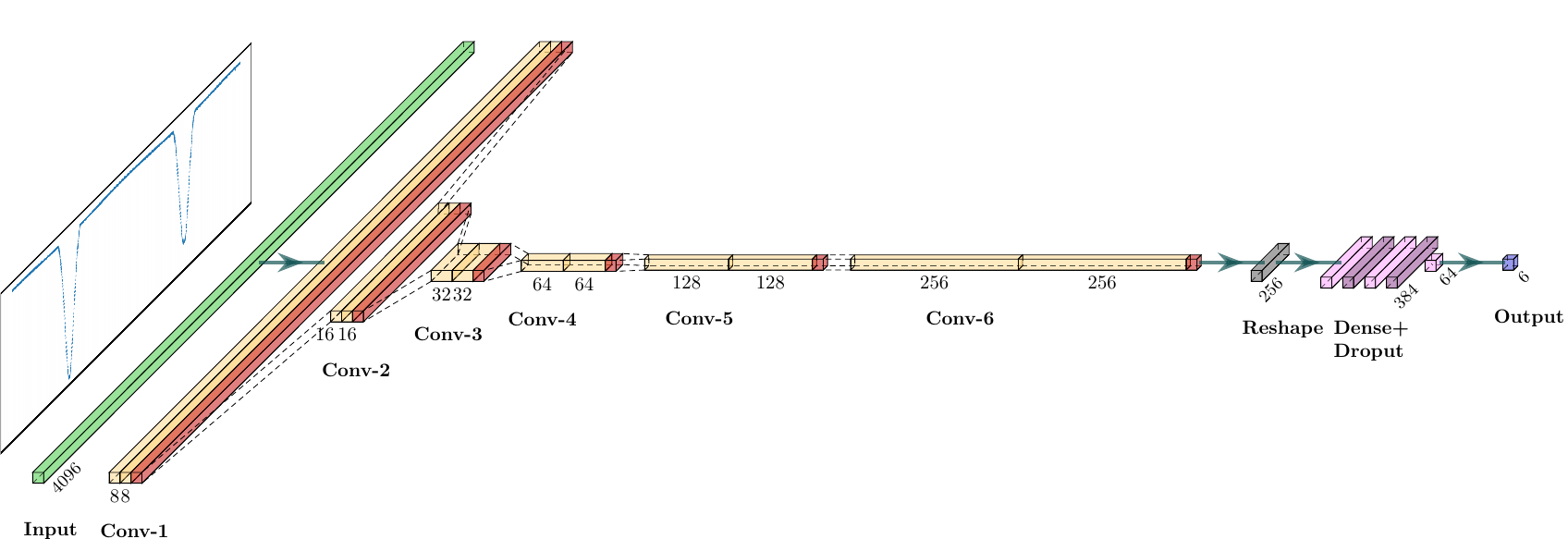}
		\caption{Structure of the \ebopmaven\ model with the input to the left and outputs to the right. The input phase-folded light curve data is processed by six consecutive convolution blocks, each block consisting of a pair of 1-d convolutional layers (with the number of filters is shown) followed by a pooling layer which reduces the size of its output by a factor of 4. The output of the convolution blocks is flattened before being interpreted by a neural network made up of three dense layers interleaved with two dropout layers. Visualisation based on the PlotNeuralNet code \citep{Iqbal88plotneuralnet}.}
		\label{fig:full-model-diagram}
	\end{figure*}

	\reff{The \ebopmaven\ model is shown in Fig.~\ref{fig:full-model-diagram} and summarised in Table~\ref{tab:model-summary}. The input data consists of a phase-folded light curve of 4096 binned relative magnitude values. These data are initially processed by six pairs of 1-dimensional convolutional layers, each pair followed by an average pooling layer. The first pair of convolutional layers have a kernel size of 9 and 8 filters each, with subsequent pairs having an unchanged kernel size but double the number of filters, giving the final pair 256 filters each. The ReLU (Rectified Linear Unit) activation function was used for the convolutional layers to ensure positive output values. The average pooling layers have strides of 4, a pool size of 5 and padding set to "same", which together have the effect of reducing the resolution of their outputs by a factor of 4. The combined effect of this structure is to give the kernels of successive convolutional layers an ever-widening receptive field.}
	
	\reff{The convolutional network is followed by an artificial neural network. The fully connected neurons in each dense layer are initialized with a He normal distribution \citep{HeZhang+15} and use the Leaky ReLU activation function. The leaky variant of ReLU mitigates for the effect of \emph{dead} neurons, which may occur during training if the product of a neuron's bias and weights drop below zero. With ReLU activation all negative outputs are rectified to zero so a neuron cannot leave this state, however leaky ReLU addresses this by \emph{leaking} a small negative output which is often sufficient for the neuron to remain active.}
	
	\reff{The neural network features Dropout layers which combat over-fitting during training by disabling a random subset of the network's neurons during each training step \citep{Hinton+12MLdropout, Srivastava+14MLdropout}. The fraction of neurons dropped is controlled by the rate hyperparameter. With this approach, the learnt input weights of each neuron are less likely to exhibit an over-dependence on a small subset of its strongest inputs as each will have been absent at times during training. The resulting trained network effectively becomes an ensemble of each of the randomly chosen sub-networks used during training. Additionally, the inclusion of dropout layers enables the use of the MC Dropout technique when making predictions \citep{Gal+Chahramani16MLmcdropout}. With MC Dropout, repeated predictions are made with dropout enabled to yield probability distributions for the predicted values.}
	
	\reff{Finally, the output layer consists of six neurons each initialized with a He normal distribution. These yield continuous values via linear activation functions for the six predicted parameters: \rAplusrB, $k$, $j$, \ecosw, \esinw\ and \bP.}
	
	\begin{table}\centering
		\caption{A summary of the \ebopmaven\ model's layers.}
		\label{tab:model-summary}
		\setlength\tabcolsep{3pt} 
		\begin{tabular}{lc}
			\hline
			layer       & configuration \\
			\hline
			Input       & 4096 units \\[9pt]
			
			\multicolumn{2}{c}{6 convolutional blocks, with filters = 8, 16, 32, 64, 128 \& 256, each} \\
			Conv1D 		& kernel size = 9, strides = 1, ReLU activation \\
			Conv1D      & kernel size = 9, strides = 1, ReLU activation \\
			AvgPool1D   & strides = 4, pool size = 5, "same" padding \\[9pt]
			
			Reshape  	& from shape (1, 256) to (256, 1) \\[9pt]
			
			Dense	    & 384 units, He normal initialization, leaky ReLU activation \\
			Dropout		& rate 0.3 \\
			Dense	    & 384 units, He normal initialization, leaky ReLU activation \\
			Dropout		& rate 0.3 \\
			Dense	    & 64 units, He normal initialization, leaky ReLU activation \\[9pt]
			Output      & 6 units, He normal initialization, linear activation \\
			\hline		
		\end{tabular}
	\end{table}
	
	\reff{Models were trained on the training dataset, with 20\% of the instances held back for use as the validation dataset. The training data were shuffled and split into 1\,000 mini-batches per training epoch. After each epoch the prediction loss was evaluated on the validation dataset, with loss function used being the mean absolute error (MAE). This is given by}
	\begin{equation}\label{eqn:mae}
		\textrm{MAE} = \frac{1}{N} \sum_{i=1}^{N} \left| y_i - \hat{y}_i \right|
	\end{equation}
	where $N$ is the number of instances being tested, $y_i$ is the $i$th instance's label value and $\hat{y}_i$ the corresponding prediction. Other loss functions were considered such as mean squared error (MSE) and the Huber loss, however the model search favoured MAE loss which is the least influenced by outliers as it avoids the use of a quadratic component. An early stopping algorithm was used where the training was stopped once the loss measured against the validation dataset failed to show an improvement after five epochs.
	

	\reff{The optimiser used during training was the Adam (adaptive moment estimation) algorithm \citep{KingmaBa17adam}. While Adam features an adaptive learning rate, faster convergence was achieved with an explicit learning rate schedule. A cosine decay schedule \citep{LoshchilovHutter17} was employed, which sets the learning rate $\eta$ for each training step $s$. This is implemented within \keras\ as
	\begin{equation}\label{eqn:cosine_decay_lr}
		\eta(s) = \eta_0 \left[ \left(1 - \alpha\right) \times \frac{1}{2} \left( 1 + \cos{\left( \frac{\pi s}{s_{\rm tot}} \right)} \right) + \alpha \right]
	\end{equation}
	where $\eta_{0}$ is the initial learning rate, $\alpha$ is the target learning rate as a fraction of $\eta_{0}$, $s$ is the current step and $s_{\rm tot}$ is the total number of steps over which to decay the learning rate to the target. The values used were $\eta_{0}=0.0008$, $\alpha=0.01$ and $s_{\rm tot}=60\,000$. Prior to the cosine decay, the learning rate was increased from zero to $\eta_{0}$ linearly over $2\,000$ warm-up steps.}
	
	\reff{The training dataset pipeline includes an augmentation step} which applies Gaussian noise and random phase and magnitude shifts to the phase-folded data to add variety to the instances and to combat over-fitting. The scale of additive noise was dictated by the $1\sigma$ value of a normal distribution with the value being randomly chosen from the uniform range of $[0.001, 0.030)$ mag. The phase and magnitude shifts values were each chosen from normal distributions with $1\sigma$ scale values of 0.066 phase and 0.030 mag respectively.

	\section{Results}
	
	\subsection{Predictions on the synthetic test dataset}\label{sec:testing-synth-test-ds}
	
	\reff{The relative errors reported here are defined as}
	\begin{equation}\label{eqn:relative-error}
		{\rm RE} = \left| \frac{{\rm label}-{\rm prediction}}{\sigma_{\rm label}}  \right|
	\end{equation}
	\reff{where $\sigma_{\rm label}$ is the standard deviation of the values of the corresponding label across the entire synthetic test dataset. This metric was adopted as it consistently scales prediction errors relative to the corresponding parameter's range of expected values.}
	
\begin{figure}
		\includegraphics[width=\linewidth]{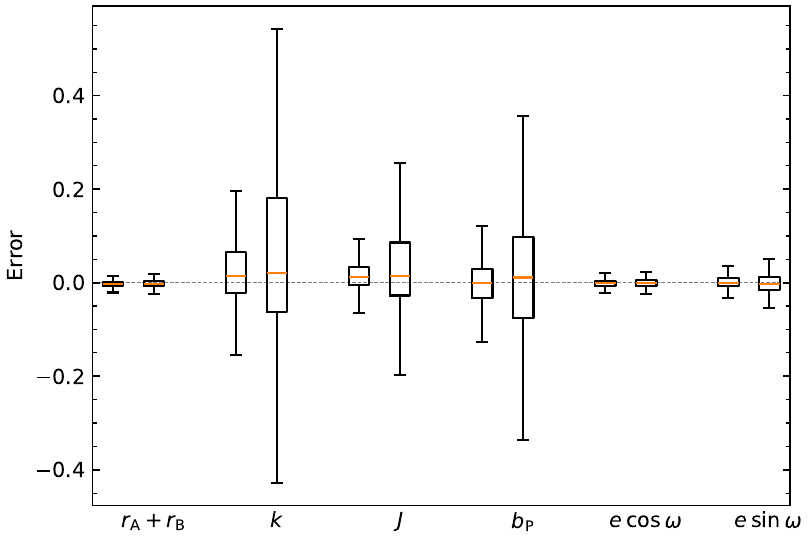}
		\caption{The distribution of the actual error in the model's predictions without MC Dropout for each predicted parameter against the synthetic test dataset's \emph{deep} (left) and \emph{shallow} subsets (right).}
		\label{fig:predictions-synth-test-ds-box-plot}
	\end{figure}

	\reff{The model was tested with the 20\,000 instances of the previously unseen synthetic test dataset, yielding mean relative error of $0.14$ (2~s.~f.) across the six directly predicted parameters. The relative error values the predictions on this dataset are summarised in Table~\ref{tab:eval-mist-tess-dataset}.} These results are given without uncertainties as both the labels, which are the exact values used to generate each test instance's light curve feature, and the predictions, which are made without the MC Dropout approach, are without uncertainties. \reff{The table additionally shows the results split into subsets for systems where at least one eclipse type is total, those with only partial eclipses, systems where both primary and secondary eclipses have a depth of at least 0.1~mag (hereafter referred to as the \emph{deep} subset), and those with at least one eclipse type is shallower than 0.1 mag (the \emph{shallow} subset).} While it is a simplistic distinction, the \emph{deep} subset is broadly representative of a population of systems with detectable dEB light curve features upon which this model is expected to operate. \reff{As such, the results for the \emph{deep} subset are of particular importance and we see an improved mean relative error of 0.086 (2~s.~f.). Fig.~\ref{fig:predictions-synth-test-ds-box-plot} shows a box plot of the actual error distribution by predicted parameter, with each subdivided into the \emph{deep} and \emph{shallow} subsets.}
	
	\begin{table}\centering
		\caption{The \reff{mean relative error of} predictions made on the full synthetic testing dataset without the MC Dropout technique.}
		\label{tab:eval-mist-tess-dataset}
		\setlength\tabcolsep{2pt} 
		\begin{tabular}{cccccc}
			\hline
			parameter   	& all instances & total eclipses & partial eclipses & \emph{deep} & \emph{shallow} \\
			(count)         & (20\,000)& (2\,831) & (17\,169) & (13\,692) & (6\,308) \\
			\hline
			\rAplusrB 		& $0.0609$ & $0.0837$ & $0.0571$ & $0.0522$ & $0.0797$ \\
			$k$             & $0.2459$ & $0.3551$ & $0.2279$ & $0.1385$ & $0.4789$ \\
			$J$             & $0.2134$ & $0.1921$ & $0.2169$ & $0.1204$ & $0.4152$ \\
			\bP				& $0.1847$ & $0.3037$ & $0.1651$ & $0.0967$ & $0.3757$ \\
			\ecosw 			& $0.0347$ & $0.0412$ & $0.0337$ & $0.0314$ & $0.0420$ \\
			\esinw			& $0.1043$ & $0.1639$ & $0.0944$ & $0.0777$ & $0.1620$ \\[3pt]
			all parameters  & $0.1406$ & $0.1899$ & $0.1325$ & $0.0862$ & $0.2590$ \\
			\hline			
		\end{tabular}
	\end{table}
	
	\reff{The predictions for \rAplusrB\ and \ecosw\ yield the best results with mean relative errors over the entire dataset of $0.061$ and $0.035$ (2~s.~f.), respectively. The predictions benefit from both parameters being well-correlated with easily isolated light curve features: the eclipse durations with \rAplusrB\ and the phase of the eclipses with $\ecosw\approx\frac{\pi}{2}(\phi_{\rm S}-\phi_{\rm P}-0.5)$.} While \esinw\ is correlated with the ratio of the eclipse durations, being approximated by $(d_{\rm S}-d_{\rm P})/(d_{\rm S}+d_{\rm P})$, it appears to be a more difficult feature for the model to resolve. This may have some grounding in the coarseness of the phase-folded light curve feature at 4096 bins, however expanding this to 8192 was found to produce negligible improvement in the predictions.
	
	The predictions for $k$, $J$ and \bP\ show the greatest mean relative error values of 0.25, 0.21 and 0.18 (2~s.f), respectively. The worse performance is likely the result of degeneracies between the ratios of the radii and brightness, and the orbital inclination in their relationship with eclipse features, especially where the eclipses are shallow and partial \citep{TorresLacy+00aj}. \reff{However, Table~\ref{tab:eval-mist-tess-dataset} shows the predictions for $k$ and \bP\ for systems with total eclipses have higher mean errors of 0.36 and 0.30, respectively, when compared with those where the eclipses are partial (0.23 and 0.17). These results are greatly affected by the difficulty the model has predicting $k$ and \bP\ on totally eclipsing systems within the \emph{shallow} subset, with mean relative errors of 0.76 and 0.36, respectively.} This is graphically demonstrated in Fig.~\ref{fig:predictions-synth-test-ds-box-plot} where we see a marked improvement in the error distributions of these parameters within the \emph{deep} subset when compared with the \emph{shallow} subset.
	
	\begin{table}\centering
		\caption{The \reff{mean relative error of} predictions made on the \emph{deep} subset of the synthetic test dataset with 1000 MC Dropout iterations.}
		\label{tab:eval-mist-tess-dataset-deep-mc}
		\setlength\tabcolsep{5pt} 
		\begin{tabular}{cr@{\,$\pm$\,}lr@{\,$\pm$\,}lr@{\,$\pm$\,}l}
			\hline
			parameter & \multicolumn{2}{c}{all instances} & \multicolumn{2}{c}{total eclipses} & \multicolumn{2}{c}{partial eclipses} \\
			(count)   & \multicolumn{2}{c}{(13\,692)} & \multicolumn{2}{c}{(1\,719)} & \multicolumn{2}{c}{(11\,973)} \\
			\hline
			\rAplusrB 		& 0.0554 & 0.0012 & 0.0585 & 0.0034 & 0.0550 & 0.0013 \\
			$k$             & 0.1350 & 0.0007 & 0.0894 & 0.0016 & 0.1416 & 0.0007 \\
			$J$             & 0.1166 & 0.0009 & 0.0831 & 0.0022 & 0.1214 & 0.0010 \\
			\bP      		& 0.0960 & 0.0005 & 0.1246 & 0.0013 & 0.0918 & 0.0005 \\
			\ecosw			& 0.0288 & 0.0006 & 0.0334 & 0.0018 & 0.0282 & 0.0006 \\
			\esinw			& 0.0745 & 0.0009 & 0.0832 & 0.0032 & 0.0733 & 0.0009 \\[3pt]
			all parameters  & 0.0844 & 0.0003 & 0.0787 & 0.0010 & 0.0852 & 0.0004 \\
			\hline
		\end{tabular}
	\end{table}
	
	Finally, in Table~\ref{tab:eval-mist-tess-dataset-deep-mc} we look more closely at the model's predictions on the \emph{deep} subset. This is a key test for the model as the \emph{deep} subset is more representative of a potential target population of dEB systems than the synthetic test dataset as a whole.  To further simulate real world usage these predictions were made with MC Dropout enabled and the uncertainties given summarize the $1\sigma$ standard deviation of 1000 predictions made on each instance. With the \emph{deep} subset as a whole we see a \reff{mean relative error of $0.0844\pm0.0003$ (3~s.~f.) across the six predicted parameters, while similar values of $0.0787\pm0.0010$ and $0.0852\pm0.0004$ are recorded when this subset is further split into groups with total eclipses and those where both eclipses are partial. Unlike the synthetic test dataset as a whole, we see the accuracy of the predictions for both $k$ and $J$ worsen for instances without total eclipses, correctly reflecting the degeneracy these parameters have on the features of partial eclipses.}
	
	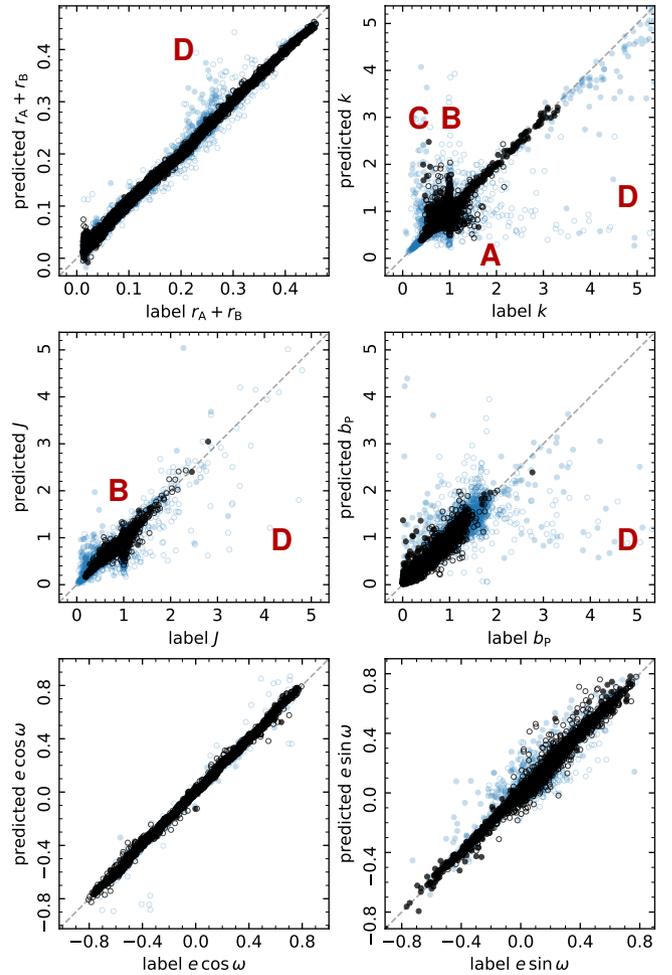
\begin{figure}
		\begin{annotationimage}{width=\linewidth}{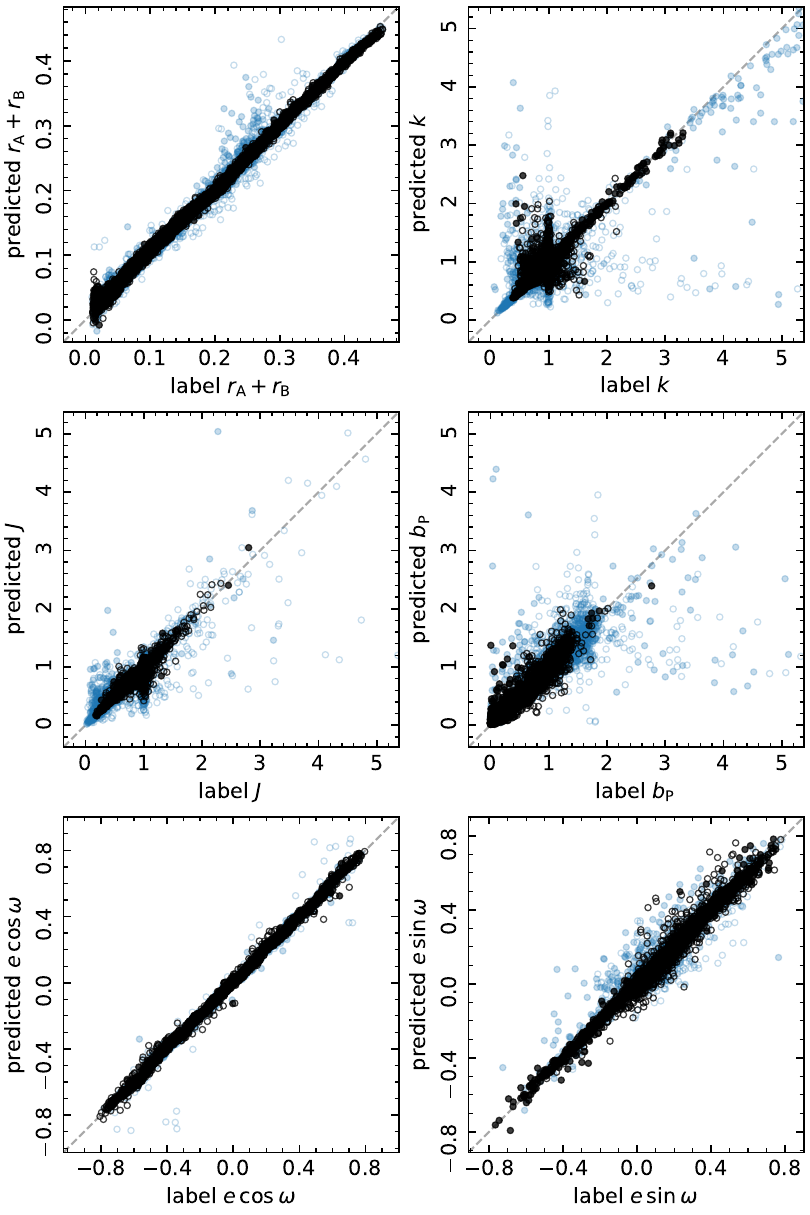}
			\begin{scope}[x={(image.south east)}, y={(image.north west)}, font={\Large\bfseries\sffamily}]
				
				\node [black!30!red] (note) at (0.74, 0.74) {A}; 
				\node [black!30!red] (note) at (0.68, 0.88) {B}; 
				\node [black!30!red] (note) at (0.17, 0.50) {B}; 
				\node [black!30!red] (note) at (0.63, 0.88) {C}; 
				
				\node [black!30!red] (note) at (0.27, 0.95) {D}; 
				\node [black!30!red] (note) at (0.95, 0.80) {D}; 
				\node [black!30!red] (note) at (0.42, 0.45) {D}; 
				\node [black!30!red] (note) at (0.95, 0.45) {D}; 
			\end{scope}
		\end{annotationimage}
		\caption{The predicted parameter values vs the label values of the synthetic testing dataset. Instances from the \emph{shallow} subset are plotted in blue and those in the \emph{deep} subset are plotted in black. \reff{Filled markers indicate systems with total eclipses and open markers those where the eclipses are partial.}}
		\label{fig:predictions-synth-test-ds-vs-labels}
	\end{figure}
	
	\subsubsection{Reviewing the scatter plot}	
	In Fig.~\ref{fig:predictions-synth-test-ds-vs-labels} we plot each  prediction made against the synthetic test dataset, revealing trends not visible in the summary statistics. The predictions for those instances within the \emph{deep} subset (shown in black) show tighter distributions than those from the \emph{shallow} subset, reflecting the general improvement seen in the summary statistics. While the predictions are generally close to the diagonal, which indicates perfect results, there are groupings of outliers.
	
	Within the plot for $k$ we find a pronounced broadening centred on $k\sim1.0$ (annotation {\sf A}). \reff{A census of the instances within this region shows the vast majority to be partially eclipsing systems.} Some 64\% of the group are within the \emph{shallow} subset where we expect predictions for $k$ and $J$ to be less accurate and this is supported by these systems also featuring in the less pronounced scatter in $J$ below 2.0. Nearly all (91\%) have one or other eclipse with a depth of less than 0.25~mag indicating that even moderately shallow eclipses are proving a challenge for the prediction of $k$.
	
	\reff{Also within the plot for $k$ we find a narrow vertical band of poor predictions centred on $k\sim1$ (annotation {\sf B}). This consists of instances which have similar shallow, or moderately shallow, partial eclipses which are also responsible for the similar but less pronounced band in the plot for $J$. These represent systems known to challenge formal analysis, having multiple potential solutions centred on $k=1$ \citep{TorresLacy+00aj, BarochGimenez+22aa}, so it is unsurprising to find the model has difficulties here. Nearby, we find a scattering of poor predictions for label $k<0.9$ (annotated {\sf C}). While most of these instances have similar shallow or moderately shallow partial eclipses, there are fourteen with total eclipses. The total eclipses are mostly very narrow and the model may be interpreting them as partial eclipses, leading to the inflated predictions for $k$ plotted.}
	
	Finally, \reff{within the \emph{shallow} subset}, we see a scattering of over-predictions in \rAplusrB\ for label values between 0.2 and 0.3, and horizontal bands of under-prediction in $k$, $J$ and \bP\ (all annotated {\sf D}). A histogram of the membership of these regions finds the majority belong to the same group of instances with small primary stars ($r_{\rm A}<0.1$) significantly larger secondary stars ($k>2.5$) and very shallow primary eclipses of $\le0.04$~mag.
	
	\subsection{Testing on the dataset of real systems}\label{sec:testing-real-test-ds}
	
	\subsubsection{Predictions}\label{sec:preds-real-test-ds}
	
	\begin{figure}
		\includegraphics[width=\linewidth]{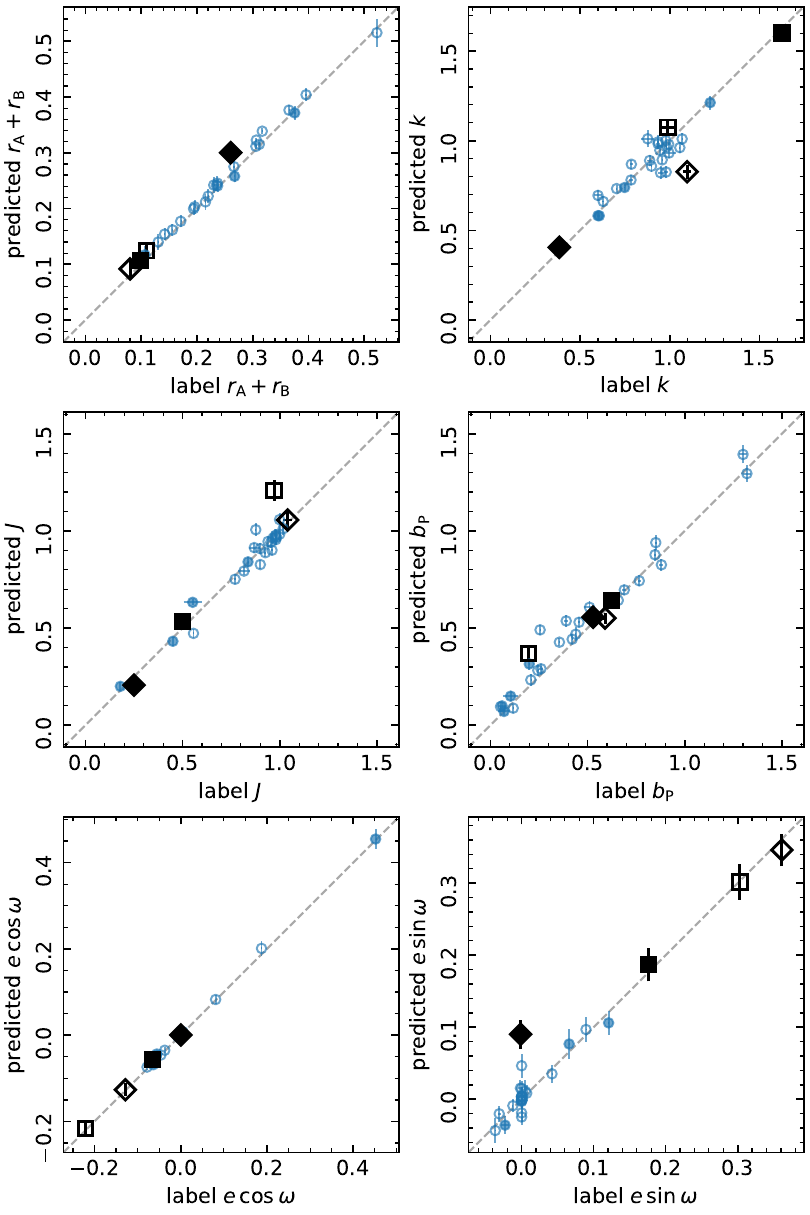}
		\caption{The predicted parameter values vs label values for the test dataset of real systems. The predictions' nominal values and error bars are the mean and $1\sigma$ values from 1000 MC Dropout iterations. \reff{Filled markers indicate systems with total eclipses and open markers those where the eclipses are partial. The highlighted systems are V436 Per (open diamond), V889 Aql (open square), V362 Pav (filled diamond) and AI Phe (filled square).}}
		\label{fig:predictions-mc-real-test-ds-vs-labels}
	\end{figure}
	
	\begin{table}\centering
		\caption{The \reff{mean relative error of predictions made} on the test dataset of real systems made without the MC Dropout technique.}
		\label{tab:eval-formal-test-dataset-nonmc}
		\setlength\tabcolsep{6pt} 
		\begin{tabular}{cccc}
			\hline
			parameter   	& all instances & total eclipses & partial eclipses \\
			(count)         & (28)     & (6)      & (22)  \\
			\hline
			\rAplusrB		& $0.0862$ & $0.1156$ & $0.0781$ \\
			$k$             & $0.0978$ & $0.0248$ & $0.1177$ \\
			$J$             & $0.1221$ & $0.1042$ & $0.1269$ \\
			\bP   			& $0.1084$ & $0.0756$ & $0.1173$ \\
			\ecosw			& $0.0166$ & $0.0197$ & $0.0157$ \\
			\esinw			& $0.0588$ & $0.1187$ & $0.0425$ \\[3pt]
			all parameters  & $0.0816$ & $0.0764$ & $0.0830$ \\
			\hline
			$i$ (\degr)     & $0.1327$ & $0.1324$ & $0.1327$ \\[3pt]
			all parameters  & $0.0857$ & $0.0859$ & $0.0856$ \\
			\multicolumn{4}{l}{($i$ replacing \bP)} \\
			\hline
		\end{tabular}
	\end{table}
	
	The model was used to predict fitting input values for the members of the test dataset of real systems. Two sets of predictions were made, the first without using the MC Dropout technique and and the second with 1000 MC Dropout iterations. Table~\ref{tab:eval-formal-test-dataset-nonmc} shows the predictions made without MC Dropout, where we find \reff{an overall mean relative error of 0.082 (2~s.~f.) across all six directly predicted values ranging from a low of 0.015 for \ecosw\ to 0.12 for $J$. This breaks down to 0.076 for those systems with total eclipses and 0.083 for those where the eclipses are partial. When \bP\ is replaced with the calculated parameter $i$ required by \jktebop\ for fitting, we see consistent MRE values of 0.086 for all three groups. The reason for the larger change to the totally eclipsing group can be found in the error for $i$ where the larger errors for \rAplusrB\ and \esinw, when compared to the set as a whole, offset any improvements in $k$ and \bP\ in these calculations (equation~\ref{eqn:inc-from-predictions}).} These non MC Dropout predictions are without uncertainty so no indication of confidence is given.
	
	\begin{table}\centering
		\caption{The \reff{mean relative error of} predictions made on the test dataset of real systems with 1000 MC Dropout iterations.}
		\label{tab:eval-formal-test-dataset-mc}
		\setlength\tabcolsep{5pt} 
		\begin{tabular}{cr@{\,$\pm$\,}lr@{\,$\pm$\,}lr@{\,$\pm$\,}l}
			\hline
			parameter & \multicolumn{2}{c}{all instances} & \multicolumn{2}{c}{total eclipses} & \multicolumn{2}{c}{partial eclipses} \\
			(count)   & \multicolumn{2}{c}{(28)} & \multicolumn{2}{c}{(6)} & \multicolumn{2}{c}{(22)} \\
			\hline
			\rAplusrB 		& 0.0890 & 0.0218 & 0.1203 & 0.0486 & 0.0805 & 0.0244 \\
			$k$             & 0.0963 & 0.0111 & 0.0315 & 0.0199 & 0.1140 & 0.0130 \\
			$J$             & 0.1183 & 0.0184 & 0.1020 & 0.0366 & 0.1227 & 0.0213 \\
			\bP      		& 0.1083 & 0.0115 & 0.0763 & 0.0258 & 0.1170 & 0.0128 \\
			\ecosw			& 0.0153 & 0.0064 & 0.0177 & 0.0178 & 0.0147 & 0.0066 \\
			\esinw			& 0.0638 & 0.0147 & 0.1232 & 0.0274 & 0.0476 & 0.0155 \\[3pt]
			all parameters  & 0.0818 & 0.0061 & 0.0785 & 0.0135 & 0.0827 & 0.0068 \\
			\hline
			$i$ (\degr)     & 0.1373 & 0.0176 & 0.1372 & 0.0376 & 0.1374 & 0.0199 \\[3pt]
			all parameters  & 0.0867 & 0.0065 & 0.0887 & 0.0142 & 0.0861 & 0.0073 \\
			\multicolumn{7}{l}{($i$ replacing \bP)} \\
			\hline
		\end{tabular}
	\end{table}
	
	The model's predictions made with 1000 MC Dropout iterations are summarised in Table~\ref{tab:eval-formal-test-dataset-mc} and each prediction is plotted against its corresponding label value in Fig.~\ref{fig:predictions-mc-real-test-ds-vs-labels}. These values are derived from the mean of the 1000 MC Dropout predictions with a corresponding uncertainty value based on the $1\sigma$ standard deviation. The label values in this dataset are complete with uncertainties and therefore the uncertainty in the reported loss is derived from both the predictions and labels. \reff{The mean relative error across the six directly predicted parameters is $0.082\pm0.006$ (2~s.~f.) which breaks down to $0.079\pm0.014$ for systems with total eclipses and $0.083\pm0.007$ for those with only partial eclipses. When \bP\ is replaced with the calculated parameter $i$ the corresponding errors are $0.087\pm0.007$, $0.089\pm0.014$ and $0.086\pm0.007$, respectively.}
	
	The overall test results for the dataset of real systems represent a significant improvement over the synthetic test dataset, however this is a much smaller dataset without the broad coverage across the parameter space of the synthetic test dataset. With membership of this dataset based on the availability of \tess\ photometry and a published characterisation, the criteria lend a bias towards bright systems with clear eclipses. \reff{As such, the similarity of these summary results with those for the synthetic test dataset's \emph{deep} subset (see Table~\ref{tab:eval-mist-tess-dataset-deep-mc}) is the more significant outcome.} Despite this, the predictions of a small number of the systems deviate from the labels and warrant further discussion.
	
	\begin{table}\centering
		\caption{The label values for V436 Per \citep{SouthworthBowman22mnras}, V889 Aql \citep{BarochGimenez+22aa} and V362 Pav \citep{GraczykPietrzynski+22aa}, and the \reff{predictions of the machine learning model with 1000 MC Dropout iterations with the calculated value for $i$}.}
		\label{tab:mc-preds-of-interest}
		\setlength\tabcolsep{5pt} 
		\begin{tabular}{cr@{\,$\pm$\,}lr@{\,$\pm$\,}lr@{\,$\pm$\,}l}
			\hline
			parameter    & \multicolumn{2}{c}{label} & \multicolumn{2}{c}{prediction} & \multicolumn{2}{c}{error} \\ 
			\hline
			\multicolumn{7}{c}{V436 Per} \\	[3pt]
			\rAplusrB    &     0.08015 &  0.00028    &     0.0917 &  0.0168    &    -0.0116 &  0.0168     \\ 
			$k$          &       1.097 &  0.022      &     0.8270 &  0.0383    &      0.270 &  0.044      \\ 
			$J$          &       1.041 &  0.022      &     1.0564 &  0.0434    &     -0.015 &  0.049      \\ 
			\bP          &        0.59 &  0.01       &     0.5513 &  0.0279    &       0.04 &  0.03       \\
			\ecosw       &    -0.12838 &  0.00011    &    -0.1267 &  0.0149    &    -0.0017 &  0.0149     \\ 
			\esinw       &      0.3614 &  0.0020     &     0.3463 &  0.0227    &     0.0151 &  0.0228     \\ [3pt]
			$i$ (\degr)  &      87.951 &  0.025      &    87.5285 &  0.4780    &      0.422 &  0.481      \\ [3pt]
			
			\hline
			\multicolumn{7}{c}{V889 Aql} \\	[3pt]		
			\rAplusrB    &     0.10932 &  0.00005    &     0.1242 &  0.0141    &    -0.0149 &  0.0141     \\
			$k$          &        0.99 &  0.04       &     1.0754 &  0.0413    &      -0.09 &  0.06       \\
			$J$          &       0.971 &  0.002      &     1.2090 &  0.0533    &     -0.238 &  0.053      \\
			\bP          &       0.197 &  0.006      &     0.3699 &  0.0362    &     -0.173 &  0.037      \\		
			\ecosw       &      -0.221 &  0.001      &    -0.2168 &  0.0185    &     -0.004 &  0.019      \\
			\esinw       &      0.3030 &  0.0008     &     0.3017 &  0.0245    &     0.0013 &  0.0245     \\ [3pt]
			$i$ (\degr)  &       89.06 &  0.02       &    88.0841 &  0.2979    &       0.98 &  0.30       \\ [3pt]
			
			\hline
			\multicolumn{7}{c}{V362 Pav} \\ [3pt]			
			\rAplusrB    &       0.2605 &  0.0008    &     0.3002 &  0.0063    &    -0.0397 &  0.0064     \\
			$k$          &       0.384  &  0.003     &     0.4060 &  0.0204    &     -0.022 &  0.021      \\
			$J$          & \multicolumn{2}{c}{ 0.25} &     0.2062 &  0.0289    &       0.04 &  0.03       \\
			\bP          &       0.528  &  0.004     &     0.5560 &  0.0295    &     -0.028 &  0.030      \\
			\ecosw       &       0.0003 &  0.0040    &     0.0005 &  0.0063    &    -0.0002 &  0.0075     \\
			\esinw       &      -0.0014 &  0.0004    &     0.0904 &  0.0202    &    -0.0918 &  0.0202     \\ [3pt]
			$i$ (\degr)  &      84.304  &  0.035     &    82.5000 &  0.4745    &      1.804 &  0.476      \\
			\hline
		\end{tabular}
		\\
		\vspace{1ex}
		{\raggedright \textbf{Notes.} \\
			The labels for V362 Pav are taken from \cite{GraczykPietrzynski+22aa} who do not publish a value for $J$ so we estimate a value from a trial fit with \jktebop.
			\\
		}
	\end{table}
	
	\reff{The predictions for V436 Per are shown in Table~\ref{tab:mc-preds-of-interest} and Fig.~\ref{fig:predictions-mc-real-test-ds-vs-labels} (represented with an open diamond) where we see a significant under-prediction in the value for $k$ when compared with the corresponding label value. However, the prediction of $0.83\pm0.04$ is a match for one of the alternative solutions found by \cite{SouthworthBowman22mnras} in their analysis of this system which may explain why the model has arrived at this value.}
	
	The predictions for V889 Aql are shown in Table~\ref{tab:mc-preds-of-interest} and Fig.~\ref{fig:predictions-mc-real-test-ds-vs-labels} (represented with an open square) where we see a significant error in the predicted values for $J$ and \bP\ when compared with the corresponding labels from \cite{BarochGimenez+22aa}. The high degree of eccentricity and similarity of the components' radii will have made predictions challenging, however these features are not unique within the dataset and other predictions fared better. The likely cause is the presence of significant third light in the \tess\ light curves, with \cite{BarochGimenez+22aa} reporting $L_{\rm 3}=0.191\pm0.002$ in their characterisation of sectors 14 and 40, and our own control fit of sector 54 time-series data yielding a similar value of $0.202\pm0.003$. This highlights a limitation of the current \ebopmaven\ model, which was trained on data with $L_{\rm 3}$ fixed at zero. This constraint may be relaxed in the future, at the expense of a more complicated model. In the meantime we find that \jktebop\ has no difficulty in fitting for third light when given the starting parameters from \ebopmaven. 
	
	\reff{The predictions shown in Table~\ref{tab:mc-preds-of-interest} for V362 Pav (Fig.~\ref{fig:predictions-mc-real-test-ds-vs-labels} filled diamond) show a significant error in the prediction of \esinw\ when compared with the labels derived from \cite{GraczykPietrzynski+22aa}. This contributes to the value calculated for the orbital inclination being 1.8\degr\ below that of the label value. The likely cause of the poor prediction is the system's very shallow secondary eclipses, which would easily qualify for membership of the \emph{shallow} subset were it part of the synthetic test dataset discussed in section~\ref{sec:testing-synth-test-ds}.}
	
	\subsubsection{Fitting the test dataset of real systems}
	
	\begin{figure}
		\includegraphics[width=\linewidth]{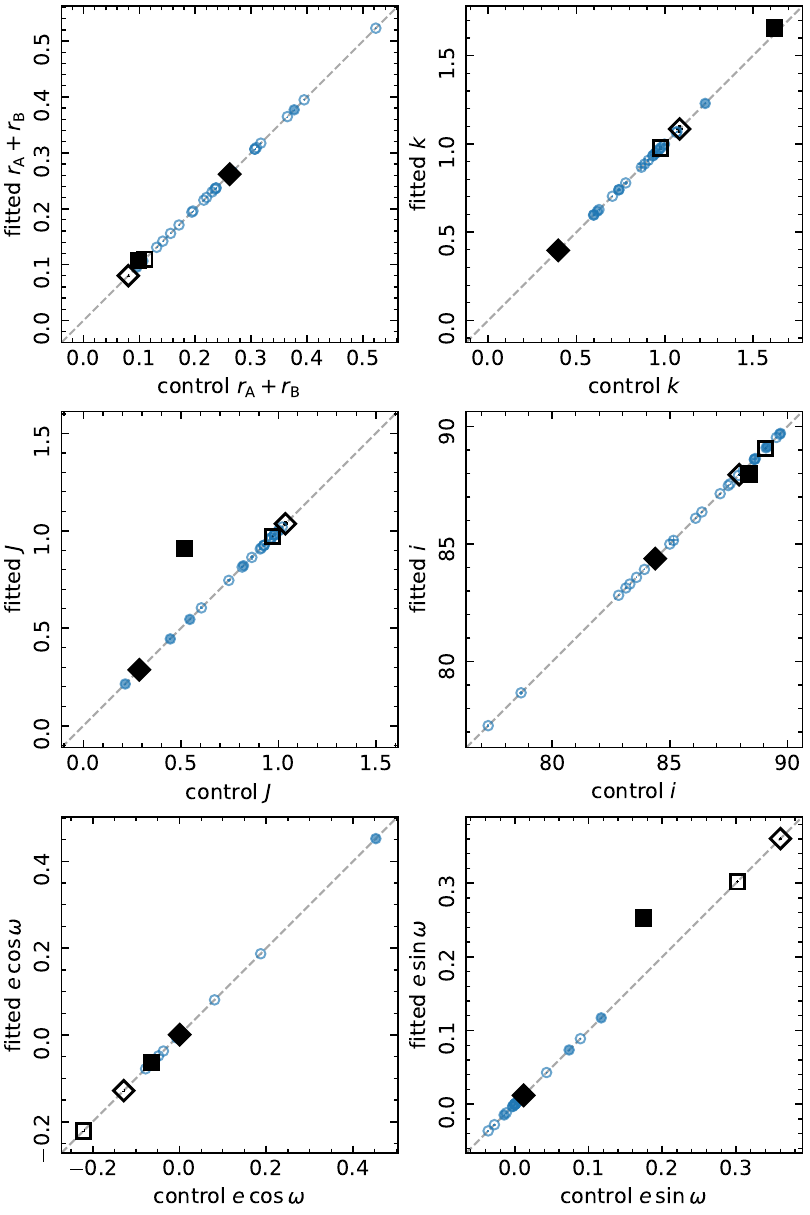}
		\caption{The fitted parameter values vs the control fitted values for the test dataset of real systems, with the error bars taken from the formal errors calculated by the \jktebop. \reff{Filled markers indicate systems with total eclipses and open markers those where the eclipses are partial. The highlighted systems are V436 Per (open diamond), V889 Aql (open square), V362 Pav (filled diamond) and AI Phe (filled square).}}
		\label{fig:fit-mc-real-test-ds-vs-control}
	\end{figure}
	
	The \reff{light curves of the systems within the test dataset of real systems were analysed with \jktebop} using the characterisation pipeline previously discussed in section~\ref{sec:real-test-ds}. For the purposes of testing, the targets' processing and fitting configuration were consistent with those for the control fitting except for the use of the model's predictions as the input values for the \rAplusrB, $k$, $J$, \ecosw, \esinw\ and $i$ parameters, where $i$ is calculated from the other predicted values with equation~\ref{eqn:inc-from-predictions}. \reff{When using input parameters taken from the mean predicted values of 1000 MC Dropout iterations, 27 of the 28 systems in the dataset yielded a set of fitted parameters which agree with those from the corresponding control fit. Generally the matches were not exact, but a fit was considered good if the differences were smaller than the corresponding uncertainties.} The fitted parameters resulting from the \jktebop\ analyses are shown plotted against the equivalent values from the control fits in Fig.~\ref{fig:fit-mc-real-test-ds-vs-control}. The phase-folded light curve input features are shown in Fig.~\ref{fig:mags-features-real-test-ds} along with corresponding light curves based on the fitting input parameters including predictions, and the resulting fitted light curves.
	
	\begin{table}\centering
		\caption{The label values for AI Phe \citep{Maxted+20mnras} and the predicted parameters, including the calculated value for $i$, from 1000 MC Dropout iterations.}
		\label{tab:preds-ai-phe}
		\setlength\tabcolsep{3.5pt} 
		\begin{tabular}{cr@{\,$\pm$\,}lr@{\,$\pm$\,}lr@{\,$\pm$\,}l}	
			\hline
			parameter    & \multicolumn{2}{c}{label} & \multicolumn{2}{c}{prediction} & \multicolumn{2}{c}{error} \\ 
			\hline
			\multicolumn{7}{c}{AI Phe} \\ [3pt]			
			\rAplusrB    &      0.09895 &  0.00008   &     0.1063 &  0.0142    &    -0.0074 &  0.0142     \\
			$k$          &       1.6255 &  0.0026    &     1.6004 &  0.0435    &     0.0251 &  0.0436     \\
			$J$          & \multicolumn{2}{c}{ 0.50} &     0.5353 &  0.0169    &      -0.04 &  0.02       \\
			\bP          &        0.623 &  0.003     &     0.6408 &  0.0268    &     -0.018 &  0.027      \\
			\ecosw       &    -0.065161 &  0.000010  &    -0.0569 &  0.0088    &    -0.0083 &  0.0088     \\
			\esinw       &       0.1765 &  0.0008    &     0.1869 &  0.0233    &    -0.0104 &  0.0234     \\ [3pt]
			$i$ (\degr)  &       88.359 &  0.007     &    88.1471 &  0.2658    &      0.212 &  0.266      \\
			\hline
		\end{tabular}
		\\
		\vspace{1ex}
		{\raggedright \textbf{Notes.} \\
			The labels are taken from \cite{Maxted+20mnras} who do not publish a value for $J$ so we use an estimate based on the effective temperatures and luminosities published by \cite{Kirkby-KentMaxted+16aa}.
			\\
		}
	\end{table}
	
	The one system where the test fit failed was AI Phe (shown as a filled square in Figs.~\ref{fig:predictions-mc-real-test-ds-vs-labels} and \ref{fig:fit-mc-real-test-ds-vs-control}), despite the generally good predictions shown in Table \ref{tab:preds-ai-phe}. In this case \jktebop\ converged on an alternative solution with $i\simeq88.0\degr$ and $J\simeq0.91$ which \reff{was found to be highly attractive during the development of this project.} The labels and control fit of this system are based on analysis run K from \cite{Maxted+20mnras} which fits with \jktebop\ using the quadratic limb darkening law. Our simplified pipeline takes hints from the approach taken, specifically fixing the orbital period and the use of the quadratic limb darkening law with only the quadratic coefficients free to fit. With the limb darkening coefficients not published, \reff{the pipeline performed} a lookup from \cite{Claret18aa} based on effective temperatures \reff{for stars A and B} of $6257$ and $5100$ K and a metallicity of 0.0 \citep{Kirkby-KentMaxted+16aa}. These coefficients yield a usable control fit but are ineffective at constraining the analysis in the event of minor deviations in the other fitting parameters. A more reliable fit can be achieved by fixing the quadratic coefficients to the output values from the control fit but such an intervention would undermine the use of this system as a test subject.
	
	In contrast to AI Phe, a good fit was achieved for \reff{V436 Per, V889 Aql and V362 Pav} despite the shortcomings in the predictions discussed in section \ref{sec:preds-real-test-ds} and shown in Table \ref{tab:mc-preds-of-interest}. \reff{For V436 Per the predicted value of $k$ is a match for an alternative solution found by \cite{SouthworthBowman22mnras} in the reference analysis of the system. Typically test fits converged on this incorrect solution unless well constrained by the remaining input parameters, with the outcome being especially sensitive to the input value of \ecosw.}
	
	The control fit for V889 Aql made use of the spectroscopic light ratio of $0.98\pm0.06$ and limb darkening treatment from the reference characterisation of \cite{BarochGimenez+22aa}. As with all of the test systems, any such constraints \reff{were carried over to the test fit with both achieving a MRE across the fitted parameters of \rAplusrB, $k$, $J$, $i$, \ecosw\ and \esinw\ of $0.0059\pm0.0116$ when compared with the reference characterisation.} In the absence of the constraints, where limb darkening is handled by the quadratic law with run-time lookup of the coefficients based on published stellar characteristics, both control and test fit show a slight degradation of the \reff{MRE to $0.0125\pm0.0130$} suggesting that the predictions are sufficiently accurate for \jktebop\ to usefully characterise this system.
	
	V362 Pav has very shallow secondary eclipses, and for the control fit a mass ratio of 0.45 from the reference characterisation \citep{GraczykPietrzynski+22aa} is used with a fixed third light value of zero to yield an overall MRE for the fitted parameters of $0.0385\pm0.0053$. The test fit achieved a similar \reff{MRE of $0.0387\pm0.0053$} with the same constraints. \reff{Removing the constraints fixed the mass ratio at the pipeline's default value of 1.0 and allowed the third light to fit. This similarly affected both control and test fitting, with the MRE across the fitted parameters slightly worse at $0.0399\pm0.0075$ in both cases. However, if the default mass ratio was set 0.0 rather than 1.0 both the control and test fits converged on an incorrect, yet plausible, solution with a MRE of $0.0818\pm0.0067$. A simple automated pipeline without access to these constraints and suitable default parameters will produce a plausible, but incorrect characterisation for this system. However, the presence of shallow eclipses is a simple criterion with which to flag the solution, and those for similar systems, as potentially unreliable.}
	
	\section{Technical considerations}
	The code for this project and supporting documentation is publicly available on GitHub\footnote{\url{https://github.com/SteveOv/ebop\_maven/tree/v1.0}}. Those wishing to replicate the results reported here or to generate their own model can clone the entire repository which includes the pre-trained model in addition to supporting code for generating the training and test datasets, for training the model and for running the tests. It is also possible to install the \texttt{ebop\_maven} python package with a {\tt pip} command\footnote{\texttt{pip install git+https://github.com/SteveOv/ebop\_maven@v1.0}} which will install the subset of the repository which supports making predictions with the pre-trained model along with any dependencies.
	
	Interacting with a \tensorflow\ machine learning model directly can be difficult as it offers minimal information on the structure and order of its inputs and outputs. This is particularly true when using the MC Dropout technique, for which there is no native support, with client code responsible for invoking the model in training mode and aggregating the results from multiple sets of predictions. For these reasons an Estimator class has been implemented in the \ebopmaven\ code base which offers an easy way to interact with the machine learning model. Once the class has been initialised with an instance of a trained \ebopmaven\ model (it defaults to using the pre-trained model if none specified), predictions can be made by simply calling its \texttt{predict()} function. Inputs and outputs are in the form of \numpy\ arrays with the input feature for each instance a 4096 bin phase-folded light curve centred on the midpoint between the eclipses and with fluxes converted to relative magnitudes, and the output predictions are a structured array of \uncertainties\ UFloat types within named columns. If MC Dropout predictions are required, the \texttt{iterations} argument may be set to the desired number of iterations.
	
	Within the GitHub repo is a requirements file listing the packages necessary to support the training and testing of the machine learning model and instructions for setting up a \python\ virtual environment with them in place. Also included are instructions for downloading and installing both the \jktebop\ code, and the MIST isochrones used in the generation of the synthetic test dataset. No training or testing data is stored in the repo, however modules are included to generate these data which are required to support model training and testing. A further module is included, {\tt make\_trained\_cnn\_model.py}, which may be run to train a new copy of the model reported here, however the pre-trained model is included so this is optional. Also provided is a module to test a model both by evaluating predictions against labels and, for the real test systems only, performing test fitting with \jktebop\ and reporting the results against control fits. These codes will also produce plots similar to those presented here. Finally, two {\tt jupyter} notebooks are included which allow a more interactive approach to testing and additionally provide annotated sample code for using the \ebopmaven\ Estimator class within a typical scenario of downloading fits files from MAST, preparing the light curve data, predicting the fitting input parameters, invoking \jktebop\ to perform a fit, and reporting on the outcome.
	
	The development, training and testing of the \ebopmaven\ model was carried out on a consumer grade laptop with an 11th generation Intel i7 processer and 64 GiB RAM. \reff{While the laptop included a GPU capable of accelerating machine learning applications with \tensorflow, it was disabled to ensure repeatable results. In this configuration, building the training and test datasets took approximately 6.5 hours, training the model responsible for the results reported here completed in approximately 6 hours, and the full suite of tests, including performing the \jktebop\ fits of the real systems' \tess\ data, took 2 hours. The training and testing datasets take up approximately 16.5 GiB of disk space.}
	
	\reff{The testing module records some rudimentary timings while conducting its tests. On the hardware described above, initializing an Estimator takes less than 1 second and predictions on the phase-folded light curve feature of a single test instance takes less than 0.05~s. Predictions with 1000 MC Dropout iterations on the same instance extends the average time taken to approximately 20~s, reflecting the 1000 implicit predictions made within the Estimator. Calls to predict scale well on multiple instances: making predictions on 1000 test instances in a single call takes approximately 0.3~s, extending to 200~s if 1000 MC Dropout iterations are requested}. This compares very favourably to the linear scaling of predicting each instance separately, however the number of instances supported by this approach will be dictated by available RAM and tends to be further reduced if MC Dropout is used. Use of more recent hardware or GPU-acceleration will likely see these timings improve.
	
	\section{Conclusion}
	This work covers the development of the \ebopmaven\ project. The core component is a one-dimensional CNN regression model trained to predict six key orbital parameters from an input feature consisting of a system's phase-folded light curve. The predictions are intended to serve as input parameters for subsequent formal analysis by the \jktebop\ code. Training was based on 1 million synthetic instances split 80:20 between training and validation datasets. To mitigate overfitting the datasets were shuffled during training, and augmentations were applied which added random noise and shifts to each instance.
	
	Testing was carried out with two datasets: the first a synthetic dataset of 20\,000 physically plausible systems generated from MIST stellar models and \tess\ photometric parameters, and the second a smaller dataset of real systems selected  on the availability of \tess\ time-series photometry and a published characterisation leading to membership of DEBCat. \reff{Predictions made on the test datasets were evaluated against the label values with the \ebopmaven\ model achieving a mean relative error of 14.1\% on the synthetic dataset, improving to 8.6\% on the subset of test instances with eclipses deeper the 0.1 mag. On the test dataset of real systems the predictions achieved a mean relative error of $8.7\pm0.7\%$. This second dataset was further tested by fitting the systems' \tess\ light curve data with \jktebop\ while using the predictions as input parameters. Of the 28 systems in the dataset, 27 yielded an analysis in agreement with a corresponding control fit.}
	
	This work has demonstrated the feasibility of using a machine learning model to perform an initial analysis of \reff{phase-folded dEB light curves in order to find an appropriate starting position for full formal analysis with the \jktebop\ code.} The next step is to incorporate this approach into a reusable dEB characterisation pipeline combining \tess\ photometry, astrometric and spectroscopic data from \gaia, and ephemerides from the TESS-EBS (TESS Eclipsing Binary Stars; \citealt{PrsaKochoska+22apjs}) or TIDAK (TIming DAtabase at Krakow; \citealt{Kreiner04acta}) catalogues or analysis with codes such as STAR SHADOW \citep{IJspeertTkachenko+24aa}. An input catalogue of candidate EBs will be required, which will be assembled from sources such as TESS-EBS and similar catalogues (e.g. \citealt{IJspeertTkachenko+21aa, JustesenAlbrecht21apj}), and by cross referencing the \tess\ input catalogue \citep{StassunOelkers+19aj} with \gaia\ data such as the DR3 catalogue of EB candidates \citep{MowlaviHoll+23aa}. Subsequently, the pipeline will be used to assemble a catalogue of sufficiently well-characterised dEB systems, covering photometric parameters and estimated physical properties, for use in selecting targets for observation with PLATO and subsequent observational campaigns.
	
	\section*{Acknowledgements}
	We would like to thank the anonymous reviewers for their reports and Zac Jennings for his extensive feedback on an early draft of this manuscript.
	
	SO gratefully acknowledges the financial support of the UK Science and Technologies Facilities Council (STFC) in the form of a PhD studentship. JS acknowledges support from STFC under grant number ST/Y002563/1. This paper includes data collected by the \tess\ mission and obtained from the MAST data archive at the Space Telescope Science Institute (STScI). Funding for the \tess\ mission is provided by NASA's Science Mission Directorate. STScI is operated by the Association of Universities for Research in Astronomy, Inc., under NASA contract NAS 5–26555.
	
	The following resources were also used in the course of this work: the NASA Astrophysics Data System; the ar$\chi$iv scientific paper preprint service operated by Cornell University; the SIMBAD database and VizieR catalogue access tool operated by the Centre des Donn\'ees Stellaires, Strasbourg, France; the \python\ programming language and Python Package Index (PyPI) maintained by the Python Software Foundation (\url{https://www.python.org}).
	
	\section*{Data availability}
	The \tess\ data used in this work are available in the MAST archive (\url{https://mast.stsci.edu/portal/Mashup/Clients/Mast/Portal.html}). The packaged theoretical stellar isochrones used are available from the MIST web site (\url{http://waps.cfa.harvard.edu/MIST/data/tarballs_v1.2/MIST_v1.2_vvcrit0.4_basic_isos.txz}). The tables of quadratic and power-2 limb darkening coefficients used are available from the VizieR catalogue (\url{https://cdsarc.cds.unistra.fr/viz-bin/cat/J/A+A/618/A20} and \url{https://cdsarc.cds.unistra.fr/viz-bin/cat/J/A+A/674/A63}). The code to generate training and testing data, and to train the \ebopmaven\ model is available on GitHub (\url{https://github.com/SteveOv/ebop_maven/tree/v1.0}). Bugs in the \ebopmaven\ code may be reported through the issue tracking facility of the GitHub repository.
	
	
	\newpage
	\bibliographystyle{rasti}
	\bibliography{bib/zotero.bib,bib/additional.bib}

\begin{thebibliography}{121}
\expandafter\ifx\csname natexlab\endcsname\relax\def\natexlab#1{#1}\fi

\bibitem[Abadi et~al.(2015)Abadi, Agarwal, Barham, Brevdo, Chen, Citro,
  Corrado, Davis, Dean, Devin, Ghemawat, Goodfellow, Harp, Irving, Isard, Jia,
  Jozefowicz, Kaiser, Kudlur, Levenberg, Man\'{e}, Monga, Moore, Murray, Olah,
  Schuster, Shlens, Steiner, Sutskever, Talwar, Tucker, Vanhoucke, Vasudevan,
  Vi\'{e}gas, Vinyals, Warden, Wattenberg, Wicke, Yu, \&
  Zheng]{Abadi+2015tensorflow}
Abadi, M., Agarwal, A., Barham, P., Brevdo, E., Chen, Z., Citro, C., Corrado,
  G.~S., Davis, A., Dean, J., Devin, M., Ghemawat, S., Goodfellow, I., Harp,
  A., Irving, G., Isard, M., Jia, Y., Jozefowicz, R., Kaiser, L., Kudlur, M.,
  Levenberg, J., Man\'{e}, D., Monga, R., Moore, S., Murray, D., Olah, C.,
  Schuster, M., Shlens, J., Steiner, B., Sutskever, I., Talwar, K., Tucker, P.,
  Vanhoucke, V., Vasudevan, V., Vi\'{e}gas, F., Vinyals, O., Warden, P.,
  Wattenberg, M., Wicke, M., Yu, Y., \& Zheng, X., 2015.
\newblock {TensorFlow}: Large-scale machine learning on heterogeneous systems,
  Software available from tensorflow.org.

\bibitem[Abt \& Levato(1978)]{AbtLevato78pasp}
Abt, H.~A. \& Levato, H., 1978.
\newblock Spectral types in the {{Pleiades}}., {\it PASP\/}, {\bf 90},
  201--203.

\bibitem[Andersen(1991)]{Andersen91aarv}
Andersen, J., 1991.
\newblock Accurate masses and radii of normal stars, {\it A\&ARv\/}, {\bf
  3}(2), 91.

\bibitem[{Astropy Collaboration} et~al.(2022){Astropy Collaboration},
  {Price-Whelan}, Lim, Earl, Starkman, Bradley, Shupe, Patil, Corrales,
  Brasseur, N{\"o}the, Donath, Tollerud, Morris, Ginsburg, Vaher, Weaver,
  Tocknell, Jamieson, {van Kerkwijk}, Robitaille, Merry, Bachetti, G{\"u}nther,
  Aldcroft, {Alvarado-Montes}, Archibald, B{\'o}di, Bapat, Barentsen,
  Baz{\'a}n, Biswas, Boquien, Burke, Cara, Cara, Conroy, Conseil, Craig, Cross,
  Cruz, D'Eugenio, Dencheva, Devillepoix, Dietrich, Eigenbrot, Erben, Ferreira,
  {Foreman-Mackey}, Fox, Freij, Garg, Geda, Glattly, Gondhalekar, Gordon,
  Grant, Greenfield, Groener, Guest, Gurovich, Handberg, Hart,
  {Hatfield-Dodds}, Homeier, Hosseinzadeh, Jenness, Jones, Joseph, Kalmbach,
  Karamehmetoglu, Ka{\l}uszy{\'n}ski, Kelley, Kern, Kerzendorf, Koch, Kulumani,
  Lee, Ly, Ma, MacBride, Maljaars, Muna, Murphy, Norman, O'Steen, Oman,
  Pacifici, Pascual, {Pascual-Granado}, Patil, Perren, Pickering, Rastogi,
  Roulston, Ryan, Rykoff, Sabater, Sakurikar, Salgado, Sanghi, Saunders,
  Savchenko, Schwardt, {Seifert-Eckert}, Shih, Jain, Shukla, Sick, Simpson,
  Singanamalla, Singer, Singhal, Sinha, Sip{\H o}cz, Spitler, Stansby,
  Streicher, {\v S}umak, Swinbank, Taranu, Tewary, Tremblay, {de Val-Borro},
  Van~Kooten, Vasovi{\'c}, Verma, {de Miranda Cardoso}, Williams, Wilson,
  Winkel, {Wood-Vasey}, Xue, Yoachim, Zhang, Zonca, \& {Astropy Project
  Contributors}]{Astropy22apj}
{Astropy Collaboration}, {Price-Whelan}, A.~M., Lim, P.~L., Earl, N., Starkman,
  N., Bradley, L., Shupe, D.~L., Patil, A.~A., Corrales, L., Brasseur, C.~E.,
  N{\"o}the, M., Donath, A., Tollerud, E., Morris, B.~M., Ginsburg, A., Vaher,
  E., Weaver, B.~A., Tocknell, J., Jamieson, W., {van Kerkwijk}, M.~H.,
  Robitaille, T.~P., Merry, B., Bachetti, M., G{\"u}nther, H.~M., Aldcroft,
  T.~L., {Alvarado-Montes}, J.~A., Archibald, A.~M., B{\'o}di, A., Bapat, S.,
  Barentsen, G., Baz{\'a}n, J., Biswas, M., Boquien, M., Burke, D.~J., Cara,
  D., Cara, M., Conroy, K.~E., Conseil, S., Craig, M.~W., Cross, R.~M., Cruz,
  K.~L., D'Eugenio, F., Dencheva, N., Devillepoix, H. A.~R., Dietrich, J.~P.,
  Eigenbrot, A.~D., Erben, T., Ferreira, L., {Foreman-Mackey}, D., Fox, R.,
  Freij, N., Garg, S., Geda, R., Glattly, L., Gondhalekar, Y., Gordon, K.~D.,
  Grant, D., Greenfield, P., Groener, A.~M., Guest, S., Gurovich, S., Handberg,
  R., Hart, A., {Hatfield-Dodds}, Z., Homeier, D., Hosseinzadeh, G., Jenness,
  T., Jones, C.~K., Joseph, P., Kalmbach, J.~B., Karamehmetoglu, E.,
  Ka{\l}uszy{\'n}ski, M., Kelley, M. S.~P., Kern, N., Kerzendorf, W.~E., Koch,
  E.~W., Kulumani, S., Lee, A., Ly, C., Ma, Z., MacBride, C., Maljaars, J.~M.,
  Muna, D., Murphy, N.~A., Norman, H., O'Steen, R., Oman, K.~A., Pacifici, C.,
  Pascual, S., {Pascual-Granado}, J., Patil, R.~R., Perren, G.~I., Pickering,
  T.~E., Rastogi, T., Roulston, B.~R., Ryan, D.~F., Rykoff, E.~S., Sabater, J.,
  Sakurikar, P., Salgado, J., Sanghi, A., Saunders, N., Savchenko, V.,
  Schwardt, L., {Seifert-Eckert}, M., Shih, A.~Y., Jain, A.~S., Shukla, G.,
  Sick, J., Simpson, C., Singanamalla, S., Singer, L.~P., Singhal, J., Sinha,
  M., Sip{\H o}cz, B.~M., Spitler, L.~R., Stansby, D., Streicher, O., {\v
  S}umak, J., Swinbank, J.~D., Taranu, D.~S., Tewary, N., Tremblay, G.~R., {de
  Val-Borro}, M., Van~Kooten, S.~J., Vasovi{\'c}, Z., Verma, S., {de Miranda
  Cardoso}, J.~V., Williams, P. K.~G., Wilson, T.~J., Winkel, B., {Wood-Vasey},
  W.~M., Xue, R., Yoachim, P., Zhang, C., Zonca, A., \& {Astropy Project
  Contributors}, 2022.
\newblock The {{Astropy Project}}: {{Sustaining}} and {{Growing}} a
  {{Community-oriented Open-source Project}} and the {{Latest Major Release}}
  (v5.0) of the {{Core Package}}, {\it ApJ\/}, {\bf 935}, 167.

\bibitem[Auvergne et~al.(2009)Auvergne, Bodin, Boisnard, Buey, Chaintreuil,
  Epstein, Jouret, {Lam-Trong}, Levacher, Magnan, Perez, Plasson, Plesseria,
  Peter, Steller, Tiph{\`e}ne, Baglin, Agogu{\'e}, Appourchaux, Barbet,
  Beaufort, Bellenger, Berlin, Bernardi, Blouin, Boumier, Bonneau, Briet,
  Butler, Cautain, Chiavassa, Costes, Cuvilho, {Cunha-Parro}, Fialho, Decaudin,
  Defise, Djalal, Docclo, Drummond, Dupuis, Exil, Faur{\'e}, Gaboriaud, Gamet,
  Gavalda, Grolleau, Gueguen, Guivarc'h, Guterman, Hasiba, Huntzinger, Hustaix,
  Imbert, Jeanville, Johlander, Jorda, Journoud, Karioty, Kerjean, Lafond,
  Lapeyrere, Landiech, Larqu{\'e}, Laudet, Merrer, Leporati, Leruyet, Levieuge,
  Llebaria, Martin, Mazy, Mesnager, Michel, Moalic, Monjoin, Naudet,
  Neukirchner, {Nguyen-Kim}, Ollivier, Orcesi, Ottacher, Oulali, Parisot,
  Perruchot, Piacentino, da~Silva, Platzer, Pontet, Pradines, Quentin, Rohbeck,
  Rolland, Rollenhagen, Romagnan, Russ, Samadi, Schmidt, Schwartz, Sebbag,
  Smit, Sunter, Tello, Toulouse, Ulmer, Vandermarcq, Vergnault, Wallner,
  Waultier, \& Zanatta]{AuvergneBodin+09aa}
Auvergne, M., Bodin, P., Boisnard, L., Buey, J.-T., Chaintreuil, S., Epstein,
  G., Jouret, M., {Lam-Trong}, T., Levacher, P., Magnan, A., Perez, R.,
  Plasson, P., Plesseria, J., Peter, G., Steller, M., Tiph{\`e}ne, D., Baglin,
  A., Agogu{\'e}, P., Appourchaux, T., Barbet, D., Beaufort, T., Bellenger, R.,
  Berlin, R., Bernardi, P., Blouin, D., Boumier, P., Bonneau, F., Briet, R.,
  Butler, B., Cautain, R., Chiavassa, F., Costes, V., Cuvilho, J.,
  {Cunha-Parro}, V., Fialho, F. D.~O., Decaudin, M., Defise, J.-M., Djalal, S.,
  Docclo, A., Drummond, R., Dupuis, O., Exil, G., Faur{\'e}, C., Gaboriaud, A.,
  Gamet, P., Gavalda, P., Grolleau, E., Gueguen, L., Guivarc'h, V., Guterman,
  P., Hasiba, J., Huntzinger, G., Hustaix, H., Imbert, C., Jeanville, G.,
  Johlander, B., Jorda, L., Journoud, P., Karioty, F., Kerjean, L., Lafond, L.,
  Lapeyrere, V., Landiech, P., Larqu{\'e}, T., Laudet, P., Merrer, J.~L.,
  Leporati, L., Leruyet, B., Levieuge, B., Llebaria, A., Martin, L., Mazy, E.,
  Mesnager, J.-M., Michel, J.-P., Moalic, J.-P., Monjoin, W., Naudet, D.,
  Neukirchner, S., {Nguyen-Kim}, K., Ollivier, M., Orcesi, J.-L., Ottacher, H.,
  Oulali, A., Parisot, J., Perruchot, S., Piacentino, A., da~Silva, L.~P.,
  Platzer, J., Pontet, B., Pradines, A., Quentin, C., Rohbeck, U., Rolland, G.,
  Rollenhagen, F., Romagnan, R., Russ, N., Samadi, R., Schmidt, R., Schwartz,
  N., Sebbag, I., Smit, H., Sunter, W., Tello, M., Toulouse, P., Ulmer, B.,
  Vandermarcq, O., Vergnault, E., Wallner, R., Waultier, G., \& Zanatta, P.,
  2009.
\newblock The {{CoRoT}} satellite in flight: Description and performance, {\it
  A\&A\/}, {\bf 506}(1), 411--424.

\bibitem[Bak{\i}{\c s} et~al.(2014)Bak{\i}{\c s}, Hensberge, Bilir, Bak{\i}{\c
  s}, Y{\i}lmaz, K{\i}ran, Demircan, Zejda, \& Mikul{\'a}{\v
  s}ek]{BakisHensberge+14aj}
Bak{\i}{\c s}, V., Hensberge, H., Bilir, S., Bak{\i}{\c s}, H., Y{\i}lmaz, F.,
  K{\i}ran, E., Demircan, O., Zejda, M., \& Mikul{\'a}{\v s}ek, Z., 2014.
\newblock Study of {{Eclipsing Binary}} and {{Multiple Systems}} in {{OB
  Associations}}. {{II}}. {{The Cygnus OB Region}}: {{V443 Cyg}}, {{V456 Cyg}},
  and {{V2107 Cyg}}, {\it AJ\/}, {\bf 147}, 149.

\bibitem[Ball \& Brunner(2010)]{BallBrunner10ijmp}
Ball, N.~M. \& Brunner, R.~J., 2010.
\newblock Data mining and machine learning in astronomy, {\it Int. J. Mod.
  Phys. D\/}, {\bf 19}(07), 1049--1106.

\bibitem[Barbara et~al.(2022)Barbara, Bedding, Fulcher, Murphy, \&
  Van~Reeth]{BarbaraBedding+22mnras}
Barbara, N.~H., Bedding, T.~R., Fulcher, B.~D., Murphy, S.~J., \& Van~Reeth,
  T., 2022.
\newblock Classifying {{Kepler}} light curves for 12\,000 {{A}} and {{F}} stars
  using supervised feature-based machine learning, {\it MNRAS\/}, {\bf 514}(2),
  2793--2804.

\bibitem[Baroch et~al.(2022)Baroch, Gim{\'e}nez, Morales, Ribas, Herrero,
  Perdelwitz, Jordi, Granzer, \& Allende~Prieto]{BarochGimenez+22aa}
Baroch, D., Gim{\'e}nez, A., Morales, J.~C., Ribas, I., Herrero, E.,
  Perdelwitz, V., Jordi, C., Granzer, T., \& Allende~Prieto, C., 2022.
\newblock Absolute dimensions and apsidal motion of the eclipsing binaries
  {{V889 Aquilae}} and {{V402 Lacertae}}, {\it A\&A\/}, {\bf 665}, A13.

\bibitem[Baron(2019)]{Baron19}
Baron, D., 2019.
\newblock Machine {{Learning}} in {{Astronomy}}: A practical overview.

\bibitem[Benz et~al.(2021)Benz, Broeg, Fortier, Rando, Beck, Beck, Queloz,
  Ehrenreich, Maxted, Isaak, Billot, Alibert, Alonso, Ant{\'o}nio, Asquier,
  Bandy, B{\'a}rczy, Barrado, Barros, Baumjohann, Bekkelien, Bergomi, Biondi,
  Bonfils, Borsato, Brandeker, Busch, Cabrera, Cessa, Charnoz, Chazelas,
  Collier~Cameron, Corral Van~Damme, Cortes, Davies, Deleuil, Deline, Delrez,
  Demangeon, Demory, Erikson, Farinato, Fossati, Fridlund, Futyan, Gandolfi,
  Garcia~Munoz, Gillon, Guterman, Gutierrez, Hasiba, Heng, Hernandez, Hoyer,
  Kiss, Kovacs, Kuntzer, Laskar, {Lecavelier des Etangs}, Lendl, L{\'o}pez,
  Lora, Lovis, L{\"u}ftinger, Magrin, Malvasio, Marafatto, Michaelis, {de
  Miguel}, Modrego, Munari, Nascimbeni, Olofsson, Ottacher, Ottensamer, Pagano,
  Palacios, Pall{\'e}, Peter, Piazza, Piotto, Pizarro, Pollaco, Ragazzoni,
  Ratti, Rauer, Ribas, Rieder, Rohlfs, Safa, Salatti, Santos, Scandariato,
  S{\'e}gransan, Simon, Smith, Sordet, Sousa, Steller, Szab{\'o}, Szoke,
  Thomas, Tschentscher, Udry, Van~Grootel, Viotto, Walter, Walton, Wildi, \&
  Wolter]{BenzBroeg+21expastron}
Benz, W., Broeg, C., Fortier, A., Rando, N., Beck, T., Beck, M., Queloz, D.,
  Ehrenreich, D., Maxted, P. F.~L., Isaak, K.~G., Billot, N., Alibert, Y.,
  Alonso, R., Ant{\'o}nio, C., Asquier, J., Bandy, T., B{\'a}rczy, T., Barrado,
  D., Barros, S. C.~C., Baumjohann, W., Bekkelien, A., Bergomi, M., Biondi, F.,
  Bonfils, X., Borsato, L., Brandeker, A., Busch, M.-D., Cabrera, J., Cessa,
  V., Charnoz, S., Chazelas, B., Collier~Cameron, A., Corral Van~Damme, C.,
  Cortes, D., Davies, M.~B., Deleuil, M., Deline, A., Delrez, L., Demangeon,
  O., Demory, B.~O., Erikson, A., Farinato, J., Fossati, L., Fridlund, M.,
  Futyan, D., Gandolfi, D., Garcia~Munoz, A., Gillon, M., Guterman, P.,
  Gutierrez, A., Hasiba, J., Heng, K., Hernandez, E., Hoyer, S., Kiss, L.~L.,
  Kovacs, Z., Kuntzer, T., Laskar, J., {Lecavelier des Etangs}, A., Lendl, M.,
  L{\'o}pez, A., Lora, I., Lovis, C., L{\"u}ftinger, T., Magrin, D., Malvasio,
  L., Marafatto, L., Michaelis, H., {de Miguel}, D., Modrego, D., Munari, M.,
  Nascimbeni, V., Olofsson, G., Ottacher, H., Ottensamer, R., Pagano, I.,
  Palacios, R., Pall{\'e}, E., Peter, G., Piazza, D., Piotto, G., Pizarro, A.,
  Pollaco, D., Ragazzoni, R., Ratti, F., Rauer, H., Ribas, I., Rieder, M.,
  Rohlfs, R., Safa, F., Salatti, M., Santos, N.~C., Scandariato, G.,
  S{\'e}gransan, D., Simon, A.~E., Smith, A. M.~S., Sordet, M., Sousa, S.~G.,
  Steller, M., Szab{\'o}, G.~M., Szoke, J., Thomas, N., Tschentscher, M., Udry,
  S., Van~Grootel, V., Viotto, V., Walter, I., Walton, N.~A., Wildi, F., \&
  Wolter, D., 2021.
\newblock The {{CHEOPS}} mission, {\it Exp. Astron.\/}, {\bf 51}(1), 109--151.

\bibitem[Bergstra et~al.(2011)Bergstra, Bardenet, Bengio, \&
  K{\'e}gl]{Bergstra+11algos}
Bergstra, J., Bardenet, R., Bengio, Y., \& K{\'e}gl, B., 2011.
\newblock Algorithms for {{Hyper-Parameter Optimization}}, in {\em Advances in
  {{Neural Information Processing Systems}}\/}, vol.~24, Curran Associates,
  Inc.

\bibitem[Bergstra et~al.(2013)Bergstra, Yamins, \& Cox]{Bergstra+2013hyperopt}
Bergstra, J., Yamins, D., \& Cox, D., 2013.
\newblock Making a {{Science}} of {{Model Search}}: {{Hyperparameter
  Optimization}} in {{Hundreds}} of {{Dimensions}} for {{Vision
  Architectures}}, in {\em Proceedings of the 30th {{International Conference}}
  on {{Machine Learning}}\/}, pp. 115--123, PMLR.

\bibitem[Blancato et~al.(2022)Blancato, Ness, Huber, Lu, \&
  Angus]{BlancatoNess+22apj}
Blancato, K., Ness, M.~K., Huber, D., Lu, Y., \& Angus, R., 2022.
\newblock Data-driven {{Derivation}} of {{Stellar Properties}} from
  {{Photometric Time Series Data Using Convolutional Neural Networks}}, {\it
  ApJ\/}, {\bf 933}, 241.

\bibitem[Borucki(2016)]{Borucki+16rpf:Kepler}
Borucki, W.~J., 2016.
\newblock {{KEPLER Mission}}: Development and overview, {\it Rep. Prog.
  Phys.\/}, {\bf 79}(3), 036901.

\bibitem[Breton et~al.(2021)Breton, Santos, Bugnet, Mathur, Garc{\'i}a, \&
  Pall{\'e}]{BretonSantos+21aa}
Breton, S.~N., Santos, A. R.~G., Bugnet, L., Mathur, S., Garc{\'i}a, R.~A., \&
  Pall{\'e}, P.~L., 2021.
\newblock {{ROOSTER}}: A machine-learning analysis tool for {{Kepler}} stellar
  rotation periods, {\it A\&A\/}, {\bf 647}, A125.

\bibitem[{Castro-Ginard} et~al.(2018){Castro-Ginard}, Jordi, Luri, Julbe,
  Morvan, {Balaguer-N{\'u}{\~n}ez}, \&
  {Cantat-Gaudin}]{Castro-Ginard+18aa:DBSCAN}
{Castro-Ginard}, A., Jordi, C., Luri, X., Julbe, F., Morvan, M.,
  {Balaguer-N{\'u}{\~n}ez}, L., \& {Cantat-Gaudin}, T., 2018.
\newblock A new method for unveiling open clusters in {{Gaia}}. {{New}} nearby
  open clusters confirmed by {{DR2}}, {\it A\&A\/}, {\bf 618}, A59.

\bibitem[Chabrier(2003)]{Chabrier03pasp}
Chabrier, G., 2003.
\newblock Galactic {{Stellar}} and {{Substellar Initial Mass Function}}, {\it
  PASP\/}, {\bf 115}, 763--795.

\bibitem[Chen et~al.(2022)Chen, Ding, Cheng, Zhang, Li, Ji, Xiong, Li, \&
  Luo]{ChenDing+22apjs}
Chen, X., Ding, X., Cheng, L., Zhang, X., Li, Y., Ji, K., Xiong, J., Li, X., \&
  Luo, C., 2022.
\newblock Detection of {$\delta$} {{Scuti Pulsators}} in the {{Eclipsing
  Binaries Observed}} by {{TESS}}, {\it ApJS\/}, {\bf 263}(2), 34.

\bibitem[Choi et~al.(2016)Choi, Dotter, Conroy, Cantiello, Paxton, \&
  Johnson]{ChoiDotter+16apj}
Choi, J., Dotter, A., Conroy, C., Cantiello, M., Paxton, B., \& Johnson, B.~D.,
  2016.
\newblock {{MESA ISOCHRONES AND S}}tel{{LAR TRACKS}} ({{MIST}}). {{I}}.
  {{SOLAR-SCALED MODELS}}, {\it ApJ\/}, {\bf 823}(2), 102.

\bibitem[Chollet et~al.(2015)]{Chollet+2015keras}
Chollet, F. et~al., 2015.
\newblock Keras, \url{https://keras.io}.

\bibitem[Claret(2018)]{Claret18aa}
Claret, A., 2018.
\newblock A new method to compute limb-darkening coefficients for stellar
  atmosphere models with spherical symmetry: The space missions {{TESS}},
  {{Kepler}}, {{CoRoT}}, and {{MOST}}, {\it A\&A\/}, {\bf 618}, A20.

\bibitem[Claret \& Southworth(2023)]{ClaretSouthworth23aa}
Claret, A. \& Southworth, J., 2023.
\newblock Power-2 limb-darkening coefficients for the uvby, {{UBVRIJHK}},
  {{SDSS}} ugriz, {{Gaia}}, {{Kepler}}, {{TESS}}, and {{CHEOPS}} photometric
  systems. {{II}}. {{PHOENIX}} spherically symmetric stellar atmosphere models,
  {\it A\&A\/}, {\bf 674}, A63.

\bibitem[Claret \& Torres(2016)]{ClaretTorres16aa}
Claret, A. \& Torres, G., 2016.
\newblock The dependence of convective core overshooting on stellar mass, {\it
  A\&A\/}, {\bf 592}, A15.

\bibitem[Claret \& Torres(2018)]{ClaretTorres18apj}
Claret, A. \& Torres, G., 2018.
\newblock The {{Dependence}} of {{Convective Core Overshooting}} on {{Stellar
  Mass}}: {{Additional Binary Systems}} and {{Improved Calibration}}, {\it
  ApJ\/}, {\bf 859}(2), 100.

\bibitem[Deleuil et~al.(2018)Deleuil, Aigrain, Moutou, Cabrera, Bouchy, Deeg,
  Almenara, H{\'e}brard, Santerne, Alonso, Bonomo, Bord{\'e}, Csizmadia,
  D{\`i}az, Erikson, Fridlund, Gandolfi, Guenther, Guillot, Guterman, Grziwa,
  Hatzes, L{\'e}ger, Mazeh, Ofir, Ollivier, P{\"a}tzold, Parviainen, Rauer,
  Rouan, Schneider, {Titz-Weider}, Tingley, \& Weingrill]{Deleuil+18aa:CoRoT}
Deleuil, M., Aigrain, S., Moutou, C., Cabrera, J., Bouchy, F., Deeg, H.~J.,
  Almenara, J.-M., H{\'e}brard, G., Santerne, A., Alonso, R., Bonomo, A.~S.,
  Bord{\'e}, P., Csizmadia, S., D{\`i}az, R.~F., Erikson, A., Fridlund, M.,
  Gandolfi, D., Guenther, E., Guillot, T., Guterman, P., Grziwa, S., Hatzes,
  A., L{\'e}ger, A., Mazeh, T., Ofir, A., Ollivier, M., P{\"a}tzold, M.,
  Parviainen, H., Rauer, H., Rouan, D., Schneider, J., {Titz-Weider}, R.,
  Tingley, B., \& Weingrill, J., 2018.
\newblock Planets, candidates, and binaries from the {{CoRoT}}/{{Exoplanet}}
  programme - {{The CoRoT}} transit catalogue, {\it A\&A\/}, {\bf 619}, A97.

\bibitem[Demircan \& Kahraman(1991)]{DemircanKahraman91apss}
Demircan, O. \& Kahraman, G., 1991.
\newblock Stellar {{Mass}} / {{Luminosity}} and {{Mass}} / {{Radius
  Relations}}, {\it Ap\&SS\/}, {\bf 181}, 313--322.

\bibitem[Dotter(2016)]{Dotter16apjs}
Dotter, A., 2016.
\newblock {{MESA Isochrones}} and {{Stellar Tracks}} ({{MIST}}) 0: {{Methods}}
  for the {{Construction}} of {{Stellar Isochrones}}, {\it ApJS\/}, {\bf 222},
  8.

\bibitem[{Gaia Collaboration}(2016)]{GaiaCollab16aa:Gaia}
{Gaia Collaboration}, 2016.
\newblock The {{Gaia}} mission, {\it A\&A\/}, {\bf 595}, A1.

\bibitem[Gal \& Ghahramani(2016)]{Gal+Chahramani16MLmcdropout}
Gal, Y. \& Ghahramani, Z., 2016.
\newblock Dropout as a {{Bayesian Approximation}}: {{Representing Model
  Uncertainty}} in {{Deep Learning}}.

\bibitem[{Garcia-Dias} et~al.(2018){Garcia-Dias}, Allende~Prieto,
  S{\'a}nchez~Almeida, \& {Ordov{\'a}s-Pascual}]{Garcia-Dias+18aa:kmeans}
{Garcia-Dias}, R., Allende~Prieto, C., S{\'a}nchez~Almeida, J., \&
  {Ordov{\'a}s-Pascual}, I., 2018.
\newblock Machine learning in {{APOGEE}}. {{Unsupervised}} spectral
  classification with {{K-means}}, {\it A\&A\/}, {\bf 612}, A98.

\bibitem[Giles \& Walkowicz(2019)]{GilesWalkowicz19mnras}
Giles, D. \& Walkowicz, L., 2019.
\newblock Systematic serendipity: A test of unsupervised machine learning as a
  method for anomaly detection, {\it MNRAS\/}, {\bf 484}(1), 834--849.

\bibitem[Giles \& Walkowicz(2020)]{GilesWalkowicz20mnras}
Giles, D.~K. \& Walkowicz, L., 2020.
\newblock Density-based outlier scoring on {{Kepler}} data, {\it MNRAS\/}, {\bf
  499}, 524--542.

\bibitem[Graczyk et~al.(2016)Graczyk, Smolec, Pavlovski, Southworth,
  Pietrzy{\'n}ski, Maxted, Konorski, Gieren, Pilecki, Taormina, Suchomska,
  Karczmarek, G{\'o}rski, Wielg{\'o}rski, \& Anderson]{GraczykSmolec+16aa}
Graczyk, D., Smolec, R., Pavlovski, K., Southworth, J., Pietrzy{\'n}ski, G.,
  Maxted, P. F.~L., Konorski, P., Gieren, W., Pilecki, B., Taormina, M.,
  Suchomska, K., Karczmarek, P., G{\'o}rski, M., Wielg{\'o}rski, P., \&
  Anderson, R.~I., 2016.
\newblock A solar twin in the eclipsing binary {{LL Aquarii}}, {\it A\&A\/},
  {\bf 594}, A92.

\bibitem[Graczyk et~al.(2022)Graczyk, Pietrzy{\'n}ski, Galan, Southworth,
  Gieren, Ka{\l}uszy{\'n}ski, Zgirski, Gallenne, G{\'o}rski, Hajdu, Karczmarek,
  Kervella, Maxted, Nardetto, Narloch, Pilecki, Pych, Rojas~Garcia, Storm,
  Suchomska, Taormina, \& Wielg{\'o}rski]{GraczykPietrzynski+22aa}
Graczyk, D., Pietrzy{\'n}ski, G., Galan, C., Southworth, J., Gieren, W.,
  Ka{\l}uszy{\'n}ski, M., Zgirski, B., Gallenne, A., G{\'o}rski, M., Hajdu, G.,
  Karczmarek, P., Kervella, P., Maxted, P. F.~L., Nardetto, N., Narloch, W.,
  Pilecki, B., Pych, W., Rojas~Garcia, G., Storm, J., Suchomska, K., Taormina,
  M., \& Wielg{\'o}rski, P., 2022.
\newblock Surface brightness-colour relations of dwarf stars from detached
  eclipsing binaries. {{I}}. {{Calibrating}} sample, {\it A\&A\/}, {\bf 666},
  A128.

\bibitem[Harris et~al.(2020)Harris, Millman, van~der Walt, Gommers, Virtanen,
  Cournapeau, Wieser, Taylor, Berg, Smith, Kern, Picus, Hoyer, van Kerkwijk,
  Brett, Haldane, del R{\'{i}}o, Wiebe, Peterson, G{\'{e}}rard-Marchant,
  Sheppard, Reddy, Weckesser, Abbasi, Gohlke, \& Oliphant]{Harris+2020numpy}
Harris, C.~R., Millman, K.~J., van~der Walt, S.~J., Gommers, R., Virtanen, P.,
  Cournapeau, D., Wieser, E., Taylor, J., Berg, S., Smith, N.~J., Kern, R.,
  Picus, M., Hoyer, S., van Kerkwijk, M.~H., Brett, M., Haldane, A., del
  R{\'{i}}o, J.~F., Wiebe, M., Peterson, P., G{\'{e}}rard-Marchant, P.,
  Sheppard, K., Reddy, T., Weckesser, W., Abbasi, H., Gohlke, C., \& Oliphant,
  T.~E., 2020.
\newblock Array programming with {NumPy}, {\it Nature\/}, {\bf 585}(7825),
  357--362.

\bibitem[He et~al.(2015)He, Zhang, Ren, \& Sun]{HeZhang+15}
He, K., Zhang, X., Ren, S., \& Sun, J., 2015.
\newblock Delving {{Deep}} into {{Rectifiers}}: {{Surpassing Human-Level
  Performance}} on {{ImageNet Classification}}.

\bibitem[{Hilditch}(2001)]{Hilditch01book}
{Hilditch}, R.~W., 2001.
\newblock {\it {An Introduction to Close Binary Stars}\/}, Cambridge University
  Press, Cambridge, UK.

\bibitem[Hinton et~al.(2012)Hinton, Srivastava, Krizhevsky, Sutskever, \&
  Salakhutdinov]{Hinton+12MLdropout}
Hinton, G.~E., Srivastava, N., Krizhevsky, A., Sutskever, I., \& Salakhutdinov,
  R.~R., 2012.
\newblock Improving neural networks by preventing co-adaptation of feature
  detectors.

\bibitem[Howell et~al.(2014)Howell, Sobeck, Haas, Still, Barclay, Mullally,
  Troeltzsch, Aigrain, Bryson, Caldwell, Chaplin, Cochran, Huber, Marcy,
  Miglio, Najita, Smith, Twicken, \& Fortney]{Howell+14pasp:K2}
Howell, S.~B., Sobeck, C., Haas, M., Still, M., Barclay, T., Mullally, F.,
  Troeltzsch, J., Aigrain, S., Bryson, S.~T., Caldwell, D., Chaplin, W.~J.,
  Cochran, W.~D., Huber, D., Marcy, G.~W., Miglio, A., Najita, J.~R., Smith,
  M., Twicken, J.~D., \& Fortney, J.~J., 2014.
\newblock The {{K2 Mission}}: {{Characterization}} and {{Early Results}}, {\it
  PASP\/}, {\bf 126}(938), 398.

\bibitem[IJspeert et~al.(2021)IJspeert, Tkachenko, Johnston, Garcia, Ridder,
  Reeth, \& Aerts]{IJspeertTkachenko+21aa}
IJspeert, L.~W., Tkachenko, A., Johnston, C., Garcia, S., Ridder, J.~D., Reeth,
  T.~V., \& Aerts, C., 2021.
\newblock An all-sky sample of intermediate- to high-mass {{OBA-type}}
  eclipsing binaries observed by {{TESS}}, {\it A\&A\/}, {\bf 652}, A120.

\bibitem[IJspeert et~al.(2024)IJspeert, Tkachenko, Johnston, Pr{\v s}a, Wells,
  \& Aerts]{IJspeertTkachenko+24aa}
IJspeert, L.~W., Tkachenko, A., Johnston, C., Pr{\v s}a, A., Wells, M.~A., \&
  Aerts, C., 2024.
\newblock Automated eccentricity measurement from raw eclipsing binary light
  curves with intrinsic variability, {\it A\&A\/}, {\bf 685}, 41.

\bibitem[Iqbal(2018)]{Iqbal88plotneuralnet}
Iqbal, H., 2018.
\newblock Plotneuralnet, \url{https://doi.org/10.5281/zenodo.2526395}.

\bibitem[Jenkins et~al.(2016)Jenkins, Twicken, McCauliff, Campbell, Sanderfer,
  Lung, {Mansouri-Samani}, Girouard, Tenenbaum, Klaus, Smith, Caldwell, Chacon,
  Henze, Heiges, Latham, Morgan, Swade, Rinehart, \&
  Vanderspek]{JenkinsTwicken+16}
Jenkins, J.~M., Twicken, J.~D., McCauliff, S., Campbell, J., Sanderfer, D.,
  Lung, D., {Mansouri-Samani}, M., Girouard, F., Tenenbaum, P., Klaus, T.,
  Smith, J.~C., Caldwell, D.~A., Chacon, A.~D., Henze, C., Heiges, C., Latham,
  D.~W., Morgan, E., Swade, D., Rinehart, S., \& Vanderspek, R., 2016.
\newblock The {{TESS}} science processing operations center, in {\em Software
  and {{Cyberinfrastructure}} for {{Astronomy IV}}\/}, vol. 9913 of {\bf
  Society of {{Photo-Optical Instrumentation Engineers}} ({{SPIE}})
  {{Conference Series}}}, p. 99133E.

\bibitem[Jennings et~al.(2023)Jennings, Southworth, Maxted, \&
  Mancini]{JenningsSouthworth+23mnras}
Jennings, Z., Southworth, J., Maxted, P. F.~L., \& Mancini, L., 2023.
\newblock Revising the properties of low mass eclipsing binary stars using
  {{TESS}} light curves, {\it MNRAS\/}, {\bf 521}(3), 3405--3420.

\bibitem[Jennings et~al.(2024)Jennings, Southworth, Rappaport, Borkovits,
  Handler, \& Kurtz]{JenningsSouthworth+24mnras}
Jennings, Z., Southworth, J., Rappaport, S.~A., Borkovits, T., Handler, G., \&
  Kurtz, D.~W., 2024.
\newblock Characterization of the {$\delta$} {{Scuti}} eclipsing binary {{KIC}}
  4851217 and its tertiary companion as well as detection of tidally tilted
  pulsations, {\it MNRAS\/}, {\bf 533}, 2705--2726.

\bibitem[Justesen \& Albrecht(2021)]{JustesenAlbrecht21apj}
Justesen, A.~B. \& Albrecht, S., 2021.
\newblock Temperature and {{Distance Dependence}} of {{Tidal Circularization}}
  in {{Close Binaries}}: {{A Catalog}} of {{Eclipsing Binaries}} in the
  {{Southern Hemisphere Observed}} by the {{TESS Satellite}}, {\it ApJ\/}, {\bf
  912}(2), 123.

\bibitem[Kingma \& Ba(2017)]{KingmaBa17adam}
Kingma, D.~P. \& Ba, J., 2017.
\newblock Adam: {{A Method}} for {{Stochastic Optimization}}, in {\em 3rd
  {{International Conference}} for {{Learning Representations}}\/}, San Diego.

\bibitem[{Kirkby-Kent} et~al.(2016){Kirkby-Kent}, Maxted, Serenelli, Turner,
  Evans, Anderson, Hellier, \& West]{Kirkby-KentMaxted+16aa}
{Kirkby-Kent}, J.~A., Maxted, P. F.~L., Serenelli, A.~M., Turner, O.~D., Evans,
  D.~F., Anderson, D.~R., Hellier, C., \& West, R.~G., 2016.
\newblock Absolute parameters for {{AI Phoenicis}} using {{WASP}} photometry,
  {\it A\&A\/}, {\bf 591}, A124.

\bibitem[Kostov et~al.(2025)Kostov, Powell, Fornear, Di~Fraia, Gagliano,
  Jacobs, {de Lambilly}, Durantini~Luca, Majewski, Omohundro, Orosz, Rappaport,
  Salik, Short, Welsh, Alexandrov, {da Silva}, Dunning, G{\"u}hne, Huten,
  Hyogo, Iannone, Lee, Magliano, Sharma, Tarr, Yablonsky, Acharya, Adams,
  Barclay, Montet, Mullally, Olmschenk, Pr{\v s}a, Quintana, Wilson, Balcioglu,
  Kruse, \& {The Eclipsing Binary Patrol Collaboration}]{KostovPowell+25apjs}
Kostov, V.~B., Powell, B.~P., Fornear, A.~U., Di~Fraia, M.~Z., Gagliano, R.,
  Jacobs, T.~L., {de Lambilly}, J.~S., Durantini~Luca, H.~A., Majewski, S.~R.,
  Omohundro, M., Orosz, J., Rappaport, S.~A., Salik, R., Short, D., Welsh, W.,
  Alexandrov, S., {da Silva}, C.~M., Dunning, E., G{\"u}hne, G., Huten, M.,
  Hyogo, M., Iannone, D., Lee, S., Magliano, C., Sharma, M., Tarr, A.,
  Yablonsky, J., Acharya, S., Adams, F., Barclay, T., Montet, B.~T., Mullally,
  S., Olmschenk, G., Pr{\v s}a, A., Quintana, E., Wilson, R., Balcioglu, H.,
  Kruse, E., \& {The Eclipsing Binary Patrol Collaboration}, 2025.
\newblock The {{TESS Ten Thousand Catalog}}: 10,001 {{Uniformly Vetted}} and
  {{Validated Eclipsing Binary Stars Detected}} in {{Full-frame Image Data}} by
  {{Machine Learning}} and {{Analyzed}} by {{Citizen Scientists}}, {\it
  ApJS\/}, {\bf 279}, 50.

\bibitem[Kreiner(2004)]{Kreiner04acta}
Kreiner, J.~M., 2004.
\newblock Up-to-{{Date Linear Elements}} of {{Eclipsing Binaries}}, {\it Acta
  Astron.\/}, {\bf 54}, 207--210.

\bibitem[Krizhevsky et~al.(2012)Krizhevsky, Sutskever, \&
  Hinton]{Krizhevsky+12alexnet}
Krizhevsky, A., Sutskever, I., \& Hinton, G.~E., 2012.
\newblock {{ImageNet Classification}} with {{Deep Convolutional Neural
  Networks}}, in {\em Advances in {{Neural Information Processing Systems}}\/},
  vol.~25, Curran Associates, Inc.

\bibitem[Lacy \& Frueh(1985)]{LacyFrueh85apj}
Lacy, C.~H. \& Frueh, M.~L., 1985.
\newblock Absolute dimensions and masses of eclipsing binaries. {{V}}. {{IQ
  Persei}}., {\it ApJ\/}, {\bf 295}, 569--579.

\bibitem[Lebigot(2016)]{Lebigot2016uncertainties}
Lebigot, E.~O., 2016.
\newblock Uncertainties: a python package for calculations with uncertainties,
  \url{http://pythonhosted.org/uncertainties/}.

\bibitem[Lecun et~al.(1998)Lecun, Bottou, Bengio, \& Haffner]{LeCun+98LeNet-5}
Lecun, Y., Bottou, L., Bengio, Y., \& Haffner, P., 1998.
\newblock Gradient-based learning applied to document recognition, {\it
  Proceedings of the IEEE\/}, {\bf 86}(11), 2278--2324.

\bibitem[{Lightkurve Collaboration} et~al.(2018){Lightkurve Collaboration},
  Cardoso, Hedges, {Gully-Santiago}, Saunders, Cody, Barclay, Hall, Sagear,
  Turtelboom, Zhang, Tzanidakis, Mighell, Coughlin, Bell, {Berta-Thompson},
  Williams, Dotson, \& Barentsen]{Lightkurve+18}
{Lightkurve Collaboration}, Cardoso, J. V. d.~M., Hedges, C., {Gully-Santiago},
  M., Saunders, N., Cody, A.~M., Barclay, T., Hall, O., Sagear, S., Turtelboom,
  E., Zhang, J., Tzanidakis, A., Mighell, K., Coughlin, J., Bell, K.,
  {Berta-Thompson}, Z., Williams, P., Dotson, J., \& Barentsen, G., 2018.
\newblock Lightkurve: {{Kepler}} and {{TESS}} time series analysis in
  {{Python}}, Astrophysics Source Code Library.

\bibitem[Loshchilov \& Hutter(2017)]{LoshchilovHutter17}
Loshchilov, I. \& Hutter, F., 2017.
\newblock {{SGDR}}: {{Stochastic Gradient Descent}} with {{Warm Restarts}}.

\bibitem[Martin et~al.(2024)Martin, Sethi, Armitage, Gilbert,
  Rodr{\'i}guez~Mart{\'i}nez, \& Gilbert]{MartinSethi+24mnras}
Martin, D.~V., Sethi, R., Armitage, T., Gilbert, G.~J.,
  Rodr{\'i}guez~Mart{\'i}nez, R., \& Gilbert, E.~A., 2024.
\newblock The benchmark {{M}} dwarf eclipsing binary {{CM Draconis}} with
  {{TESS}}: Spots, flares, and ultra-precise parameters, {\it MNRAS\/}, {\bf
  528}, 963--975.

\bibitem[Maschberger(2013)]{Maschberger13mnras}
Maschberger, T., 2013.
\newblock On the function describing the stellar initial mass function, {\it
  MNRAS\/}, {\bf 429}(2), 1725--1733.

\bibitem[Matchev et~al.(2022)Matchev, Matcheva, \& Roman]{Matchev+22psj:PCA}
Matchev, K.~T., Matcheva, K., \& Roman, A., 2022.
\newblock Unsupervised {{Machine Learning}} for {{Exploratory Data Analysis}}
  of {{Exoplanet Transmission Spectra}}, {\it The Planetary Science Journal\/},
  {\bf 3}, 205.

\bibitem[Matijevi{\v c} et~al.(2012)Matijevi{\v c}, Pr{\v s}a, Orosz, Welsh,
  Bloemen, \& Barclay]{MatijevicPrsa+12aj}
Matijevi{\v c}, G., Pr{\v s}a, A., Orosz, J.~A., Welsh, W.~F., Bloemen, S., \&
  Barclay, T., 2012.
\newblock Kepler {{Eclipsing Binary Stars}}. {{III}}. {{Classification}} of
  {{Kepler Eclipsing Binary Light Curves}} with {{Locally Linear Embedding}},
  {\it AJ\/}, {\bf 143}, 123.

\bibitem[Maxted et~al.(2020)Maxted, Gaulme, Graczyk, He{\l}miniak, Johnston,
  Orosz, Pr{\v s}a, Southworth, Torres, Davies, Ball, \&
  Chaplin]{Maxted+20mnras}
Maxted, P. F.~L., Gaulme, P., Graczyk, D., He{\l}miniak, K.~G., Johnston, C.,
  Orosz, J.~A., Pr{\v s}a, A., Southworth, J., Torres, G., Davies, G.~R., Ball,
  W., \& Chaplin, W.~J., 2020.
\newblock The {{TESS}} light curve of {{AI Phoenicis}}, {\it MNRAS\/}, {\bf
  498}(1), 332--343.

\bibitem[Mitchell(1997)]{Mitchell97mlbook}
Mitchell, T.~M., 1997.
\newblock {\it Machine learning\/}, McGraw-Hill series in computer science,
  McGraw-Hill, New York.

\bibitem[Moe \& Di~Stefano(2017)]{MoeDiStefano17apjs}
Moe, M. \& Di~Stefano, R., 2017.
\newblock Mind {{Your Ps}} and {{Qs}}: {{The Interrelation}} between {{Period}}
  ({{P}}) and {{Mass-ratio}} ({{Q}}) {{Distributions}} of {{Binary Stars}},
  {\it ApJS\/}, {\bf 230}, 15.

\bibitem[Morales et~al.(2009)Morales, Ribas, Jordi, Torres, Gallardo, Guinan,
  Charbonneau, Wolf, Latham, {Anglada-Escud{\'e}}, Bradstreet, Everett,
  O'Donovan, Mandushev, \& Mathieu]{MoralesRibas+09apj}
Morales, J.~C., Ribas, I., Jordi, C., Torres, G., Gallardo, J., Guinan, E.~F.,
  Charbonneau, D., Wolf, M., Latham, D.~W., {Anglada-Escud{\'e}}, G.,
  Bradstreet, D.~H., Everett, M.~E., O'Donovan, F.~T., Mandushev, G., \&
  Mathieu, R.~D., 2009.
\newblock Absolute properties of the low-mass eclipsing binary {{CM Draconis}},
  {\it ApJ\/}, {\bf 691}(2), 1400.

\bibitem[Mowlavi et~al.(2023)Mowlavi, Holl, {Lecoeur-Ta{\"i}bi}, Barblan,
  Kochoska, Pr{\v s}a, Mazeh, Rimoldini, Gavras, Audard, {Jevardat de
  Fombelle}, Nienartowicz, {Garc{\'i}a-Lario}, \& Eyer]{MowlaviHoll+23aa}
Mowlavi, N., Holl, B., {Lecoeur-Ta{\"i}bi}, I., Barblan, F., Kochoska, A.,
  Pr{\v s}a, A., Mazeh, T., Rimoldini, L., Gavras, P., Audard, M., {Jevardat de
  Fombelle}, G., Nienartowicz, K., {Garc{\'i}a-Lario}, P., \& Eyer, L., 2023.
\newblock Gaia {{Data Release}} 3. {{The}} first {{Gaia}} catalogue of
  eclipsing-binary candidates, {\it A\&A\/}, {\bf 674}, A16.

\bibitem[Nelson \& Davis(1972)]{NelsonDavis72apj:ebop}
Nelson, B. \& Davis, W.~D., 1972.
\newblock Eclipsing-{{Binary Solutions}} by {{Sequential Optimization}} of the
  {{Parameters}}, {\it ApJ\/}, {\bf 174}, 617.

\bibitem[North et~al.(1997)North, Studer, \& Kunzli]{NorthStuder+97aa}
North, P., Studer, M., \& Kunzli, M., 1997.
\newblock Eclipsing binaries with candidate {{CP}} stars. {{I}}. {{Parameters}}
  of the systems {{HD}} 143654, {{HD}} 184035 and {{HD}} 185257., {\it A\&A\/},
  {\bf 324}, 137--154.

\bibitem[Overall \& Southworth(2024)]{OverallSouthworth24obsR17}
Overall, S. \& Southworth, J., 2024.
\newblock Rediscussion of eclipsing binaries. {{Paper XVII}}. {{The F-type}}
  twin system {{CW Eridani}}, {\it The Observatory\/}, {\bf 144}, 71--85.

\bibitem[Pashchenko et~al.(2018)Pashchenko, Sokolovsky, \&
  Gavras]{Pashchenko+18mnras}
Pashchenko, I.~N., Sokolovsky, K.~V., \& Gavras, P., 2018.
\newblock Machine learning search for variable stars, {\it MNRAS\/}, {\bf 475},
  2326--2343.

\bibitem[Pedregosa et~al.(2011)Pedregosa, Varoquaux, Gramfort, Michel, Thirion,
  Grisel, Blondel, Prettenhofer, Weiss, Dubourg, Vanderplas, Passos,
  Cournapeau, Brucher, Perrot, \& Duchesnay]{Pedregosa+2011skl}
Pedregosa, F., Varoquaux, G., Gramfort, A., Michel, V., Thirion, B., Grisel,
  O., Blondel, M., Prettenhofer, P., Weiss, R., Dubourg, V., Vanderplas, J.,
  Passos, A., Cournapeau, D., Brucher, M., Perrot, M., \& Duchesnay, E., 2011.
\newblock Scikit-learn: Machine learning in {P}ython, {\it Journal of Machine
  Learning Research\/}, {\bf 12}, 2825--2830.

\bibitem[{Pietrzy{\'n}ski} et~al.(2019){Pietrzy{\'n}ski}, {Graczyk},
  {Gallenne}, {Gieren}, {Thompson}, {Pilecki}, {Karczmarek}, {G{\'o}rski},
  {Suchomska}, {Taormina}, {Zgirski}, {Wielg{\'o}rski}, {Ko{\l}aczkowski},
  {Konorski}, {Villanova}, {Nardetto}, {Kervella}, {Bresolin}, {Kudritzki},
  {Storm}, {Smolec}, \& {Narloch}]{PietrzynskiGraczyk+19nat:distance}
{Pietrzy{\'n}ski}, G., {Graczyk}, D., {Gallenne}, A., {Gieren}, W., {Thompson},
  I.~B., {Pilecki}, B., {Karczmarek}, P., {G{\'o}rski}, M., {Suchomska}, K.,
  {Taormina}, M., {Zgirski}, B., {Wielg{\'o}rski}, P., {Ko{\l}aczkowski}, Z.,
  {Konorski}, P., {Villanova}, S., {Nardetto}, N., {Kervella}, P., {Bresolin},
  F., {Kudritzki}, R.~P., {Storm}, J., {Smolec}, R., \& {Narloch}, W., 2019.
\newblock {A distance to the Large Magellanic Cloud that is precise to one per
  cent}, {\it \nat\/}, {\bf 567}(7747), 200--203.

\bibitem[Popper \& Etzel(1981)]{PopperEtzel81apj:ebop}
Popper, D.~M. \& Etzel, P.~B., 1981.
\newblock Photometric orbits of seven detached eclipsing binaries., {\it AJ\/},
  {\bf 86}, 102--120.

\bibitem[{Press} et~al.(1992){Press}, {Vetterling}, {Teukolsky}, \&
  {Flannery}]{Press+92book}
{Press}, W.~H., {Vetterling}, W.~T., {Teukolsky}, S.~A., \& {Flannery}, B.~P.,
  1992.
\newblock {\it Numerical Recipes in FORTRAN 77\/}, The art of scientific
  computing, Cambridge University Press, Cambridge, UK, 2nd edn.

\bibitem[Prisinzano et~al.(2022)Prisinzano, Damiani, Sciortino, Flaccomio,
  Guarcello, Micela, Tognelli, Jeffries, \& Alcal{\'a}]{Prisinzano+22aa:DBSCAN}
Prisinzano, L., Damiani, F., Sciortino, S., Flaccomio, E., Guarcello, M.~G.,
  Micela, G., Tognelli, E., Jeffries, R.~D., \& Alcal{\'a}, J.~M., 2022.
\newblock Low-mass young stars in the {{Milky Way}} unveiled by {{DBSCAN}} and
  {{Gaia EDR3}}: {{Mapping}} the star forming regions within 1.5 kpc, {\it
  A\&A\/}, {\bf 664}, A175.

\bibitem[Pr{\v s}a \& Zwitter(2005)]{PrsaZwitter05apj}
Pr{\v s}a, A. \& Zwitter, T., 2005.
\newblock A {{Computational Guide}} to {{Physics}} of {{Eclipsing Binaries}}.
  {{I}}. {{Demonstrations}} and {{Perspectives}}, {\it ApJ\/}, {\bf 628}(1),
  426.

\bibitem[Pr{\v s}a et~al.(2008)Pr{\v s}a, Guinan, Devinney, DeGeorge,
  Bradstreet, Giammarco, Alcock, \& Engle]{PrsaGuinan+08apj}
Pr{\v s}a, A., Guinan, E.~F., Devinney, E.~J., DeGeorge, M., Bradstreet, D.~H.,
  Giammarco, J.~M., Alcock, C.~R., \& Engle, S.~G., 2008.
\newblock Artificial {{Intelligence Approach}} to the {{Determination}} of
  {{Physical Properties}} of {{Eclipsing Binaries}}. {{I}}. {{The EBAI
  Project}}, {\it ApJ\/}, {\bf 687}(1), 542.

\bibitem[Pr{\v s}a et~al.(2022)Pr{\v s}a, Kochoska, Conroy, Eisner, Hey,
  IJspeert, Kruse, Fleming, Johnston, Kristiansen, LaCourse, Mortensen, Pepper,
  Stassun, Torres, {Abdul-Masih}, Chakraborty, Gagliano, Guo, Hambleton, Hong,
  Jacobs, Jones, Kostov, Lee, Omohundro, Orosz, Page, Powell, Rappaport, Reed,
  Schnittman, Schwengeler, Shporer, Terentev, Vanderburg, Welsh, Caldwell,
  Doty, Jenkins, Latham, Ricker, Seager, Schlieder, Shiao, Vanderspek, \&
  Winn]{PrsaKochoska+22apjs}
Pr{\v s}a, A., Kochoska, A., Conroy, K.~E., Eisner, N., Hey, D.~R., IJspeert,
  L., Kruse, E., Fleming, S.~W., Johnston, C., Kristiansen, M.~H., LaCourse,
  D., Mortensen, D., Pepper, J., Stassun, K.~G., Torres, G., {Abdul-Masih}, M.,
  Chakraborty, J., Gagliano, R., Guo, Z., Hambleton, K., Hong, K., Jacobs, T.,
  Jones, D., Kostov, V., Lee, J.~W., Omohundro, M., Orosz, J.~A., Page, E.~J.,
  Powell, B.~P., Rappaport, S., Reed, P., Schnittman, J., Schwengeler, H.~M.,
  Shporer, A., Terentev, I.~A., Vanderburg, A., Welsh, W.~F., Caldwell, D.~A.,
  Doty, J.~P., Jenkins, J.~M., Latham, D.~W., Ricker, G.~R., Seager, S.,
  Schlieder, J.~E., Shiao, B., Vanderspek, R., \& Winn, J.~N., 2022.
\newblock {{TESS Eclipsing Binary Stars}}. {{I}}. {{Short-cadence
  Observations}} of 4584 {{Eclipsing Binaries}} in {{Sectors}} 1--26, {\it
  ApJS\/}, {\bf 258}(1), 16.

\bibitem[Rauer et~al.(2024)Rauer, Aerts, Cabrera, Deleuil, Erikson, Gizon,
  Goupil, Heras, {Lorenzo-Alvarez}, Marliani, {Martin-Garcia}, {Mas-Hesse},
  O'Rourke, Osborn, Pagano, Piotto, Pollacco, Ragazzoni, Ramsay, Udry,
  Appourchaux, Benz, Brandeker, G{\"u}del, {Janot-Pacheco}, Kabath, Kjeldsen,
  Min, Santos, Smith, Suarez, Werner, Aboudan, Abreu, Acu{\~n}a, Adams,
  Adibekyan, Affer, Agneray, Agnor, {B{\o}rsen-Koch}, Ahmed, Aigrain,
  {Al-Bahlawan}, Gil, Alei, Alencar, Alexander, {Alfonso-Garz{\'o}n}, Alibert,
  Prieto, Almeida, Sobrino, Altavilla, Althaus, Trujillo, Amarsi, Eiff,
  Am{\^o}res, Andrade, {Antoniadis-Karnavas}, Ant{\'o}nio, {del Moral},
  Appolloni, Arena, Armstrong, Aliaga, Asplund, Audenaert, Auricchio, Avelino,
  Baeke, Bailli{\'e}, Balado, Balestra, Ball, Ballans, Ballot, Barban, Barbary,
  Barbieri, Forteza, Barker, Barklem, Barnes, Navascues, Barragan, Baruteau,
  Basu, Baudin, Baumeister, Bayliss, Bazot, Beck, Bedding, Belkacem, Bellinger,
  Benatti, Benomar, B{\'e}rard, Bergemann, Bergomi, Bernardo, Biazzo,
  Bignamini, Bigot, Billot, Binet, Biondi, Biondi, Birch, Bitsch, Ceballos,
  B{\'o}di, Bogn{\'a}r, Boisse, Bolmont, Bonanno, Bonavita, Bonfanti, Bonfils,
  Bonito, Bonomo, B{\"o}rner, Saikia, Mart{\'i}n, Borsa, Borsato, Bossini,
  Bouchy, Bou{\'e}, Boufleur, Boumier, Bourrier, Bowman, Bozzo, Bradley, Bray,
  Bressan, Breton, Brienza, Brito, Brogi, Brown, Brown, Brun, Bruno, Bruns,
  Buchhave, Bugnet, Buldgen, Burgess, Busatta, Busso, Buzasi, Caballero,
  Cabral, Calderone, Cameron, Cameron, Campante, Martins, Cara, Carone,
  Carrasco, Casagrande, Casewell, Cassisi, Castellani, Castro, Catala,
  Fern{\'a}ndez, Catelan, Cegla, Cerruti, Cessa, Chadid, Chaplin, Charpinet,
  Chiappini, Chiarucci, Chiavassa, Chinellato, Chirulli,
  {Christensen-Dalsgaard}, Church, Claret, Clarke, Claudi, Clermont, Coelho,
  Coelho, Cogato, Colom{\'e}, Condamin, Conseil, Corbard, Correia, Corsaro,
  Cosentino, Costes, Cottinelli, Covone, Creevey, Crida, Csizmadia, Cunha,
  Curry, {da Costa}, {da Silva}, Dalal, Damasso, Damiani, Damiani, das Chagas,
  Davies, Davies, Davies, Davison, {de Almeida}, {de Angeli}, {de Barros},
  Le{\~a}o, {de Freitas}, {de Freitas}, De~Martino, {de Medeiros}, {de Paula},
  {de Plaa}, De~Ridder, Deal, Decin, Deeg, Degl'Innocenti, Deheuvels, {del
  Burgo}, Del~Sordo, {Delgado-Mena}, Demangeon, Denk, Derekas, Desidera, Dexet,
  Di~Criscienzo, Di~Giorgio, Di~Mauro, Rial, {D{\'i}az-Garc{\'i}a}, Dima,
  Dinuzzi, Dionatos, Distefano, do~Nascimento~Jr., Domingo, D'Orazi, Dorn,
  Doyle, Duarte, Ducellier, Dumaye, Dumusque, Dupret, Eggenberger, Ehrenreich,
  Eigm{\"u}ller, Eising, Emilio, Eriksson, Ermocida, Giribaldi, Eschen,
  Estrela, Evans, Fabbian, Fabrizio, Faria, Farina, Farinato, Feliz, Feltzing,
  Fenouillet, Ferrari, {Ferraz-Mello}, Fialho, Fienga, Figueira, Fiori,
  Flaccomio, Focardi, Foley, Fontignie, Ford, Fornazier, Forveille, Fossati,
  Franca, {da Silva}, Frasca, Fridlund, Furlan, Gabler, Gaido, Gallagher,
  Galli, Garcia, Hern{\'a}ndez, Munoz, {Garc{\'i}a-V{\'a}zquez}, Haba, Gaulme,
  Gauthier, Gehan, Gent, Georgieva, Ghigo, Giana, Gill, Girardi, Winter, Giusi,
  {da Silva}, Zazo, {Gomez-Lopez}, Hern{\'a}ndez, Murillo, Gorius, Gouel,
  Goulty, Granata, Grenfell, Grie{\ss}bach, Grolleau, Grouffal, Grziwa,
  Guarcello, Gueguen, Guenther, Guilhem, Guillerot, Guiot, Guterman,
  Guti{\'e}rrez, {Guti{\'e}rrez-Canales}, Hagelberg, Haldemann, Hall, Handberg,
  Harrison, Harrison, Hasiba, Haswell, Hatalova, Hatzes, Haywood, H{\'e}brard,
  Heckes, Heiter, Hekker, Heller, Helling, Helminiak, Hemsley, Heng, Hermans,
  Hermes, Torres, Hinkel, Hobbs, Hodgkin, Hofmann, Hojjatpanah, Houdek, Huber,
  Huesler, {Hui-Bon-Hoa}, Huygen, Huynh, Iro, Irwin, Irwin, Izidoro, Jacquinod,
  Jannsen, Janson, Jeszenszky, Jiang, Mancebo, Jofre, Johansen, Johnston,
  Jones, Kallinger, K{\'a}lm{\'a}n, Kanitz, Karjalainen, Karjalainen, Karoff,
  Kawaler, Kawata, Keereman, Keiderling, Kennedy, Kenworthy, Kerschbaum,
  Kidger, Kiefer, Kintziger, Kislyakova, Kiss, Klagyivik, Klahr, Klevas,
  Kochukhov, K{\"o}hler, Kolb, Koncz, Korth, Kostogryz, Kov{\'a}cs, Kov{\'a}cs,
  Kozhura, Krivova, Ku{\v c}inskas, Kuhlemann, Kupka, Laauwen, Labiano,
  Lagarde, Laget, Laky, Lam, Lambrechts, Lammer, Lanza, Lanzafame, Martiz,
  Laskar, Latter, Lavanant, Lawrenson, Lazzoni, Lebre, Lebreton, des Etangs,
  Leinhardt, Leleu, Lendl, Leto, Levillain, Libert, Lichtenberg, Ligi,
  Lignieres, {Lillo-Box}, Linsky, Liu, Loidolt, Longval, Lopes, Lorenzani,
  Ludwig, Lund, Lundkvist, Luri, Maceroni, Madden, Madhusudhan, Maggio,
  Magliano, Magrin, Mahy, Maibaum, {Malac-Allain}, Malapert, Malavolta,
  Maldonado, Mamonova, Manchon, Mann, Mantovan, Marafatto, Marconi, Mardling,
  Marigo, Marinoni, Marques, Marques, Marrese, Marshall, Perales, Mary,
  Marzari, Masana, Mascher, Mathis, Mathur, Figueiredo, Maxted, Mazeh, Mazevet,
  Mazzei, McCormac, McMillan, Menou, Merle, Meru, Mesa, Messina,
  M{\'e}sz{\'a}ros, Meunier, Meunier, Micela, Michaelis, Michel, Michielsen,
  Michtchenko, Miglio, Miguel, Milligan, Mirouh, Mitchel, Moedas, Molendini,
  Moln{\'a}r, Mombarg, Montalban, Montalto, Monteiro, Morales,
  {Morales-Calderon}, Morbidelli, Mordasini, Moreau, Morel, Morello, Morin,
  Mortier, Mosser, Mourard, Mousis, Moutou, Mowlavi, Moya, Muehlmann, Muirhead,
  Munari, Musella, Mustill, Nardetto, Nardiello, Narita, Nascimbeni, Nash,
  Neiner, Nelson, Nettelmann, Nicolini, Nielsen, Niemi, Noack, {Noels-Grotsch},
  Noll, Norazman, Norton, Nsamba, Ofir, Ogilvie, Olander, Olivetto, Olofsson,
  Ong, Ortolani, Oshagh, Ottacher, Ottensamer, Ouazzani, Paardekooper, Pace,
  Pajas, Palacios, Palandri, Palle, Paproth, Parro, Parviainen, Granado,
  Passegger, {Pastor-Morales}, P{\"a}tzold, Pedersen, Hidalgo, Pepe, Pereira,
  Persson, Pertenais, Peter, Petit, Petit, Pezzuto, Pichierri, Pietrinferni,
  Pinheiro, Pinsonneault, Plachy, Plasson, Plez, Poppenhaeger, Poretti,
  Portaluri, Portell, {de Mello}, Poyatos, Pozuelos, Moroni, Pricopi,
  Prisinzano, Quade, Quirrenbach160, Reina6, Soares, Raimondo, Rainer,
  Rod{\'o}n, {Ram{\'o}n-Ballesta}, Zapata, R{\"a}tz, Rauterberg, Redman,
  Redmer, Reese, Regibo, Reiners, Reinhold, Renie, Ribas, Ribeiro, Ricciardi,
  Rice, Richard, Riello, Rieutord, Ripepi, Rixon, Rockstein, Rodr{\'i}guez,
  D{\'i}az, Garcia, {Rodriguez-Gomez}, Roehlly, Roig, {Rojas-Ayala}, Rolf,
  R{\o}rsted, Rosado, Rosotti, Roth, Roth, Rousseau, Roxburgh, Roy, Royer,
  Ruane, Mastropasqua, {de Galarreta}, Russi, Saar, Saillenfest, Salaris,
  Salmon, Saltas, Samadi, Samadi, Samra, {da Silva}, Carrasco, Santerne,
  Santoli, Santos, Mesa, Sarro, Scandariato, Sch{\"a}fer, Schlafly, Schmider,
  Schneider, Schou, Schunker, Schwarzkopf, Serenelli, Seynaeve, Shan, Shapiro,
  Shipman, Sicilia, Sanmartin, Sigot, Silliman, Silvotti, Simon, Napoli,
  Skarka, Smalley, Smiljanic, Smit, Smith, Smith, Snellen, S{\'o}dor, Sohl,
  Solanki, Sortino, Sousa, Southworth, Souto, Sozzetti, Stamatellos, Stassun,
  Steller, Stello, Stelzer, Stiebeler, Stokholm, Storelvmo, Strassmeier,
  Str{\o}m, Strugarek, Sulis, {\v S}vanda, Szabados, Szab{\'o}, Szab{\'o},
  Szuszkiewicz, Talens, Teti, Theisen, Th{\'e}venin, Thoul, Tiphene,
  {Titz-Weider}, Tkachenko, Tomecki, Tonfat, Tosi, Trampedach, Traven, Triaud,
  Tr{\o}nnes, Tsantaki, Tschentscher, Turin, Tvaruzka, Ulmer, {Ulmer-Moll},
  Ulusoy, Umbriaco, Valencia, Valentini, Valio, Guijarro, Van~Eylen,
  Van~Grootel, {van Kempen}, Van~Reeth, Van~Zelst, Vandenbussche, Vasiliou,
  Vasilyev, {de Mascarenhas}, Vazan, Nunez, Velloso, Ventura, Ventura,
  Venturini, Trallero, Veras, Verdugo, Verma, Vibert, Martinez, Vida, Vigan,
  Villacorta, Villaver, Aparicio, Viotto, Vorobyov, Vorontsov, Wagner,
  Walloschek, Walton, Walton, Wang, Waters, Watson, Wedemeyer, Weeks, Weingril,
  Weiss, Wendler, West, Westerdorff, Westphal, Wheatley, White, Whittaker,
  Wickhusen, Wilson, Windsor, Winter, Winther, Winton, Witteck, Witzke, Woitke,
  Wolter, Wuchterl, Wyatt, Yang, Yu, Sanchez, Osorio, Zechmeister, Zhou,
  Ziemke, \& Zwintz]{RauerAerts+24}
Rauer, H., Aerts, C., Cabrera, J., Deleuil, M., Erikson, A., Gizon, L., Goupil,
  M., Heras, A., {Lorenzo-Alvarez}, J., Marliani, F., {Martin-Garcia}, C.,
  {Mas-Hesse}, J.~M., O'Rourke, L., Osborn, H., Pagano, I., Piotto, G.,
  Pollacco, D., Ragazzoni, R., Ramsay, G., Udry, S., Appourchaux, T., Benz, W.,
  Brandeker, A., G{\"u}del, M., {Janot-Pacheco}, E., Kabath, P., Kjeldsen, H.,
  Min, M., Santos, N., Smith, A., Suarez, J.-C., Werner, S.~C., Aboudan, A.,
  Abreu, M., Acu{\~n}a, L., Adams, M., Adibekyan, V., Affer, L., Agneray, F.,
  Agnor, C., {B{\o}rsen-Koch}, V.~A., Ahmed, S., Aigrain, S., {Al-Bahlawan},
  A., Gil, M. d. l. A.~A., Alei, E., Alencar, S., Alexander, R.,
  {Alfonso-Garz{\'o}n}, J., Alibert, Y., Prieto, C.~A., Almeida, L., Sobrino,
  R.~A., Altavilla, G., Althaus, C., Trujillo, L. A.~A., Amarsi, A., Eiff, M.
  A.-v., Am{\^o}res, E., Andrade, L., {Antoniadis-Karnavas}, A., Ant{\'o}nio,
  C., {del Moral}, B.~A., Appolloni, M., Arena, C., Armstrong, D., Aliaga,
  J.~A., Asplund, M., Audenaert, J., Auricchio, N., Avelino, P., Baeke, A.,
  Bailli{\'e}, K., Balado, A., Balestra, A., Ball, W., Ballans, H., Ballot, J.,
  Barban, C., Barbary, G., Barbieri, M., Forteza, S.~B., Barker, A., Barklem,
  P., Barnes, S., Navascues, D.~B., Barragan, O., Baruteau, C., Basu, S.,
  Baudin, F., Baumeister, P., Bayliss, D., Bazot, M., Beck, P.~G., Bedding, T.,
  Belkacem, K., Bellinger, E., Benatti, S., Benomar, O., B{\'e}rard, D.,
  Bergemann, M., Bergomi, M., Bernardo, P., Biazzo, K., Bignamini, A., Bigot,
  L., Billot, N., Binet, M., Biondi, D., Biondi, F., Birch, A.~C., Bitsch, B.,
  Ceballos, P. V.~B., B{\'o}di, A., Bogn{\'a}r, Z., Boisse, I., Bolmont, E.,
  Bonanno, A., Bonavita, M., Bonfanti, A., Bonfils, X., Bonito, R., Bonomo,
  A.~S., B{\"o}rner, A., Saikia, S.~B., Mart{\'i}n, E.~B., Borsa, F., Borsato,
  L., Bossini, D., Bouchy, F., Bou{\'e}, G., Boufleur, R., Boumier, P.,
  Bourrier, V., Bowman, D.~M., Bozzo, E., Bradley, L., Bray, J., Bressan, A.,
  Breton, S., Brienza, D., Brito, A., Brogi, M., Brown, B., Brown, D., Brun,
  A.~S., Bruno, G., Bruns, M., Buchhave, L.~A., Bugnet, L., Buldgen, G.,
  Burgess, P., Busatta, A., Busso, G., Buzasi, D., Caballero, J.~A., Cabral,
  A., Calderone, F., Cameron, R., Cameron, A., Campante, T., Martins, B. L.~C.,
  Cara, C., Carone, L., Carrasco, J.~M., Casagrande, L., Casewell, S.~L.,
  Cassisi, S., Castellani, M., Castro, M., Catala, C., Fern{\'a}ndez, I.~C.,
  Catelan, M., Cegla, H., Cerruti, C., Cessa, V., Chadid, M., Chaplin, W.,
  Charpinet, S., Chiappini, C., Chiarucci, S., Chiavassa, A., Chinellato, S.,
  Chirulli, G., {Christensen-Dalsgaard}, J., Church, R., Claret, A., Clarke,
  C., Claudi, R., Clermont, L., Coelho, H., Coelho, J., Cogato, F., Colom{\'e},
  J., Condamin, M., Conseil, S., Corbard, T., Correia, A. C.~M., Corsaro, E.,
  Cosentino, R., Costes, J., Cottinelli, A., Covone, G., Creevey, O.~L., Crida,
  A., Csizmadia, S., Cunha, M., Curry, P., {da Costa}, J., {da Silva}, F.,
  Dalal, S., Damasso, M., Damiani, C., Damiani, F., das Chagas, M.~L., Davies,
  M., Davies, G., Davies, B., Davison, G., {de Almeida}, L., {de Angeli}, F.,
  {de Barros}, S. C.~C., Le{\~a}o, I. d.~C., {de Freitas}, D.~B., {de Freitas},
  M.~C., De~Martino, D., {de Medeiros}, J.~R., {de Paula}, L.~A., {de Plaa},
  J., De~Ridder, J., Deal, M., Decin, L., Deeg, H., Degl'Innocenti, S.,
  Deheuvels, S., {del Burgo}, C., Del~Sordo, F., {Delgado-Mena}, E., Demangeon,
  O., Denk, T., Derekas, A., Desidera, S., Dexet, M., Di~Criscienzo, M.,
  Di~Giorgio, A.~M., Di~Mauro, M.~P., Rial, F. J.~D., {D{\'i}az-Garc{\'i}a},
  J.-J., Dima, M., Dinuzzi, G., Dionatos, O., Distefano, E., do~Nascimento~Jr.,
  J.-D., Domingo, A., D'Orazi, V., Dorn, C., Doyle, L., Duarte, E., Ducellier,
  F., Dumaye, L., Dumusque, X., Dupret, M.-A., Eggenberger, P., Ehrenreich, D.,
  Eigm{\"u}ller, P., Eising, J., Emilio, M., Eriksson, K., Ermocida, M.,
  Giribaldi, R. I.~E., Eschen, Y., Estrela, I., Evans, D.~W., Fabbian, D.,
  Fabrizio, M., Faria, J.~P., Farina, M., Farinato, J., Feliz, D., Feltzing,
  S., Fenouillet, T., Ferrari, L., {Ferraz-Mello}, S., Fialho, F., Fienga, A.,
  Figueira, P., Fiori, L., Flaccomio, E., Focardi, M., Foley, S., Fontignie,
  J., Ford, D., Fornazier, K., Forveille, T., Fossati, L., Franca, R. d.~M.,
  {da Silva}, L.~F., Frasca, A., Fridlund, M., Furlan, M., Gabler, S.-M.,
  Gaido, M., Gallagher, A., Galli, E., Garcia, R.~A., Hern{\'a}ndez, A.~G.,
  Munoz, A.~G., {Garc{\'i}a-V{\'a}zquez}, H., Haba, R.~G., Gaulme, P.,
  Gauthier, N., Gehan, C., Gent, M., Georgieva, I., Ghigo, M., Giana, E., Gill,
  S., Girardi, L., Winter, S.~G., Giusi, G., {da Silva}, J.~G., Zazo, L. J.~G.,
  {Gomez-Lopez}, J.~M., Hern{\'a}ndez, J. I.~G., Murillo, K.~G., Gorius, N.,
  Gouel, P.-V., Goulty, D., Granata, V., Grenfell, J.~L., Grie{\ss}bach, D.,
  Grolleau, E., Grouffal, S., Grziwa, S., Guarcello, M.~G., Gueguen, L.,
  Guenther, E.~W., Guilhem, T., Guillerot, L., Guiot, P., Guterman, P.,
  Guti{\'e}rrez, A., {Guti{\'e}rrez-Canales}, F., Hagelberg, J., Haldemann, J.,
  Hall, C., Handberg, R., Harrison, I., Harrison, D.~L., Hasiba, J., Haswell,
  C.~A., Hatalova, P., Hatzes, A., Haywood, R., H{\'e}brard, G., Heckes, F.,
  Heiter, U., Hekker, S., Heller, R., Helling, C., Helminiak, K., Hemsley, S.,
  Heng, K., Hermans, A., Hermes, J.~J., Torres, N.~H., Hinkel, N., Hobbs, D.,
  Hodgkin, S., Hofmann, K., Hojjatpanah, S., Houdek, G., Huber, D., Huesler,
  J., {Hui-Bon-Hoa}, A., Huygen, R., Huynh, D.-D., Iro, N., Irwin, J., Irwin,
  M., Izidoro, A., Jacquinod, S., Jannsen, N.~E., Janson, M., Jeszenszky, H.,
  Jiang, C., Mancebo, A. J.~J., Jofre, P., Johansen, A., Johnston, C., Jones,
  G., Kallinger, T., K{\'a}lm{\'a}n, S., Kanitz, T., Karjalainen, M.,
  Karjalainen, R., Karoff, C., Kawaler, S., Kawata, D., Keereman, A.,
  Keiderling, D., Kennedy, T., Kenworthy, M., Kerschbaum, F., Kidger, M.,
  Kiefer, F., Kintziger, C., Kislyakova, K., Kiss, L., Klagyivik, P., Klahr,
  H., Klevas, J., Kochukhov, O., K{\"o}hler, U., Kolb, U., Koncz, A., Korth,
  J., Kostogryz, N., Kov{\'a}cs, G., Kov{\'a}cs, J., Kozhura, O., Krivova, N.,
  Ku{\v c}inskas, A., Kuhlemann, I., Kupka, F., Laauwen, W., Labiano, A.,
  Lagarde, N., Laget, P., Laky, G., Lam, K. W.~F., Lambrechts, M., Lammer, H.,
  Lanza, A.~F., Lanzafame, A., Martiz, M.~L., Laskar, J., Latter, H., Lavanant,
  T., Lawrenson, A., Lazzoni, C., Lebre, A., Lebreton, Y., des Etangs, A.~L.,
  Leinhardt, Z., Leleu, A., Lendl, M., Leto, G., Levillain, Y., Libert, A.-S.,
  Lichtenberg, T., Ligi, R., Lignieres, F., {Lillo-Box}, J., Linsky, J., Liu,
  J.~S., Loidolt, D., Longval, Y., Lopes, I., Lorenzani, A., Ludwig, H.-G.,
  Lund, M., Lundkvist, M.~S., Luri, X., Maceroni, C., Madden, S., Madhusudhan,
  N., Maggio, A., Magliano, C., Magrin, D., Mahy, L., Maibaum, O.,
  {Malac-Allain}, L., Malapert, J.-C., Malavolta, L., Maldonado, J., Mamonova,
  E., Manchon, L., Mann, A., Mantovan, G., Marafatto, L., Marconi, M.,
  Mardling, R., Marigo, P., Marinoni, S., Marques, {\'E}., Marques, J.~P.,
  Marrese, P.~M., Marshall, D., Perales, S.~M., Mary, D., Marzari, F., Masana,
  E., Mascher, A., Mathis, S., Mathur, S., Figueiredo, A. C.~M., Maxted, P.
  F.~L., Mazeh, T., Mazevet, S., Mazzei, F., McCormac, J., McMillan, P., Menou,
  L., Merle, T., Meru, F., Mesa, D., Messina, S., M{\'e}sz{\'a}ros, S.,
  Meunier, N., Meunier, J.-C., Micela, G., Michaelis, H., Michel, E.,
  Michielsen, M., Michtchenko, T., Miglio, A., Miguel, Y., Milligan, D.,
  Mirouh, G., Mitchel, M., Moedas, N., Molendini, F., Moln{\'a}r, L., Mombarg,
  J., Montalban, J., Montalto, M., Monteiro, M. J. P. F.~G., Morales, J.~C.,
  {Morales-Calderon}, M., Morbidelli, A., Mordasini, C., Moreau, C., Morel, T.,
  Morello, G., Morin, J., Mortier, A., Mosser, B., Mourard, D., Mousis, O.,
  Moutou, C., Mowlavi, N., Moya, A., Muehlmann, P., Muirhead, P., Munari, M.,
  Musella, I., Mustill, A.~J., Nardetto, N., Nardiello, D., Narita, N.,
  Nascimbeni, V., Nash, A., Neiner, C., Nelson, R.~P., Nettelmann, N.,
  Nicolini, G., Nielsen, M., Niemi, S.-M., Noack, L., {Noels-Grotsch}, A.,
  Noll, A., Norazman, A., Norton, A.~J., Nsamba, B., Ofir, A., Ogilvie, G.,
  Olander, T., Olivetto, C., Olofsson, G., Ong, J., Ortolani, S., Oshagh, M.,
  Ottacher, H., Ottensamer, R., Ouazzani, R.-M., Paardekooper, S.-J., Pace, E.,
  Pajas, M., Palacios, A., Palandri, G., Palle, E., Paproth, C., Parro, V.,
  Parviainen, H., Granado, J.~P., Passegger, V.~M., {Pastor-Morales}, C.,
  P{\"a}tzold, M., Pedersen, M.~G., Hidalgo, D.~P., Pepe, F., Pereira, F.,
  Persson, C.~M., Pertenais, M., Peter, G., Petit, A.~C., Petit, P., Pezzuto,
  S., Pichierri, G., Pietrinferni, A., Pinheiro, F., Pinsonneault, M., Plachy,
  E., Plasson, P., Plez, B., Poppenhaeger, K., Poretti, E., Portaluri, E.,
  Portell, J., {de Mello}, G. F.~P., Poyatos, J., Pozuelos, F.~J., Moroni, P.
  G.~P., Pricopi, D., Prisinzano, L., Quade, M., Quirrenbach160, n., Reina6, J.
  A.~R., Soares, M. C.~R., Raimondo, G., Rainer, M., Rod{\'o}n, J.~R.,
  {Ram{\'o}n-Ballesta}, A., Zapata, G.~R., R{\"a}tz, S., Rauterberg, C.,
  Redman, B., Redmer, R., Reese, D., Regibo, S., Reiners, A., Reinhold, T.,
  Renie, C., Ribas, I., Ribeiro, S., Ricciardi, T.~P., Rice, K., Richard, O.,
  Riello, M., Rieutord, M., Ripepi, V., Rixon, G., Rockstein, S.,
  Rodr{\'i}guez, M. T.~R., D{\'i}az, L. F.~R., Garcia, J. P.~R.,
  {Rodriguez-Gomez}, J., Roehlly, Y., Roig, F., {Rojas-Ayala}, B., Rolf, T.,
  R{\o}rsted, J.~L., Rosado, H., Rosotti, G., Roth, O., Roth, M., Rousseau, A.,
  Roxburgh, I., Roy, F., Royer, P., Ruane, K., Mastropasqua, S.~R., {de
  Galarreta}, C.~R., Russi, A., Saar, S., Saillenfest, M., Salaris, M., Salmon,
  S., Saltas, I., Samadi, R., Samadi, A., Samra, D., {da Silva}, T.~S.,
  Carrasco, M. A.~S., Santerne, A., Santoli, F., Santos, {\^A}. R.~G., Mesa,
  R.~S., Sarro, L.~M., Scandariato, G., Sch{\"a}fer, M., Schlafly, E.,
  Schmider, F.-X., Schneider, J., Schou, J., Schunker, H., Schwarzkopf, G.~J.,
  Serenelli, A., Seynaeve, D., Shan, Y., Shapiro, A., Shipman, R., Sicilia, D.,
  Sanmartin, M. A.~S., Sigot, A., Silliman, K., Silvotti, R., Simon, A.~E.,
  Napoli, R.~S., Skarka, M., Smalley, B., Smiljanic, R., Smit, S., Smith, A.,
  Smith, L., Snellen, I., S{\'o}dor, {\'A}., Sohl, F., Solanki, S.~K., Sortino,
  F., Sousa, S., Southworth, J., Souto, D., Sozzetti, A., Stamatellos, D.,
  Stassun, K., Steller, M., Stello, D., Stelzer, B., Stiebeler, U., Stokholm,
  A., Storelvmo, T., Strassmeier, K., Str{\o}m, P.~A., Strugarek, A., Sulis,
  S., {\v S}vanda, M., Szabados, L., Szab{\'o}, R., Szab{\'o}, G.~M.,
  Szuszkiewicz, E., Talens, G.~J., Teti, D., Theisen, T., Th{\'e}venin, F.,
  Thoul, A., Tiphene, D., {Titz-Weider}, R., Tkachenko, A., Tomecki, D.,
  Tonfat, J., Tosi, N., Trampedach, R., Traven, G., Triaud, A., Tr{\o}nnes, R.,
  Tsantaki, M., Tschentscher, M., Turin, A., Tvaruzka, A., Ulmer, B.,
  {Ulmer-Moll}, S., Ulusoy, C., Umbriaco, G., Valencia, D., Valentini, M.,
  Valio, A., Guijarro, {\'A}. L.~V., Van~Eylen, V., Van~Grootel, V., {van
  Kempen}, T.~A., Van~Reeth, T., Van~Zelst, I., Vandenbussche, B., Vasiliou,
  K., Vasilyev, V., {de Mascarenhas}, D.~V., Vazan, A., Nunez, M.~V., Velloso,
  E.~N., Ventura, R., Ventura, P., Venturini, J., Trallero, I.~V., Veras, D.,
  Verdugo, E., Verma, K., Vibert, D., Martinez, T.~V., Vida, K., Vigan, A.,
  Villacorta, A., Villaver, E., Aparicio, M.~V., Viotto, V., Vorobyov, E.,
  Vorontsov, S., Wagner, F.~W., Walloschek, T., Walton, N., Walton, D., Wang,
  H., Waters, R., Watson, C., Wedemeyer, S., Weeks, A., Weingril, J., Weiss,
  A., Wendler, B., West, R., Westerdorff, K., Westphal, P.-A., Wheatley, P.,
  White, T., Whittaker, A., Wickhusen, K., Wilson, T., Windsor, J., Winter, O.,
  Winther, M.~L., Winton, A., Witteck, U., Witzke, V., Woitke, P., Wolter, D.,
  Wuchterl, G., Wyatt, M., Yang, D., Yu, J., Sanchez, R.~Z., Osorio, M. R.~Z.,
  Zechmeister, M., Zhou, Y., Ziemke, C., \& Zwintz, K., 2024.
\newblock The {{PLATO Mission}}.

\bibitem[Ricker et~al.(2015)Ricker, Winn, Vanderspek, Latham, Bakos, Bean,
  {Berta-Thompson}, Brown, Buchhave, Butler, Butler, Chaplin, Charbonneau,
  {Christensen-Dalsgaard}, Clampin, Deming, Doty, Lee, Dressing, Dunham, Endl,
  Fressin, Ge, Henning, Holman, Howard, Ida, Jenkins, Jernigan, Johnson,
  Kaltenegger, Kawai, Kjeldsen, Laughlin, Levine, Lin, Lissauer, MacQueen,
  Marcy, McCullough, Morton, Narita, Paegert, Palle, Pepe, Pepper, Quirrenbach,
  Rinehart, Sasselov, Sato, Seager, Sozzetti, Stassun, Sullivan, Szentgyorgyi,
  Torres, Udry, \& Villasenor]{Ricker+15jatis:Tess}
Ricker, G.~R., Winn, J.~N., Vanderspek, R., Latham, D.~W., Bakos, G.~{\'A}.,
  Bean, J.~L., {Berta-Thompson}, Z.~K., Brown, T.~M., Buchhave, L., Butler,
  N.~R., Butler, R.~P., Chaplin, W.~J., Charbonneau, D.~B.,
  {Christensen-Dalsgaard}, J., Clampin, M., Deming, D., Doty, J.~P., Lee,
  N.~D., Dressing, C., Dunham, E.~W., Endl, M., Fressin, F., Ge, J., Henning,
  T., Holman, M.~J., Howard, A.~W., Ida, S., Jenkins, J.~M., Jernigan, G.,
  Johnson, J.~A., Kaltenegger, L., Kawai, N., Kjeldsen, H., Laughlin, G.,
  Levine, A.~M., Lin, D., Lissauer, J.~J., MacQueen, P., Marcy, G., McCullough,
  P.~R., Morton, T.~D., Narita, N., Paegert, M., Palle, E., Pepe, F., Pepper,
  J., Quirrenbach, A., Rinehart, S.~A., Sasselov, D., Sato, B., Seager, S.,
  Sozzetti, A., Stassun, K.~G., Sullivan, P., Szentgyorgyi, A., Torres, G.,
  Udry, S., \& Villasenor, J., 2015.
\newblock Transiting {{Exoplanet Survey Satellite}}, {\it JATIS\/}, {\bf 1}(1),
  014003.

\bibitem[Russell(1912)]{Russell12apjI}
Russell, H.~N., 1912.
\newblock On the {{Determination}} of the {{Orbital Elements}} of {{Eclipsing
  Variable Stars}}. {{I}}., {\it ApJ\/}, {\bf 35}, 315.

\bibitem[Shallue \& Vanderburg(2018)]{ShallueVanderburg18aj}
Shallue, C.~J. \& Vanderburg, A., 2018.
\newblock Identifying {{Exoplanets}} with {{Deep Learning}}: {{A Five-planet
  Resonant Chain}} around {{Kepler-80}} and an {{Eighth Planet}} around
  {{Kepler-90}}, {\it AJ\/}, {\bf 155}(2), 94.

\bibitem[Southworth(2008)]{Southworth08mnras}
Southworth, J., 2008.
\newblock Homogeneous studies of transiting extrasolar planets -- {{I}}.
  {{Light-curve}} analyses, {\it MNRAS\/}, {\bf 386}(3), 1644--1666.

\bibitem[Southworth(2015)]{Southworth15debcat}
Southworth, J., 2015.
\newblock {{DEBCat}}: {{A Catalog}} of {{Detached Eclipsing Binary Stars}}, in
  {\em Living {{Together}}: {{Planets}}, {{Host Stars}} and {{Binaries}}\/},
  vol. 496 of {\bf Astronomical {{Society}} of the {{Pacific Conference
  Series}}}, p. 164, eprint: arXiv:1411.1219.

\bibitem[Southworth(2020)]{Southworth20obsR1}
Southworth, J., 2020.
\newblock Rediscussion of eclipsing binaries. {{Paper I}}: The
  totally-eclipsing {{B-type}} system zeta {{Phoenicis}}, {\it The
  Observatory\/}, {\bf 140}, 247--262.

\bibitem[Southworth(2021{\natexlab{a}})]{Southworth21obsR4}
Southworth, J., 2021{\natexlab{a}}.
\newblock Rediscussion of eclipsing binaries. {{Paper IV}}: {{The}} evolved
  {{O-type}} system {{AN Camelopardalis}}, {\it The Observatory\/}, {\bf 141},
  122--133.

\bibitem[Southworth(2021{\natexlab{b}})]{Southworth21obsR5}
Southworth, J., 2021{\natexlab{b}}.
\newblock Rediscussion of eclipsing binaries. {{Paper V}}: {{The}} triple
  system {{V455 Aurigae}}, {\it The Observatory\/}, {\bf 141}, 190--203.

\bibitem[Southworth(2021{\natexlab{c}})]{Southworth21obsR6}
Southworth, J., 2021{\natexlab{c}}.
\newblock Rediscussion of eclipsing binaries. {{Paper VI}}: {{The F-type}}
  system {{V505 Persei}}, {\it The Observatory\/}, {\bf 141}, 234--245.

\bibitem[Southworth(2021{\natexlab{d}})]{Southworth21obsR7}
Southworth, J., 2021{\natexlab{d}}.
\newblock Rediscussion of eclipsing binaries. {{Paper VII}}: {{Delta Scuti}},
  {{Gamma Doradus}}, and tidally-perturbed pulsations in {{RR Lyncis}}, {\it
  The Observatory\/}, {\bf 141}, 282--295.

\bibitem[Southworth(2021{\natexlab{e}})]{Southworth21univ}
Southworth, J., 2021{\natexlab{e}}.
\newblock Space-{{Based Photometry}} of {{Binary Stars}}: {{From Voyager}} to
  {{TESS}}, {\it Universe\/}, {\bf 7}(10), 369.

\bibitem[Southworth(2022)]{Southworth22obsR11}
Southworth, J., 2022.
\newblock Rediscussion of eclipsing binaries. {{Paper XI}}: {{ZZ Ursae
  Majoris}}, a solar-type system showing total eclipses and a radius
  discrepancy, {\it The Observatory\/}, {\bf 142}, 267--284.

\bibitem[Southworth(2023{\natexlab{a}})]{Southworth23obsR12}
Southworth, J., 2023{\natexlab{a}}.
\newblock Rediscussion of {{Eclipsing Binaries}}. {{Paper XII}}: {{The F-type
  Twin System ZZ Bootis}}, {\it The Observatory\/}, {\bf 143}, 19.

\bibitem[Southworth(2023{\natexlab{b}})]{Southworth23obsR13}
Southworth, J., 2023{\natexlab{b}}.
\newblock Rediscussion of {{Eclipsing Binaries}}. {{Paper XIII}}: {{The F-Type
  Twin System IT Cassiopeiae}}, {\it The Observatory\/}, {\bf 143}, 120.

\bibitem[Southworth(2023{\natexlab{c}})]{Southworth23obsR14}
Southworth, J., 2023{\natexlab{c}}.
\newblock Rediscussion of {{Eclipsing Binaries}}. {{Paper XIV}}: {{The F-Type
  System V570 Persei}}, {\it The Observatory\/}, {\bf 143}, 165.

\bibitem[Southworth(2024{\natexlab{a}})]{Southworth24obsR18}
Southworth, J., 2024{\natexlab{a}}.
\newblock Rediscussion of eclipsing binaries. {{Paper XVIII}}. {{The F-type}}
  system {{OO Pegasi}}, {\it The Observatory\/}, {\bf 144}, 133--143.

\bibitem[Southworth(2024{\natexlab{b}})]{Southworth24obsR21}
Southworth, J., 2024{\natexlab{b}}.
\newblock Rediscussion of eclipsing binaries. {{Paper XXI}}. {{The}}
  totally-eclipsing {{B-type}} system {{IQ Persei}}, {\it The Observatory\/},
  {\bf 144}, 278--289.

\bibitem[Southworth(2025{\natexlab{a}})]{Southworth25obsR22}
Southworth, J., 2025{\natexlab{a}}.
\newblock Rediscussion of eclipsing binaries. {{Paper XXII}}. {{The B-type}}
  system {{MU Cassiopeiae}}, {\it The Observatory\/}, {\bf 145}, 26--37.

\bibitem[Southworth(2025{\natexlab{b}})]{Southworth25obsR23}
Southworth, J., 2025{\natexlab{b}}.
\newblock Rediscussion of eclipsing binaries. {{Paper XXIII}}. {{The F-type}}
  twin system {{RZ Chamaeleontis}}, {\it The Observatory\/}, {\bf 145}, 52--64.

\bibitem[Southworth(2025{\natexlab{c}})]{Southworth25obsR24}
Southworth, J., 2025{\natexlab{c}}.
\newblock Rediscussion of eclipsing binaries. {{Paper XXIV}}. {{The}} delta
  {{Scuti}} pulsator {{V596 Pup}} (formerly known as {{VV Pyx}}), {\it The
  Observatory\/}, {\bf 145}, 102--116.

\bibitem[Southworth(2025{\natexlab{d}})]{Southworth25obsR25}
Southworth, J., 2025{\natexlab{d}}.
\newblock Rediscussion of eclipsing binaries. {{Paper XXV}}. {{The}}
  chemically-peculiar system {{AR Aurigae}}, {\it The Observatory\/}, {\bf
  145}, 138--149.

\bibitem[Southworth \& Bowman(2022)]{SouthworthBowman22mnras}
Southworth, J. \& Bowman, D.~M., 2022.
\newblock High-mass pulsators in eclipsing binaries observed using {{TESS}},
  {\it MNRAS\/}, {\bf 513}(3), 3191--3209.

\bibitem[Southworth \& Van~Reeth(2022)]{SouthworthVanReeth22mnras}
Southworth, J. \& Van~Reeth, T., 2022.
\newblock Four bright eclipsing binaries with {$\gamma$}\,{{Doradus}} pulsating
  components: {{CM}}\,{{Lac}}, {{MZ}}\,{{Lac}}, {{RX}}\,{{Dra}}, and
  {{V2077}}\,{{Cyg}}, {\it MNRAS\/}, {\bf 515}(2), 2755--2765.

\bibitem[Southworth et~al.(2004)Southworth, Maxted, \&
  Smalley]{Southworth+04a:mnras:jktebop}
Southworth, J., Maxted, P. F.~L., \& Smalley, B., 2004.
\newblock Eclipsing binaries in open clusters -- {{I}}. {{V615 Per}} and {{V618
  Per}} in h {{Persei}}, {\it MNRAS\/}, {\bf 349}(2), 547--559.

\bibitem[Southworth et~al.(2005)Southworth, Smalley, Maxted, Claret, \&
  Etzel]{SouthworthSmalley+05mnras}
Southworth, J., Smalley, B., Maxted, P. F.~L., Claret, A., \& Etzel, P.~B.,
  2005.
\newblock Absolute dimensions of detached eclipsing binaries - {{I}}. {{The}}
  metallic-lined system {{WW Aurigae}}, {\it MNRAS\/}, {\bf 363}, 529--542.

\bibitem[Southworth et~al.(2021)Southworth, Bowman, \&
  Pavlovski]{SouthworthBowman+21mnras}
Southworth, J., Bowman, D.~M., \& Pavlovski, K., 2021.
\newblock A {$\beta$} {{Cephei}} pulsator and a changing orbital inclination in
  the high-mass eclipsing binary system {{VV Orionis}}, {\it MNRAS\/}, {\bf
  501}, L65--L70.

\bibitem[Southworth et~al.(2023)Southworth, Murphy, \&
  Pavlovski]{SouthworthMurphy+23mnrasl}
Southworth, J., Murphy, S.~J., \& Pavlovski, K., 2023.
\newblock {$\delta$}\,{{Scuti}} pulsations in the bright {{Pleiades}} eclipsing
  binary {{HD}}\,23642, {\it MNRASL\/}, {\bf 520}(1), L53--L57.

\bibitem[Spada et~al.(2013)Spada, Demarque, Kim, \& Sills]{SpadaDemarque+13apj}
Spada, F., Demarque, P., Kim, Y.~C., \& Sills, A., 2013.
\newblock The {{Radius Discrepancy}} in {{Low-mass Stars}}: {{Single}} versus
  {{Binaries}}, {\it ApJ\/}, {\bf 776}, 87.

\bibitem[Srivastava et~al.(2014)Srivastava, Hinton, Krizhevsky, Sutskever, \&
  Salakhutdinov]{Srivastava+14MLdropout}
Srivastava, N., Hinton, G., Krizhevsky, A., Sutskever, I., \& Salakhutdinov,
  R., 2014.
\newblock Dropout: {{A Simple Way}} to {{Prevent Neural Networks}} from
  {{Overfitting}}, {\it Journal of Machine Learning Research\/}, {\bf 15}(56),
  1929--1958.

\bibitem[Stassun et~al.(2019)Stassun, Oelkers, Paegert, Torres, Pepper, Lee,
  Collins, Latham, Muirhead, Chittidi, {Rojas-Ayala}, Fleming, Rose, Tenenbaum,
  Ting, Kane, Barclay, Bean, Brassuer, Charbonneau, Ge, Lissauer, Mann, McLean,
  Mullally, Narita, Plavchan, Ricker, Sasselov, Seager, Sharma, Shiao,
  Sozzetti, Stello, Vanderspek, Wallace, \& Winn]{StassunOelkers+19aj}
Stassun, K.~G., Oelkers, R.~J., Paegert, M., Torres, G., Pepper, J., Lee,
  N.~D., Collins, K., Latham, D.~W., Muirhead, P.~S., Chittidi, J.,
  {Rojas-Ayala}, B., Fleming, S.~W., Rose, M.~E., Tenenbaum, P., Ting, E.~B.,
  Kane, S.~R., Barclay, T., Bean, J.~L., Brassuer, C.~E., Charbonneau, D., Ge,
  J., Lissauer, J.~J., Mann, A.~W., McLean, B., Mullally, S., Narita, N.,
  Plavchan, P., Ricker, G.~R., Sasselov, D., Seager, S., Sharma, S., Shiao, B.,
  Sozzetti, A., Stello, D., Vanderspek, R., Wallace, G., \& Winn, J.~N., 2019.
\newblock The {{Revised TESS Input Catalog}} and {{Candidate Target List}},
  {\it AJ\/}, {\bf 158}(4), 138.

\bibitem[Swayne et~al.(2021)Swayne, Maxted, Triaud, Sousa, Broeg, Flor{\'e}n,
  Guterman, Simon, Boisse, Bonfanti, Martin, Santerne, Salmon, Standing,
  Van~Grootel, Wilson, Alibert, Alonso, Anglada~Escud{\'e}, Asquier,
  B{\'a}rczy, Barrado, Barros, Battley, Baumjohann, Beck, Beck, Bekkelien,
  Benz, Billot, Bonfils, Brandeker, Busch, Cabrera, Charnoz, Collier~Cameron,
  Csizmadia, Davies, Deleuil, Deline, Delrez, Demangeon, Demory, Dransfield,
  Ehrenreich, Erikson, Fortier, Fossati, Fridlund, Futyan, Gandolfi, Gillon,
  Guedel, H{\'e}brard, Heidari, Hellier, Heng, Hobson, Hoyer, Isaak, Kiss,
  Kunovac~Hod{\v z}i{\'c}, Lalitha, Laskar, {Lecavelier~des~Etangs}, Lendl,
  Lovis, Magrin, Marafatto, McCormac, Miller, Nascimbeni, Olofsson, Ottensamer,
  Pagano, Pall{\'e}, Peter, Piotto, Pollacco, Queloz, Ragazzoni, Rando, Rauer,
  Ribas, Santos, Scandariato, S{\'e}gransan, Smith, Steinberger, Steller,
  Szab{\'o}, Thomas, Udry, Walter, Walton, \& Willett]{SwayneMaxted+21mnras}
Swayne, M.~I., Maxted, P. F.~L., Triaud, A. H. M.~J., Sousa, S.~G., Broeg, C.,
  Flor{\'e}n, H.-G., Guterman, P., Simon, A.~E., Boisse, I., Bonfanti, A.,
  Martin, D., Santerne, A., Salmon, S., Standing, M.~R., Van~Grootel, V.,
  Wilson, T.~G., Alibert, Y., Alonso, R., Anglada~Escud{\'e}, G., Asquier, J.,
  B{\'a}rczy, T., Barrado, D., Barros, S. C.~C., Battley, M., Baumjohann, W.,
  Beck, M., Beck, T., Bekkelien, A., Benz, W., Billot, N., Bonfils, X.,
  Brandeker, A., Busch, M.-D., Cabrera, J., Charnoz, S., Collier~Cameron, A.,
  Csizmadia, S., Davies, M.~B., Deleuil, M., Deline, A., Delrez, L., Demangeon,
  O. D.~S., Demory, B.-O., Dransfield, G., Ehrenreich, D., Erikson, A.,
  Fortier, A., Fossati, L., Fridlund, M., Futyan, D., Gandolfi, D., Gillon, M.,
  Guedel, M., H{\'e}brard, G., Heidari, N., Hellier, C., Heng, K., Hobson, M.,
  Hoyer, S., Isaak, K.~G., Kiss, L., Kunovac~Hod{\v z}i{\'c}, V., Lalitha, S.,
  Laskar, J., {Lecavelier~des~Etangs}, A., Lendl, M., Lovis, C., Magrin, D.,
  Marafatto, L., McCormac, J., Miller, N., Nascimbeni, V., Olofsson, G.,
  Ottensamer, R., Pagano, I., Pall{\'e}, E., Peter, G., Piotto, G., Pollacco,
  D., Queloz, D., Ragazzoni, R., Rando, N., Rauer, H., Ribas, I., Santos,
  N.~C., Scandariato, G., S{\'e}gransan, D., Smith, A. M.~S., Steinberger, M.,
  Steller, M., Szab{\'o}, G.~M., Thomas, N., Udry, S., Walter, I., Walton,
  N.~A., \& Willett, E., 2021.
\newblock The {{EBLM}} project -- {{VIII}}. {{First}} results for {{M-dwarf}}
  mass, radius, and effective temperature measurements using {{CHEOPS}} light
  curves, {\it MNRAS\/}, {\bf 506}(1), 306--322.

\bibitem[Tkachenko et~al.(2020)Tkachenko, Pavlovski, Johnston, Pedersen,
  Michielsen, Bowman, Southworth, Tsymbal, \& Aerts]{TkachenkoPavlovski+20aa}
Tkachenko, A., Pavlovski, K., Johnston, C., Pedersen, M.~G., Michielsen, M.,
  Bowman, D.~M., Southworth, J., Tsymbal, V., \& Aerts, C., 2020.
\newblock The mass discrepancy in intermediate- and high-mass eclipsing
  binaries: {{The}} need for higher convective core masses, {\it A\&A\/}, {\bf
  637}, A60.

\bibitem[Torres et~al.(2000)Torres, Lacy, Claret, \& Sabby]{TorresLacy+00aj}
Torres, G., Lacy, C. H.~S., Claret, A., \& Sabby, J.~A., 2000.
\newblock Absolute {{Dimensions}} of the {{Unevolved B-Type Eclipsing Binary GG
  Orionis}}, {\it AJ\/}, {\bf 120}, 3226--3243.

\bibitem[Torres et~al.(2010)Torres, Andersen, \&
  Gim{\'e}nez]{TorresAndersen+10aarv}
Torres, G., Andersen, J., \& Gim{\'e}nez, A., 2010.
\newblock Accurate masses and radii of normal stars: Modern results and
  applications, {\it A\&ARv\/}, {\bf 18}(1/2), 67--126.

\bibitem[Torres et~al.(2014)Torres, Sandberg~Lacy, Pavlovski, Feiden, Sabby,
  Bruntt, \& Viggo~Clausen]{TorresSandbergLacy+14apj}
Torres, G., Sandberg~Lacy, C.~H., Pavlovski, K., Feiden, G.~A., Sabby, J.~A.,
  Bruntt, H., \& Viggo~Clausen, J., 2014.
\newblock The {{G}}+{{M Eclipsing Binary V530 Orionis}}: {{A Stringent Test}}
  of {{Magnetic Stellar Evolution Models}} for {{Low-mass Stars}}, {\it ApJ\/},
  {\bf 797}, 31.

\bibitem[Torres et~al.(2019)Torres, Cantero, {Rebassa-Mansergas}, Skorobogatov,
  {Jim{\'e}nez-Esteban}, \& Solano]{TorresCantero+19mnras}
Torres, S., Cantero, C., {Rebassa-Mansergas}, A., Skorobogatov, G.,
  {Jim{\'e}nez-Esteban}, F.~M., \& Solano, E., 2019.
\newblock Random {{Forest}} identification of the thin disc, thick disc, and
  halo {{Gaia-DR2}} white dwarf population, {\it MNRAS\/}, {\bf 485},
  5573--5589.

\bibitem[Van~Reeth et~al.(2022)Van~Reeth, Southworth, Van~Beeck, \&
  Bowman]{VanReethSouthworth+22aa}
Van~Reeth, T., Southworth, J., Van~Beeck, J., \& Bowman, D.~M., 2022.
\newblock V456 {{Cyg}}: {{An}} eclipsing binary with tidally perturbed g-mode
  pulsations, {\it A\&A\/}, {\bf 659}, A177.

\bibitem[{Wang} et~al.(2021){Wang}, {Fu}, {Niu}, {Pan}, {Li}, {Zong}, \&
  {Hou}]{WangFu+21mnras:starspots}
{Wang}, J., {Fu}, J., {Niu}, H., {Pan}, Y., {Li}, C., {Zong}, W., \& {Hou}, Y.,
  2021.
\newblock {KIC 5359678: a detached eclipsing binary with starspots}, {\it
  \mnras\/}, {\bf 504}(3), 4302--4311.

\bibitem[{Wang} et~al.(2022){Wang}, {Fu}, {Zong}, {Pan}, {Niu}, {Zhang}, \&
  {Zhang}]{WangFu+22mnras:starspots}
{Wang}, J., {Fu}, J., {Zong}, W., {Pan}, Y., {Niu}, H., {Zhang}, B., \&
  {Zhang}, Y., 2022.
\newblock {Properties and evolutions of starspots on three detached eclipsing
  binaries in the LAMOST-Kepler survey}, {\it \mnras\/}, {\bf 511}(2),
  2285--2301.

\bibitem[Wang et~al.(2024)Wang, Ding, Li, Xiong, Cheng, \& Ji]{WangDing+24apjs}
Wang, J., Ding, X., Li, J., Xiong, J., Cheng, Q., \& Ji, K., 2024.
\newblock A {{Method}} of {{Rapidly Deriving Late-type Contact Binary
  Parameters}} and {{Its Application}} in the {{Catalina Sky Survey}}, {\it
  ApJS\/}, {\bf 273}(2), 31.

\bibitem[Wells \& Pr{\v s}a(2021)]{WellsPrsa21apjs}
Wells, M.~A. \& Pr{\v s}a, A., 2021.
\newblock Building and {{Calibrating}} the {{Binary Star Population Using
  Kepler Data}}, {\it ApJS\/}, {\bf 253}, 32.

\bibitem[Wrona \& Pr{\v s}a(2025)]{WronaPrsa25apjs}
Wrona, M. \& Pr{\v s}a, A., 2025.
\newblock The {{Eclipsing Binaries}} via {{Artificial Intelligence}}. {{II}}.
  {{Need}} for {{Speed}} in {{PHOEBE Forward Models}}, {\it ApJS\/}, {\bf 277},
  1.

\end{thebibliography}

	
	\appendix
	
	\section*{Appendix A: Additional figures}\label{sec:appendix-a}
\begin{figure*}
		\includegraphics[width=\linewidth]{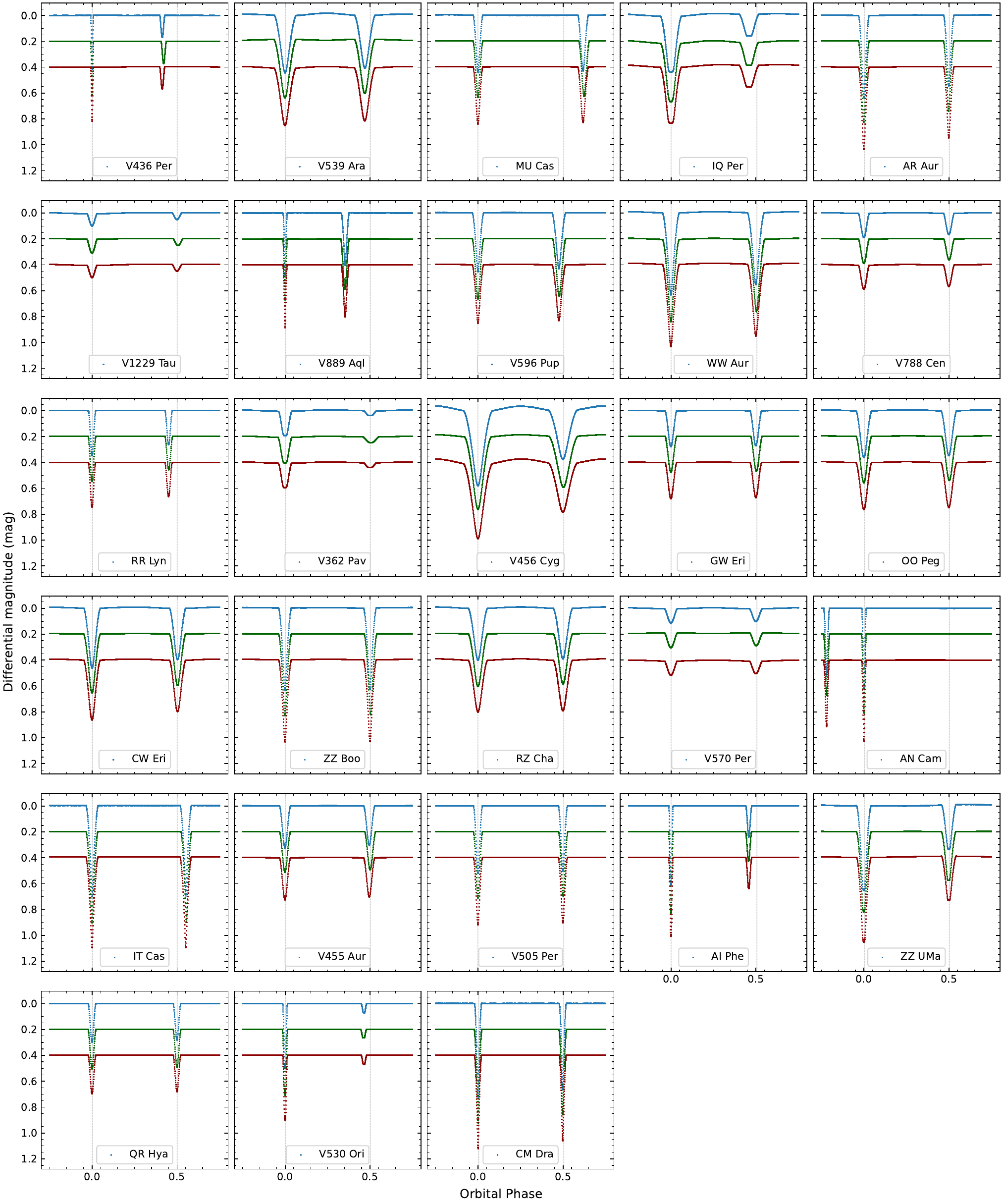}
		\caption{The phase-folded \tess\ light curve input features of each instance within the test dataset of real systems (top) and corresponding light curves generated with the fitting input parameters from 1000 MC Dropout predictions (middle) and the subsequent fitted characterisation from \jktebop\ (bottom). The latter two sets of light curves have been vertically shifted by 0.2 and 0.4 mag, respectively.}
		\label{fig:mags-features-real-test-ds}
	\end{figure*}

	Fig.~\ref{fig:mags-features-real-test-ds} shows plots of the processed and phase-folded light curve feature, from which predictions are made, for each member of the test dataset of real systems. On the same axes as each input light curve are two corresponding light curves, the first generated with the fitting input parameters, complete with the predicted values for $k$, \rAplusrB, $J$, $i$, \ecosw\ and \esinw\ made with 1000 MC Dropout iterations, and the second generated with the subsequent characterisation from fitting with \jktebop.
	
	\bsp	
	\label{lastpage}
\end{document}